\documentclass[10pt,journal,compsoc]{IEEEtran}
\usepackage{booktabs} % For formal tables
\usepackage{bm}
\usepackage{amsfonts}
\usepackage[pdftex]{graphicx}
\usepackage{subfigure}
\usepackage{amssymb}
\usepackage{amsmath}
\usepackage{amsthm}
\usepackage[ruled,vlined]{algorithm2e}
\usepackage{array}
\usepackage{tabularx}
\usepackage{multirow}
\DeclareMathOperator*{\argmin}{arg\,mim}
\usepackage{verbatim}
\usepackage{color}
\usepackage{url}
\usepackage{ragged2e}
\usepackage{diagbox}
\newcommand{\partitle}[1]{\medskip \noindent \textbf{#1.}}
\begin{document}

\title{Efficient Contour Computation of Group-based Skyline}

\author{Wenhui~Yu, Jinfei~Liu,~\IEEEmembership{Member,~IEEE,} Jian~Pei,~\IEEEmembership{Fellow,~IEEE,} \\Li~Xiong,~\IEEEmembership{Member,~IEEE,} Xu~Chen, and Zheng~Qin
\IEEEcompsocitemizethanks{
\IEEEcompsocthanksitem Wenhui Yu, Xu Chen, and Zheng Qin are with the School of Software, Tsinghua University, Beijing, P.R. China. E-mail: \{yuwh16, xu-ch14\}@mails.tsinghua.edu.cn, qingzh@mail.tsinghua.edu.cn. (Corresponding author: Zheng Qin)
\IEEEcompsocthanksitem Jinfei Liu and Li Xiong are with Department of Mathematics and Computer Science, Emory University. E-mail: \{jinfei.liu, lxiong\}@emory.edu.
\IEEEcompsocthanksitem Jian Pei is with School of Computing Science, Simon Fraser University. E-mail: jpei@cs.sfu.ca.
\IEEEcompsocthanksitem Wenhui Yu and Jinfei Liu contribute equally in this work
}% <-this % stops an unwanted space
\thanks{Manuscript received 24 Jun. 2018; revised 20 Nov. 2018; Accepted 03 Mar. 2019.}}

\markboth{Journal of \LaTeX\ Class Files,~Vol.~14, No.~8, August~2015}%
{Shell \MakeLowercase{\textit{et al.}}: Bare Demo of IEEEtran.cls for Computer Society Journals}

\IEEEtitleabstractindextext{%
\begin{abstract}
\justifying
Skyline, aiming at finding a Pareto optimal subset of points in a multi-dimensional dataset, has gained great interest due to its extensive use for multi-criteria analysis and decision making. The skyline consists of all points that are not dominated by any other points. It is a candidate set of the optimal solution, which depends on a specific evaluation criterion for optimum. However, conventional skyline queries, which return individual points, are inadequate in group querying case since optimal combinations are required. To address this gap, we study the skyline computation in the group level and propose efficient methods to find the Group-based skyline (G-skyline). For computing the front $l$ skyline layers, we lay out an efficient approach that does the search concurrently on each dimension and investigates each point in the subspace. After that, we present a novel structure to construct the G-skyline with a queue of combinations of the first-layer points. We further demonstrate that the G-skyline is a complete candidate set of top-$l$ solutions, which is the main superiority over previous group-based skyline definitions. However, as G-skyline is complete, it contains a large number of groups which can make it impractical. To represent the ``contour'' of the G-skyline, we define the Representative G-skyline (RG-skyline). Then, we propose a Group-based clustering (G-clustering) algorithm to find out RG-skyline groups. Experimental results show that our algorithms are several orders of magnitude faster than the previous work.
\end{abstract}
}

\maketitle
\IEEEdisplaynontitleabstractindextext
\IEEEpeerreviewmaketitle

\IEEEraisesectionheading{\section{Introduction}\label{sec:introduction}}
\IEEEPARstart{S}{kyline}, known as \textit{Maxima} in computational geometry or \textit{Pareto} in business management field, is of great significance for many applications. Skyline returns a subset of points that are Pareto optimal \cite{borzsony2001the}, indicating that these points cannot be dominated by any other points in the dataset. Though the exact optimal point depends on the specific criterion for optimum, skyline can provide a candidate set so that we can prune the non-candidates from the dataset when the optimal solution is queried.

Given a dataset $S$ with $n$ points, each point $p$ has $d$ numeric attributes and can be represented as a $d$-dimensional vector $(p[1],p[2],$ $\cdots,p[d])\in\mathbb{Rd}^{d}$, where $p[i]$ is the $i$-th attribute. Given two points $p=(p[1],p[2],\cdots,p[d])$ and $p^\prime=(p^\prime[1],p^\prime[2],\cdots,$ $p^\prime[d])$, point $p$ dominates point $p^\prime$ if $p[i] < p^\prime[i]$ for at least one attribute and $p[i]\leq p^\prime[i]$ for the others ($1 \leq i \leq d$). The skyline of dataset $S$ is defined as a subset consisting of all points that are not dominated by any other points in $S$. Evidently, all the points in the skyline are Pareto optimal solutions. 

\begin{figure}[ht!]
  \centering
  \subfigure[]{
    \label{fig:examp:a} %% label for first subfigure
    \includegraphics[scale = 0.58]{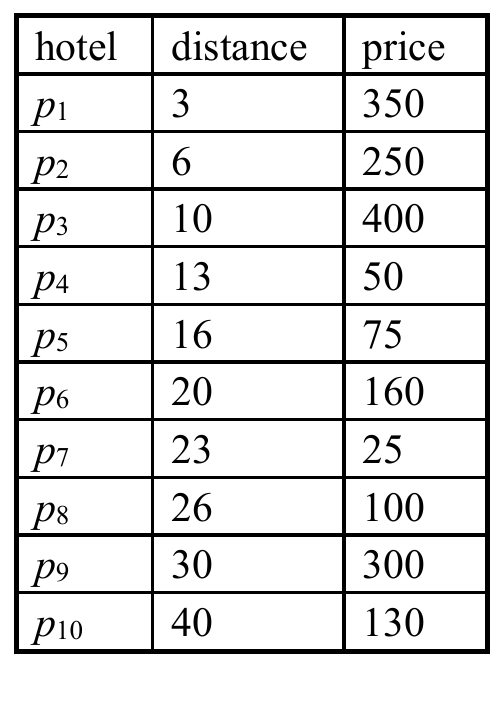}}
  \subfigure[]{
    \label{fig:examp:b} %% label for second subfigure
    \includegraphics[scale = 0.28]{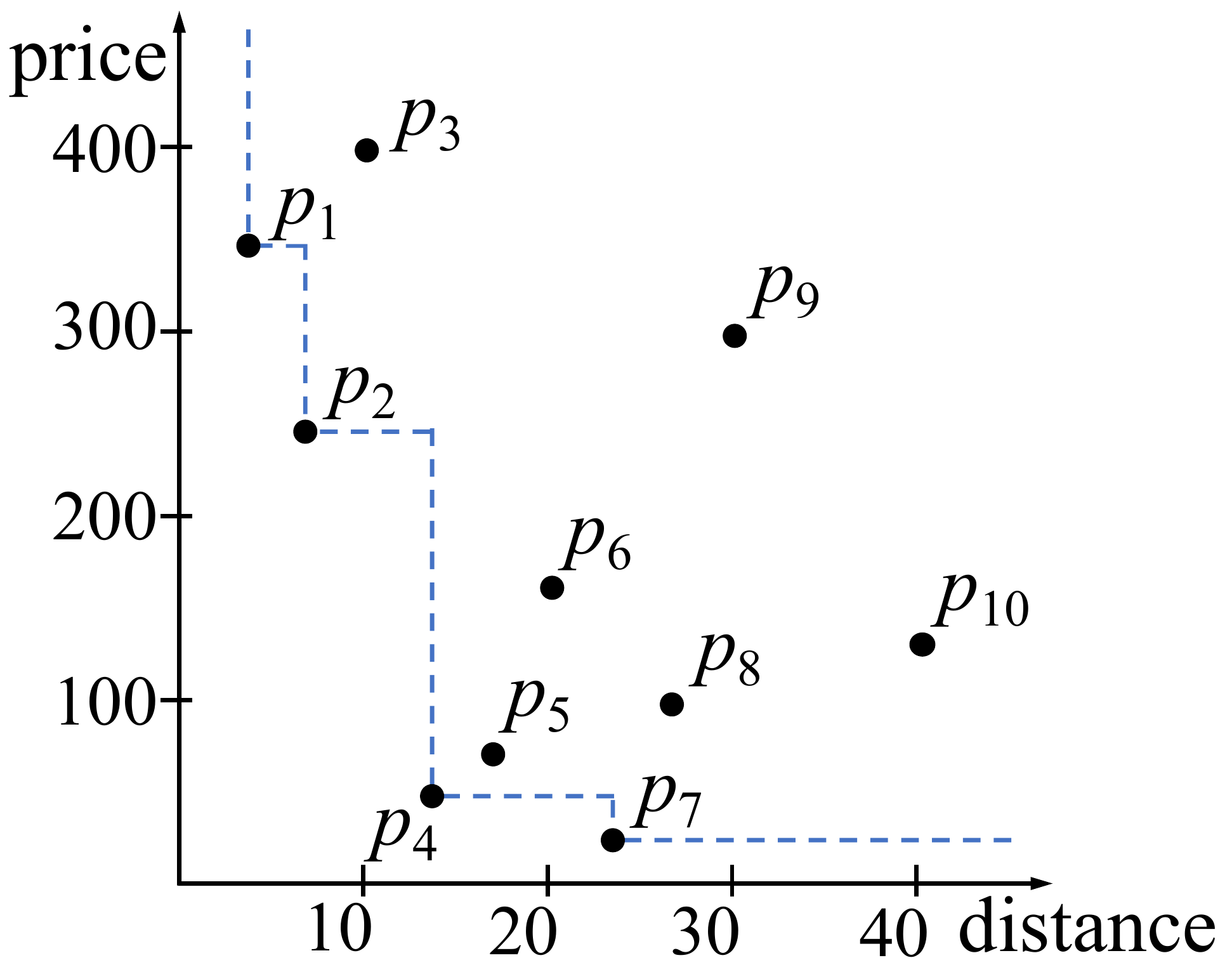}}
  \caption{A skyline example of hotels.}
  \label{fig:examp} %% label for entire figure
\end{figure}

Figure \ref{fig:examp:a} illustrates ten hotels with two attributes (the price and the distance to a given destination) in the table. Travelers desire to choose a hotel with both low price and short distance. Figure \ref{fig:examp:b} represents ten points in $2$-dimensional space and each point represents a hotel in Figure \ref{fig:examp:a} correspondingly. When choosing hotels, travelers will not benefit by choosing $p_{5}$, $p_{6}$, $p_{8}$, or $p_{10}$ because they are dominated by $p_{4}$ and are in worse situation in both distance and price attributes. Hotels $p_{1}$, $p_{2}$, $p_{4}$, and $p_{7}$ might be chosen since they are not dominated by any other hotels. The final choice depends on the travelers' criteria or the weights of attributes. For example, if the travelers are wealthy, they may attach more importance to the distance and choose $p_{1}$ or $p_{2}$. If they want to be cost-effective, they may prefer $p_{4}$ or $p_{7}$ because of the lower price. We can see that the skyline is a candidate set of the optimal solution, and the final decision depends on travelers' specific criteria.

\partitle{Motivation} Extensively studied in the database community, the skyline has been extended with many variants to provide a candidate set in different situations. However, traditional skyline, which focuses top-$1$ solutions, is inadequate when optimal groups, i.e., top-$l$ solutions, are searched rather than individual points. This is a very important problem with many real-world applications that is surprisingly neglected. To address this gap, group-based skyline (with size $l$), which is a candidate set of all top-$l$ solutions, is proposed. However, the few existing works either do not return complete candidate set \cite{im2012group, zhang2014on} or are too computationally expensive \cite{liu2015finding}, so an efficient and complete method is greatly desired.

Recalling our hotel example, we suppose that a travel agency wants to collaborate with the three best hotels. If they are mainly targeting low-end travelers, they may choose hotels $\{p_{4}, p_{7}, p_{5}\}$ or \{$p_{4}$, $p_{5}$, $p_{6}$\}. If for high-end travelers, they may choose \{$p_{1}$, $p_{2}$, $p_{3}$\}. And they may choose \{$p_{1}$, $p_{2}$, $p_{4}$\}, \{$p_{2}$, $p_{4}$, $p_{7}$\}, or \{$p_{1}$, $p_{4}$, $p_{7}$\} if they want to provide a more extensive service. In some groups mentioned above, all points contained are skyline points, e.g., \{$p_{1}$, $p_{2}$, $p_{4}$\}, \{$p_{2}$, $p_{4}$, $p_{7}$\}, \{$p_{1}$, $p_{4}$, $p_{7}$\}, but in the others, non-skyline points may be included, e.g., \{$p_{4}$, $p_{7}$, $p_{5}$\}, \{$p_{4}$, $p_{5}$, $p_{6}$\}, \{$p_{1}$, $p_{2}$, $p_{3}$\}. It is apparent that the non-skyline points can also be chosen in this problem. However, the components of these groups are not arbitrary and they need to satisfy certain requirements. For instance, groups \{$p_{4}$, $p_{7}$, $p_{8}$\} or \{$p_{4}$, $p_{7}$, $p_{6}$\} cannot be the best choice since $p_{5}$ dominates $p_{6}$ and $p_{8}$, so \{$p_{4}$, $p_{7}$, $p_{5}$\} is better than them. We can see that it becomes more complicated when finding top-$l$ solutions.

A Group-based skyline (G-skyline\footnote[1]{Note that we only use ``G-skyline'' to indicate the Pareto optimal group-based skyline proposed in \cite{liu2015finding}. For other kinds of group-based skyline \cite{im2012group,zhang2014on}, we do not use the abbreviation.}) is defined to address this problem. Consider a dataset $S$ of $n$ points in $d$-dimensional space. $G = \{p_{1}, p_{2},\cdots, p_{l}\}$ and $G' = \{p_{1}', p_{2}',\cdots, p_l'\}$ are two different groups with $l$ points. If we can find two permutations of the $l$ points for $G$ and $G'$, $G = \{p_{u_1}, p_{u_2},\cdots, p_{u_l}\}$ and $G' = \{p_{v_1}', p_{v_2}',\cdots, p_{v_l}'\}$, letting $p_{u_i}\preceq p_{v_i}'$ for all $i$ and $p_{u_i}\prec p_{v_i}'$ for at least one $i$ ($1 \leq i \leq l$), we say $G$ \textbf{g-dominates} $G'$. All groups containing $l$ points that are not g-dominated by any other groups with the same size compose the G-skyline. Similar to conventional skyline, G-skyline is a candidate set of top-$l$ solutions. Recalling our hotel example, agencies with different criteria has different optimal groups and G-skyline contains all possible choices for all agencies. We construct G-skyline to support agencies for decision making.

G-skyline is a complete candidate set for top-$l$ solutions (proved in Section \ref{sec:gsky}), but there are two critical issues, (1) it is really time-consuming and (2) it returns an oversized candidate set. To address these two problems, we (1) propose fast algorithms for the G-skyline and (2) define the Representative G-skyline (RG-skyline).

To reduce the time complexity, we optimize the algorithms proposed in \cite{liu2015finding}. We first define Multiple Skyline Layers (MSL) as the first $l$ skyline layers, which is very important in both traditional skyline and G-skyline problems. To construct MSL efficiently, our framework searches concurrently in each dimension to reduce the redundant search, and the search strategy utilizes the property of the skyline in subspace to improve efficiency. Then, based on MSL, we propose fast algorithms to construct G-skyline groups. Considering the observation that skyline points contribute more to G-skyline groups compared to non-skyline points, we use a queue to enumerate all combinations of skyline points and then find the groups that contain non-skyline points by the queue efficiently.

To reduce the output size, we introduce a distance-based representative G-skyline, RG-skyline. The key idea is to represent similar G-skyline groups with a representative group, and RG-skyline contains all these representative groups. To measure the similarity between groups, we match the points of each pair of groups and calculate the Euclidean distance between these groups. We then extend the clustering algorithm from the point level to the group level to propose a novel Group-based clustering (G-clustering) algorithm. Finally, we take the set of all cluster centers as RG-skyline.

\partitle{Contributions} 
We briefly summarize our contributions as follows:

\begin{itemize}
\item{
We propose a new algorithm to construct MSL, which is significantly more efficient than existing algorithms.}

\item{
To address the issue of high time complexity, we propose two fast algorithms to construct G-skyline groups with the combination queue.
}

\item{
To address the issue of oversized output, we define RG-skyline to represent the ``contour'' of the whole G-skyline and propose a novel G-clustering algorithm to construct it.
}

\item{We devise comprehensive experiments in synthetic and real datasets. The experimental results show that our algorithms are highly efficient and scalable.}
\end{itemize}

\partitle{Organization} The rest of the paper is organized as follows. Section \ref{sec:related_work} presents the related work. Section \ref{sec:MSL} presents the definition of MSL and the algorithmic technique to construct it. Two algorithms for finding G-skyline groups are presented in Section \ref{sec:gsky}. The definition and algorithm for the RG-skyline are illustrated in Section \ref{sec:repre}. We evaluate the performance of each algorithm in Section \ref{sec:experiments}. Section \ref{sec:conclusion} concludes the paper.

\section{Related Work}
\label{sec:related_work}
In this section, we briefly explore previous works on skyline computation. After Kung's original work that proposed the in-memory algorithms to handle the skyline computation problem \cite{kung1975on}, skyline query has attracted extensive attention over the last decade due to its significance in computational geometry and database fields \cite{kirkpatrick1985output-size,kung1975on}. \cite{borzsony2001the} first studied how to process skyline queries in database systems and devised the skyline operator. Since then, the research includes improved algorithms \cite{chomicki2003skyline,godfrey2005maximal}, progressive skyline computation \cite{papadias2005progressive,tan2001efficient}, query optimization \cite{Balke2004Efficient,jinfei_new}, top-$l$ dominating queries \cite{M1,topk1,topk2}, and variants of skyline queries \cite{chen2007privacy, Liu2017Secure1, pei2006towards, morse2007efficient}. There are also works focusing on different types of data, for example, dynamic data \cite{gomaa10computing} and uncertain data \cite{liu2015finding1,pei2007probabilistic}.

\partitle{Group-based Skyline}
In recent years, there are increasing works focusing on group-based skyline \cite{zhang2014on,im2012group,li2012on,liu2015finding}. The definition of dominance between groups in these works varies greatly. \cite{zhang2014on, im2012group, li2012on} used a single aggregate point to represent a group and find the dominance relationship of groups using the aggregate points in a traditional way. An aggregate function is responsible for generating the aggregate point, whose attribute values are aggregated over the corresponding attribute value of all points in the group. Though many aggregate functions can be considered in calculating aggregate points, only several of them, such as SUM, MIN, MAX, have been discussed in previous works. \cite{im2012group} used SUM to aggregate the points and \cite{zhang2014on} utilized MAX and MIN. Intuitively, SUM captures the collective strength of a group, while MIN/MAX compares groups by their weakest/strongest member on each attribute. The skyline groups constructed by these methods are not complete because not all Pareto optimal groups can be captured. Instead of aggregating data, \cite{liu2015finding} proposed a different notion of group-based skyline called G-skyline, the dominance relationship between two groups is defined based on the pair-wise dominance between points in the two groups. Compared with previous works, G-skyline gives a more complete solution. In fact, group-based skyline proposed in \cite{im2012group} is a subset of G-skyline in \cite{liu2015finding}. However, completeness also means high time consumption in computation thus efficient algorithms are desired. This paper is an extended version of our previous work \cite{yu2017fast}, which proposed several efficient methods for G-skyline.

\partitle{Representative Skyline} 
Due to the large output size of skyline, many research focused on representative skyline that represents the principal tradeoffs to users. \cite{max-dominance} first proposed the representative skyline by finding the skyline points that dominate the maximum number of points, which can be called max-dominance representative skyline. Similarly, \cite{max-dom-area, sliding_window} computed skyline points that maximize the dominated area or volume to show the ``contour'' of the entire skyline. \cite{distance_based} first proposed a distance-based representative skyline.  Following that, there are several works on efficient distance-based methods \cite{eff_dis_1,eff_dis_2}. Their goal is to minimize the distance between representative skyline points and non-representative skyline points (representation error).

At the group level, representative skyline has a more pressing demand due to the even larger output size. There is only one existing paper for representative skyline at the group level \cite{group_repre}. It proposed the $k$-SGQ, which is a max-dominance representative skyline, by finding $k$ groups with maximum number of dominated points. Though $k$-SGQ is efficient and easy to construct, it tends to return very balanced groups, which is against the goal of skyline to show different tradeoffs to users. To address this gap, we propose a distance-based method, RG-skyline, to show the ``contour" of the original skyline better.

\section{Constructing Multiple Skyline Layers}
\label{sec:MSL}
We define Multiple Skyline Layers (MSL) of a dataset as the first $l$ layers of skyline, where the first layer is the skyline of the original dataset and the second layer is the skyline of the remaining points after the first layer is removed from the dataset and so on. MSL is indispensable in all group-based skyline algorithms, including G-skyline \cite{liu2015finding} and aggregation approaches \cite{im2012group, zhang2014on}. It is proved that only the first $l$ layers are necessary to find group-based skyline rather than the whole dataset \cite{im2012group, zhang2014on, liu2015finding}. However, very few papers study this issue, they usually focus on how to establish the group-based skyline. MSL is also useful in traditional skyline problems \cite{lu2011flexible, li2011an, blunck2010in-place}. It can be computed layer by layer using a traditional skyline method but it is ineffective. In fact, \cite{kung1975on} showed that $O\left(n\log{n}\right)$ time is needed to construct one layer. So, constructing $l$ layers costs $O\left(ln\log{n}\right)$ running time. 

\cite{liu2015finding, blunck2010in-place, lu2011flexible} laid out several new methods that can find all layers in one enumeration. \cite{liu2015finding} proposed a Binary Search (BS) algorithm with $O\left(n+\mathbb{S}_l\log{l}\right)$ time complexity for two- and $O\left(n\mathbb{S}_l\right)$ for higher-dimensional spaces, which is the state-of-the-art algorithm for MSL. BS algorithm constructs MSL by ordering the points by one dimension and determine each point's layer in this order. It is much better but there is still great room for improvement. In this section, we investigate a high-efficiency approach to construct the first $l$ skyline layers, defined as MSL, with $O\left(\mathbb{T}_l \left(n^{\frac{d-1}{d}}l + \mathbb{S}_l \log{l} \right) \right)$ time complexity, where $\mathbb{S}_l$ is the number of the points in the first $l$ layers and $\mathbb{T}_l$ is the subspace skyline size of the $l$-th layer. Noticing $\mathbb{T}_l$ is 1 in the $2$-dimensional space, the time complexity becomes $O\left(\sqrt{n}l+\mathbb{S}_l\log{l}\right)$.

\begin{table}[ht!]
  \caption{The summary of notations.}
  \centering
  \label{tab:nontation}
  \scalebox{1}{
  \begin{tabular}{cc}
    \toprule
    Notation &Definition\\
    \midrule
    $S\backslash{layer_i}$ & points in $S$ but not in $layer_i$\\
    $p \prec/\nprec p'$ & $p$ dominates/dose not dominate $p'$\\
    $p \preceq p'$ & $p$ dominates or equals to $p'$\\
    $skyline(S)$ & the skyline of dataset S\\
    $layer(p_i)$ & the skyline layer of $p_i$\\
    $D(layer_i)$ & the dominance domain of $layer_i$ \\
    $skyline_{s}(S')$ & the skyline of $\,S^\prime$ in the subspace\\
    $D_{s}(layer_i')$ & the dominance domain of $layer_i'$ in the subspace \\
    $T_S(p)$/$T_S(G)$ & tail set of point $p$/group $G$ in sequence S\\
    $G \prec_g G'$ & group $G$ g-dominates group $G'$\\
    $GS$ & the G-skyline\\
    $RS$ & the RG-skyline\\
  \bottomrule
\end{tabular}}
\end{table}

A summary of notations is given in Table \ref{tab:nontation}. We preprocess the dataset to make the definitions and proofs later in this article concise and normative. (1) We assume the criterion function $f(\,\,)$ monotonically decreases with each attribute, or we take the opposite of this attribute. That is, for certain point $p$, the smaller $p[i]$ ($i=1,\cdots,d$) are, the better $p$ is. (2) All attributes are normalized to $[0,a]$. (3) We regard the coincidence points as one point since they are equivalent when making decision.

\newtheorem{myDef}{Definition}
\begin{myDef}[Skyline]
\label{def:skyline}
Given a dataset $S$ of $n$ points in $d$-dimensional space. Let $p$ and $p^\prime$ to be two different points in $S$, $p$ dominates $p^\prime$, denoted by $p \prec p^\prime$, if for all i, $p[i] \leq p^\prime[i]$ and for at least one i, $p[i] < p^\prime[i]$, where $p[i]$ is the $i$-th dimension of $p$ and $1 \leq i \leq d$. The skyline consists of the points that are not dominated by any other points in $S$.
\end{myDef}

\begin{myDef}[Multiple Skyline Layers (MSL)]
\label{def:MSL}
Given a dataset $S$ with $n$ points in $d$-dimensional space, MSL is a multi-layer structure. $layer_1$ is the skyline of $S$, i.e., $layer_1 = skyline\left(S\right)$ and $layer_2$ is the skyline of the complementary set of $layer_1$ in $S$, i.e., $layer_2 = skyline\left(S\backslash{layer_1}\right)$. Generally, $layer_i$ is the skyline of the complementary set of the first i-1 layers in $S$, i.e., $layer_i\!\!=\!skyline\!\left(S\backslash{\bigcup_{j=1}^{i-1}layer_j}\right)$. And MSL is a multi-layer structure defined as $\{layer_j\}_{j=1,\cdots,l}$, where $l$ is the size of the group.
\end{myDef}

\begin{figure}[ht!]
\centering
\includegraphics[scale = 0.3]{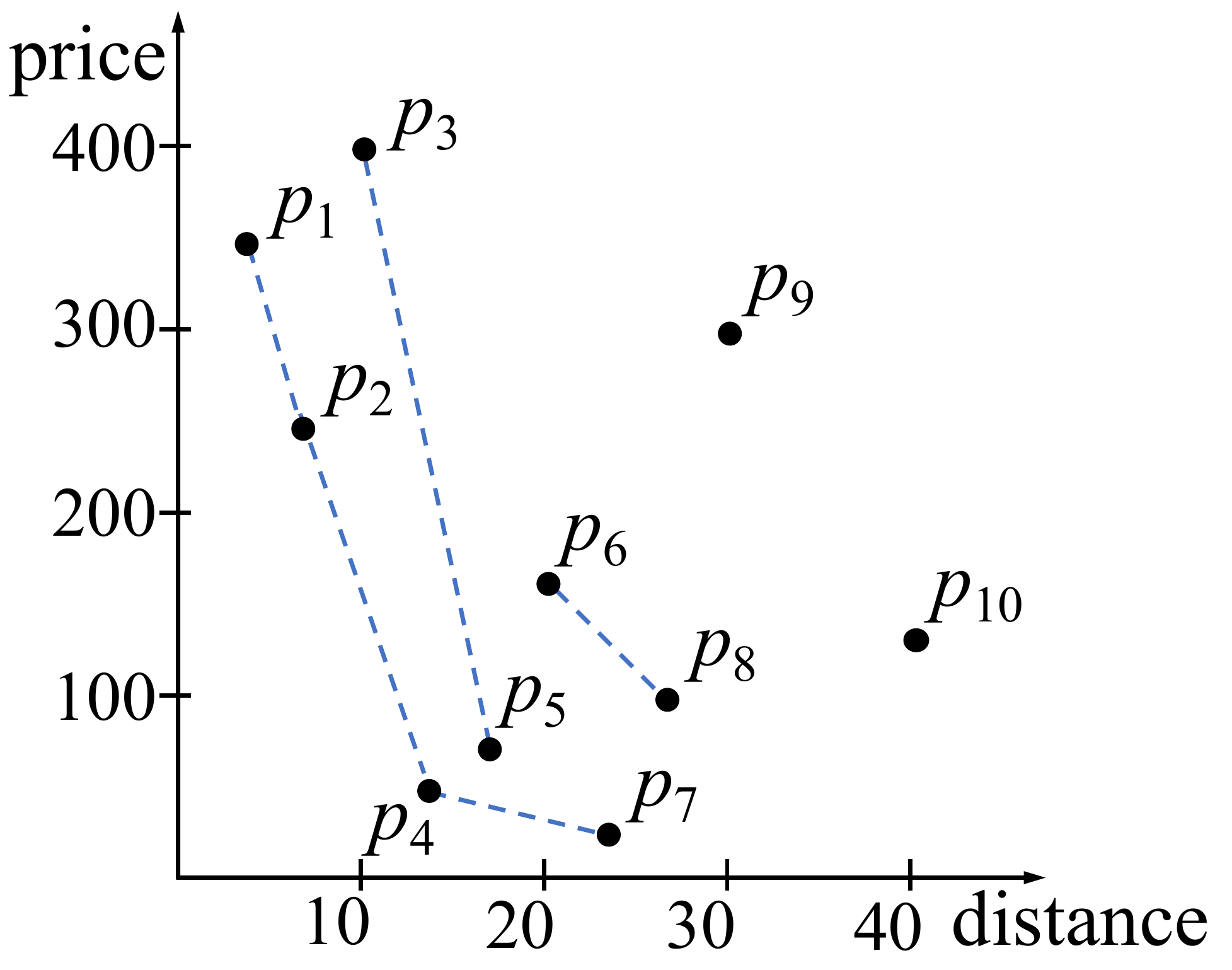}
\caption{MSL ($n = 10, d = 2, l = 3$).}
\label{fig:MSL}
\vspace{-4mm}
\end{figure}

\subsection{Simultaneous Search in Each Dimension}
\label{subsec:simul}

According to the definition of MSL, we need to establish the first $l$ layers by computing skyline $l$ times. To replace this ineffective way, we devise a novel method to find each layer's points by searching the dataset one time. Noticing that the information of each layer's points is equivalent to the information of each point's layer, we can construct each layer by finding which layer each point belongs to. The detailed procedure is introduced below.

\begin{myDef}[Dominance Domain]
\label{def:dominace_domain}
Given a dataset $S$ and a skyline layer $layer_i$, dominance domain of $layer_i$ is defined as: $D\left(layer_i\right)=S\backslash{\bigcup_{j=1}^{i}layer_j}$, or $D\left(layer_i\right)=\bigcup_{j=i+1}^{l_m}layer_j$, where $l_m$ is the maximum layer number. In $2$-dimensional case, dominance domain of $layer_i$ is the set of points in the upper right area of $layer_i$ intuitively. 
\end{myDef}

\newtheorem{myPRO}{Property}
\begin{myPRO}
\label{pro:dominace_domain}
If a point $p\in D(layer_i)$, there is at least one point $p^\prime \in layer_i$ making $p^\prime \prec p$ and if $p\notin D(layer_i)$, there is no point $p^\prime \in layer_i $, $p^\prime \prec p$. 
\end{myPRO}

\newtheorem*{myPRF}{Proof}
\begin{myPRF}
When $i>1$, $layer_i$ is the skyline of $D(layer_{i-1})$. For certain $p\in D(layer_i)$, there is at least one sequence $\{p_1, p_2, \cdots, p_m\}$ in $D(layer_{i-1})$, satisfying $p_m\preceq\cdots\preceq p_1\preceq p$. Since no point in $D(layer_{i-1})$ can dominate $p_m$, $p_m\in layer_i$. For certain $p\notin D(layer_i)$, then $p\in layer_j(j\leq i)$ and no points in $layer_i$ can dominate it. We can get the same result when $i=1$.
\end{myPRF}

\begin{myPRO}
\label{pro:point_layer}
The layer of point $p$ is equal to the maximal layer of the points that dominate $p$ plus 1, i.e., $layer\left(p \right)=\max \{layer\left(p^\prime \right) \arrowvert p^\prime \prec p \}+1$, and $layer(p) = 1$ if $\{p^\prime \arrowvert p^\prime \prec p \}=\varnothing$. 
\end{myPRO}

\begin{myPRF}
According to Definition \ref{def:dominace_domain} and Property \ref{pro:dominace_domain}, for any point $p$ in $layer_i$, where $2\leq i\leq l$. For each $m_1$ ($1\leq m_1\leq i-1$), since $p\in D(layer_{m_1})$, there must be at least one point in $layer_{m_1}$ that dominates $p$. And for each ${m_2}$ ($i\leq {m_2}\leq l$), since $p\notin D(layer_{m_2})$, no point in $layer_{m_2}$ can dominates $p$. We can see that, all points that can dominate $p$ belong to the first i-1 layers and at least one in each. So, $\max \{layer\left(p^\prime \right) \arrowvert p^\prime \prec p \}=i-1$.
\end{myPRF}

\newtheorem{myLAM}{Lamma}
\begin{myLAM}
To determine the layer of certain point $p$, we just need to compare it with all points in a hyper-cuboid space region, which is in the range $[0, p[i]]$ for each $i$ ($1 \leq i \leq d$). This region contains all points that can dominate $p$ (Taking Figure \ref{fig:examp:b} as an example, the lower left area of the current point $p$).
\end{myLAM}

\newtheorem{myEXP}{Example}
\begin{myEXP}
Considering the point $p_{8}$ in Figure \ref{fig:MSL}, we just need to search the points in its lower left area ($p_{4}$, $p_{5}$, $p_{7}$) to determine $layer(p_{8})$. According to Property \ref{pro:point_layer}, $layer(p_{8}) = \max \limits_{ i=4,5,7 }\left\{layer\left(p_i\right)\right\} + 1 = 3$.
\end{myEXP}

\begin{algorithm}[ht]
\caption{Concurrent MSL algorithm}%算法名字
\label{alg:msl_1}
\LinesNumbered %要求显示行号
\small\KwIn{a set of $n$ points in $d$ dimensional space ($ S = \{p_1, p_2, \cdots,p_n\}$) and the group size $l$ .}%输入参数
\KwOut{$l$-layer $MSL$.}%输出

\For{$i = 1$ to $n$}{
    $p_i.times = 0$\;
    $p_i.layer = 1$\;
}
$maxlayer(1,\cdots,d) = 1$\; %$\varnothing$\;
$MSL(1,\cdots,d)(1,\cdots,l) = \varnothing$\;
\For{$i = 1$ to $d$}{
    sort the $n$ points by the \textit{i}-th dimension in ascending order and record the order of the indexes $O_i = \{o_{i1}, o_{i2},\cdots,o_{in}\}$\;
}
$flag = 0$\;
\For{$i = 1$ to $n$}{
    \For{$j = 1$ to $d$}{
        $p_{o_{ji}}.times \;+\!\!= 1$\;
        \If{$p_{o_{ji}}.times == d$ \&\&\ $p_{o_{ji}}.layer == l$}{
            $flag=1$\;
            \textbf{break}\;
        }
        \If{${p_{o_{ji}}.times == 1 }$ }
        {
            \For{$layer\_num = 0$ to $maxlayer(j)-1$}{
                \If{there is a point in $MSL(j)(maxlayer(j)-$
                $layer\_num)$ can dominate $p_{o_{ji}}$}{
                    $p_{o_{ji}}.layer = maxlayer(j)$ $-$ $layer\_num$ $+$ $1$\;
                    \textbf{break}\;
                }
            }
        }
        \If{$p_{o_{ji}}.layer \leq l$}{
            store $p_{o_{ji}}$ to $MSL(j)(p_{o_{ji}}.layer)$\;
            \If{$p_{o_{ji}}.layer>maxlayer(j)$}{
                $maxlayer(j) = p_{o_{ji}}.layer$\;
            }
        }
    }
    \If{$flag == 1$}{
        \textbf{break}\;
    }
}
\Return $MSL$\;
\end{algorithm}

We propose a framework for MSL based on the above properties. We determine a given point's layer by Property \ref{pro:point_layer} and in order to find all points that can dominate $p$, we only need to check the hyper-cuboid region (the lower-left area in two dimensional case) according to Lemma 1. So our main idea is to slide a hyperplane over the points along each dimension concurrently and for each point visited, determine its layer and store it into the corresponding layer. The detailed algorithm is shown in Algorithm \ref{alg:msl_1}. We order all points $d$ times by all $d$ attributes (lines 6-7) and for each attribute, slide a hyperplane along the axis point by point in an increasing order (lines 9-25). When a point is visited by a hyperplane, compare it with all points that are already visited by this hyperplane to determine if it is dominated by any of them, and then label it with the maximal layer number of the dominating points plus one (lines 15-19). The point is then stored in the corresponding dimension and layer (lines 20-23). When a point in the $l$-th layer has been visited by all $d$ hyperplanes, the iteration stops (lines 12-14 and 24-25). We show example steps in Example \ref{exam_1}.

\begin{figure}[ht!]\vspace{-1mm}
  \centering
  \subfigure[]{
    \label{fig:ESA1:a} %% label for first subfigure
    \includegraphics[scale = 0.3]{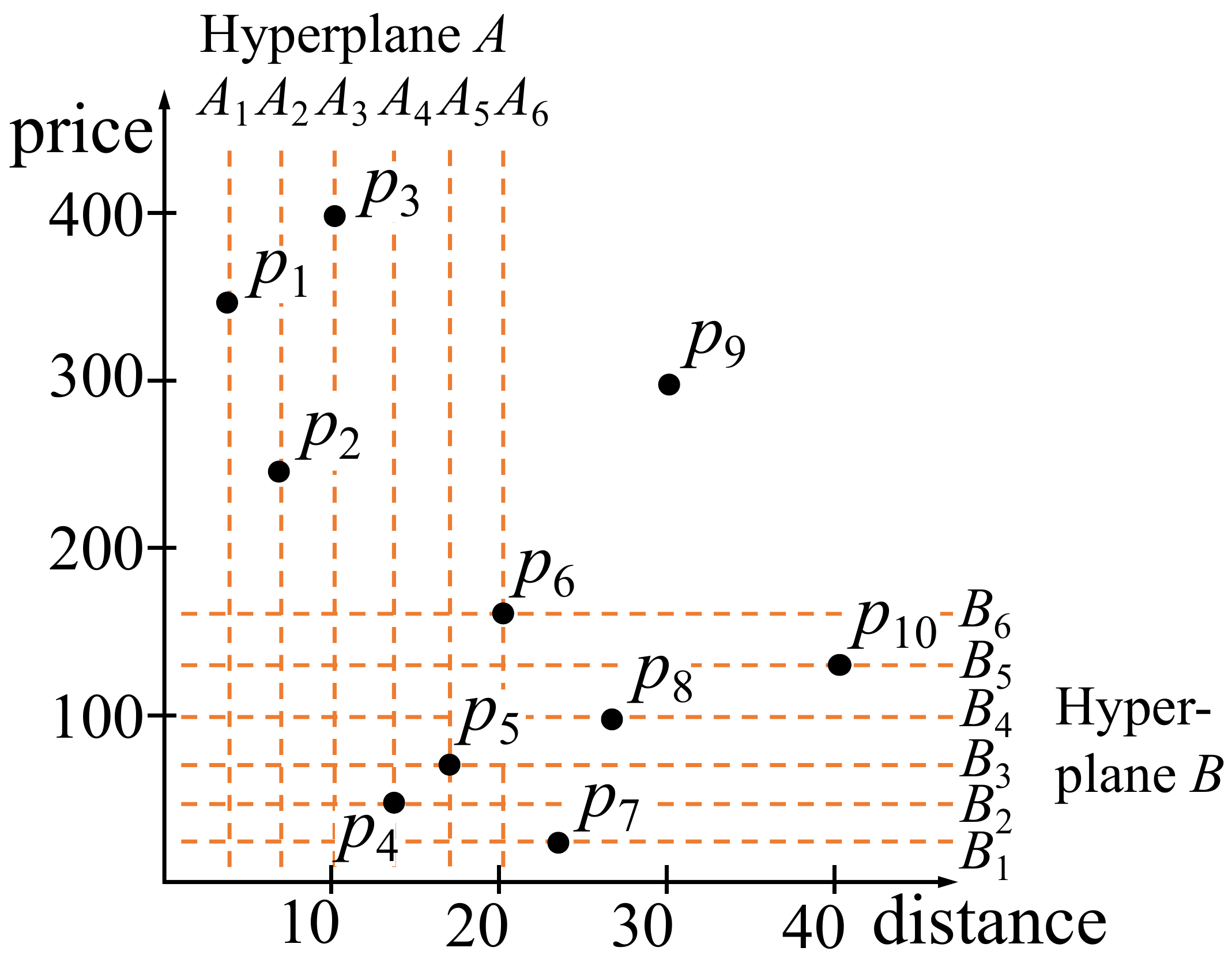}}
  \hspace{0.12in}
  \subfigure[]{
    \label{fig:ESA1:b} %% label for second subfigure
    \includegraphics[scale = 0.7]{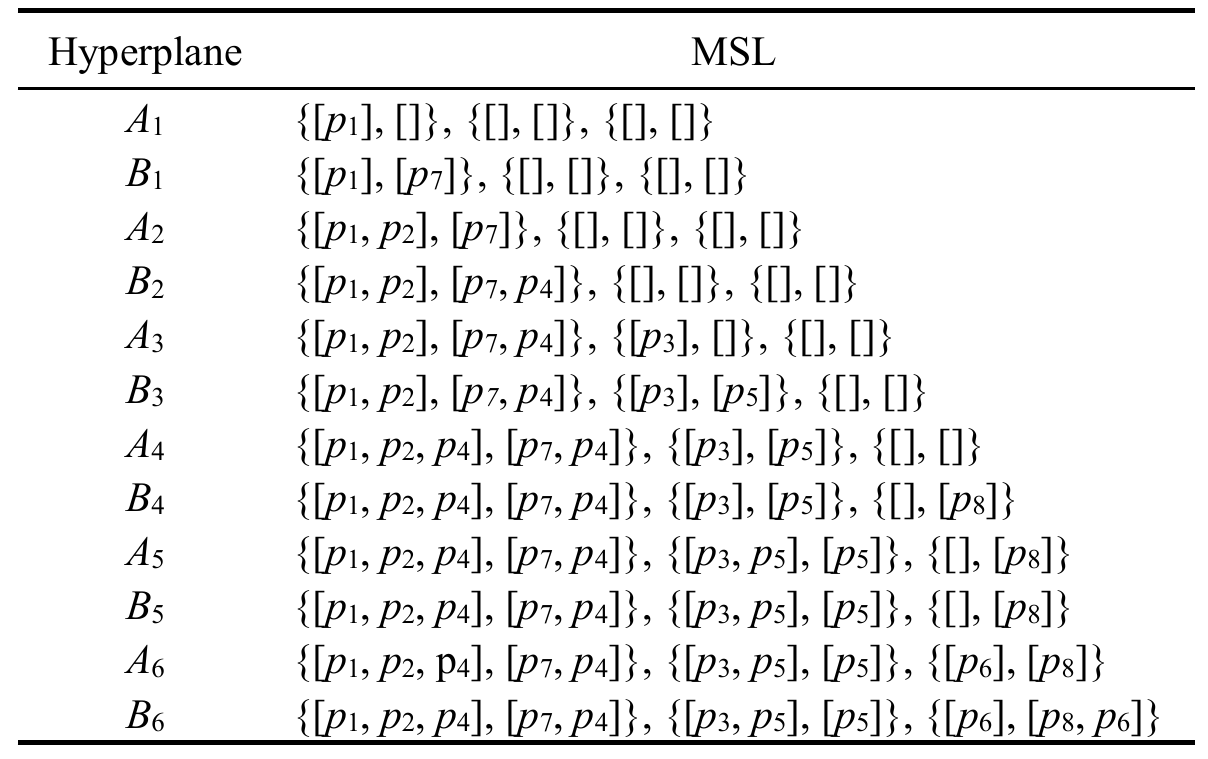}}
  \caption{Example steps of Algorithm 1.}
  \label{fig:ESA1} %% label for entire figure
\end{figure}

\begin{myEXP}
\label{exam_1}
Figure \ref{fig:ESA1} illustrates the steps of constructing the MSL presented in Figure \ref{fig:MSL}. Figure \ref{fig:ESA1:a} shows the location of the hyperplanes during each iteration. Figure \ref{fig:ESA1:b} shows the intermediate result at each iteration ($\left[ \,\, \right]$ indicates one dimension and $\{ \,\, \}$ indicates one layer). The algorithm stops at $B_6$ since the current point $p_6$ is in the $l$-th layer and has been scanned $d$ times by all hyperplanes. From this example, we can see that some points, e.g., $p_{4}$, $p_{5}$, and $p_{6}$, are visited twice by both hyperplanes and can be compared against different points for dominance. Will these comparisons return the same and correct result? Taking $p_{6}$ as an example, when visited by $A_6$, it is compared with the points in the left area, notated with $S_1 = \{p_{1}, p_{2}, p_{3}, p_{4}, p_{5}\}$. Similarly, when visited by $B_6$, it is compared with points in the area below, notated with $S_2 = \{p_{4}, p_{5}, p_{7}, p_{8}, p_{10}\}$. While according to Lemma 1, we only need to check the points in the lower left area, notated with $S_0 = \{p_{4}, p_{5}\}$. Because $S_0 \in S_1$ and $S_0 \in S_2$, we can draw the same conclusion by checking either of the two sets. Consequently, we only check when a point is first visited by a hyperplane. When it is visited again by other hyperplanes, we only save it in the corresponding dimension and layer without checking again.
\end{myEXP}

\partitle{Advancement} Benefiting from our novel structure, there are two main advancements in the proposed algorithm.

One advancement is that we reduce the number of points that need to be compared significantly. Similar to BS algorithm, our algorithm also prunes the comparison set of each point by ordering all points at the beginning. Besides this, we implement concurrent comparison in all dimensions, hence the number of points to investigate can be reduced significantly.

Another advancement is that our work provides an explicit condition to stop the iterations. It is proved that only the points in the first $l$ layers are necessary to construct group-based skylines, so the procedure can stop when points in the first $l$ layers are all found and labeled. In previous works, it is very complicated to determine when to stop. In fact, in BS algorithm \cite{liu2015finding}, authors do not terminate the procedure until all points are enumerated in dataset $S$. But in our work, we have an explicit condition to stop the algorithm: when a point in the $l$-th layer is labeled by all $d$ hyperplanes, we can terminate the procedure.

\begin{myEXP}
Figure \ref{fig:comp} gives several examples of $3$-layer MSL in an independent (INDE) synthetic dataset (points are even-distributed and independent among all attributes) in $2$-dimensional space. Shadow areas in Figures \ref{fig:comp}(a)(b)(c) represent the points need to be investigated in BS method and shadow areas in Figures \ref{fig:comp}(d)(e)(f) represent the points need to be investigated in our works. The investigation areas of our method are two narrow bands. When there is a point labeled by the two hyperplanes appearing in the third layer, the procedure can be terminated.
\end{myEXP}

%\vspace{-3mm}
\partitle{Efficiency Comparison} We estimate the number of points that need to be investigated in BS method \cite{liu2015finding} and in our concurrent MSL algorithm to demonstrate the efficiency improvement. To simplify the problem, we assume the dataset is INDE. A simple case in $2$-dimensional space is investigated first and then the conclusion is extended to high-dimensional spaces. We use $A_1$ to represent the area of shadow regions in Figures \ref{fig:comp}(a)(b)(c) and $A_2$ to represent that in Figures \ref{fig:comp}(d)(e)(f). We use $b$ and $c$ to represent the width of them (shown in Figure \ref{fig:comp}). And the value of each attribute is normalized into $[0, a]$ in the dataset preprocessing. 

First, we give an analysis of BS method in $2$-dimensional space. We define a random variable $X_1$ to indicate the maximum of the first attribute of the points in $layer_1$. According to Definition \ref{def:MSL}, we know that $X_1\!\!=\!\!\{p_i[1]|i\!\!=\!\!\argmin \limits_j p_j[2],p_j \!\!\in \!\! S\}$, so $X_1\!\!\sim\!\! U(0,a)$ due to the independence of the two dimensions. We discuss in the most optimistic situation for the lower bound of $b$. We can ignore the influence of removing $layer_1$ when $n \!\!\gg\!\! l$ and simply regard the complementary set $D(layer_1)$ as an even-distributed and independent dataset too. Similarly, we notate $X_2\!\!=\!\!\{p_i[1]|i\!\!=\!\!\argmin \limits_j p_j[2],p_j \!\!\in\!\! D(layer_1)\}$ to present the maximum of the first attribute of the points in $layer_2$. It obeys uniform distribution approximately. To extend this, we deem that all $X_i\!\!\sim\!\! U(0,a)$, $i\!=\!1,2,\dots,l$ when $n \!\gg\! l$. $X$ is a random variable defined as $X\!\!=\!\!\max\{X_1,X_2,\dots,X_l\}$ and $b=E(X)=\frac{l}{l+1}a$.

\begin{myPRF}
The probability distribution function of random variable $X$ can be represented as
%\vspace{-2mm}
\begin{align*}
F_X(x)=&P(X\leq x) = P(\max\{X_1,X_2,\dots,X_l\}\leq x)\\
=&P(X_1\leq x)P(X_2\leq x)\dots P(X_l\leq x)=\left(\frac{x}{a}\right)^l,
\end{align*}
accordingly the probability density function is
$$f_X(x)=F'_X(x)=\frac{lx^{l-1}}{a^l},$$
then we can get the expectation by integrating 
\begin{align*}
b=&E(x)=\int_0^a xf(x)dx=\int_0^a \frac{lx^l}{a^l}dx\\=&\frac{l}{a^l}\left(\frac{x^{l+1}}{l+1}\right)\bigg\arrowvert_0^a=\frac{l}{l+1}a.
\end{align*}
\end{myPRF}

We then discuss in the most pessimistic situation to get the upper bound. $n$ is not big enough that we cannot just ignore the influence of removing $layer_1$ when investigating $X_2$. We consider that $layer_1$ captures most points with a small value of the second attribute and $X_2$ is less likely to be located in the left area that loses these points after removing $layer_1$. In the most extreme situation, $X_2$ has no possibility of falling in the left area and we can get $X_2\sim U(x_1, a)$. Similarly, $Xi\sim U(x_{i-1}, a)$, $1<i\leq l$. Let $X=\max\{X_1,X_2,\dots,X_l\}=X_l$, we can get $b = E(X) = (1-0.5^l)a$.

In a more general situation, $b\in \left[\frac{l}{l+1}a,(1-0.5^l)a\right]$. Since $A_1 = a \times b$, $A_1\in \left[\frac{l}{l+1}a^2,(1-0.5^l)a^2\right]$. The conclusion is then extended to the high-dimensional situation. Using $A_1$ to present the volume of the hyper-cuboid area swept by the hyperplane, we can easily get $A_1\in \left[\frac{l}{l+1}a^d,(1-0.5^l)a^d\right]$.

\begin{figure}[ht!]
\setlength{\abovecaptionskip}{2mm}
  \vspace{-2mm}
  \centering
  \includegraphics[scale = 0.3]{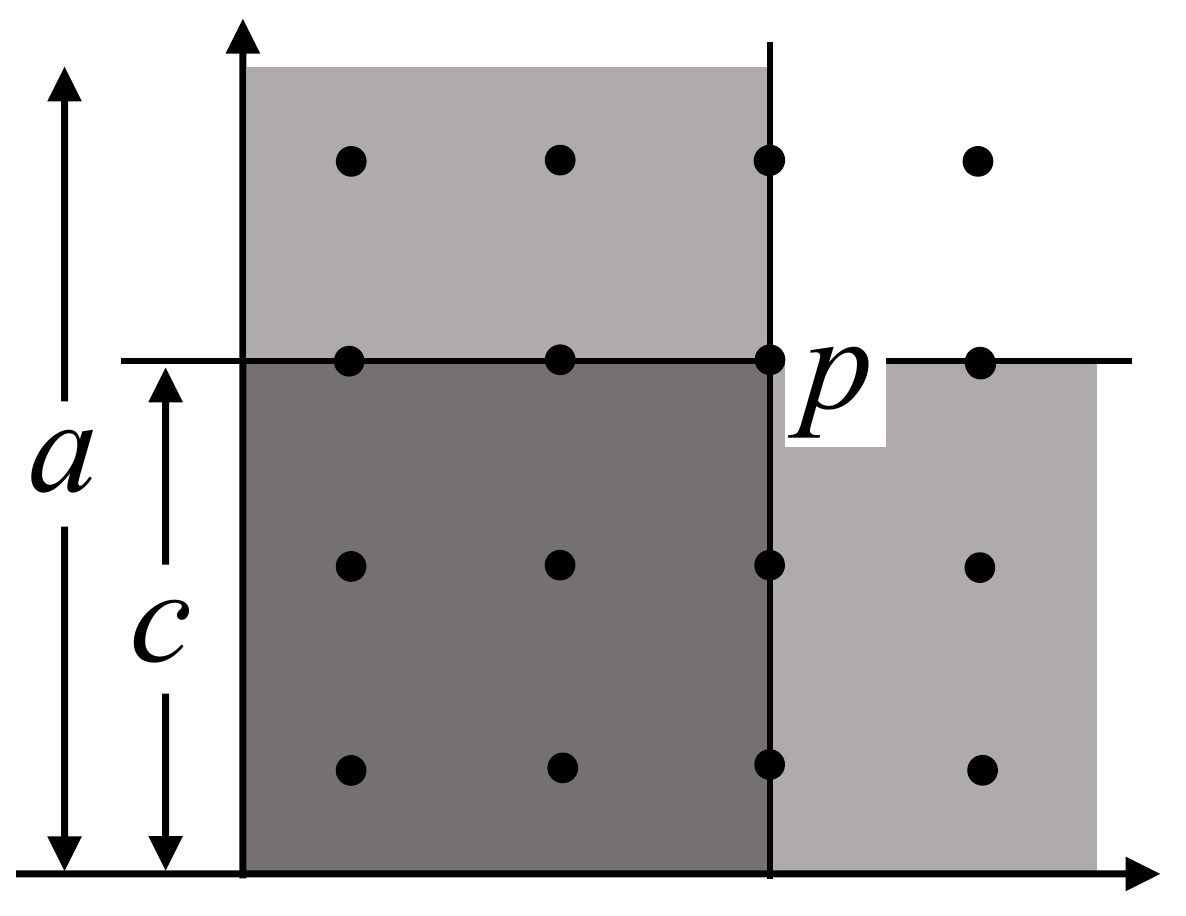}
  \caption{A model for efficiency analysis.}
  \label{fig:model}
  \vspace{-2mm}
\end{figure}

Calculating the shadow areas in Figures \ref{fig:comp}(d)(e)(f) is a very tough task, we demonstrate a very simple model shown in Figure \ref{fig:model} to give a rough estimation. Also, we first study the problem in a $2$-dimensional situation and then extend it to high-dimensional space. The shadow area is swept by the two hyperplanes, which intersect each other at point $p$. Obviously, the value of $c$ is crucial in this problem and the case of the $c\times c$ square area depends on $l$ only. Assuming there are $m$ points on the edge of this area and we get the relationship $2(m-1)+1=l$ when $l$ is odd ($m=3$, $l=5$ in this example). Considering that points on the boundary just have 50\% probability to fall in the area, the average number of points in the $c\times c$ area can be expressed as $n'= m^2 - (m-1) = \frac{l^2+3}{4} $. Because the density of points in the whole $a\times a$ area is constant, $c = \sqrt{\frac{n'}{n}}a $. When $l\ll n$, the investigation area in our work $A_2=2ac-c^2\approx \sqrt{\frac{l^2+3}{n}}a^2$. We can get the same conclusion when $l$ is even.

Now considering a high-dimensional case, we need to study the $c^d$ hypercube area. $l$ is the number of points on the edge of this area and we have $d(m-1)+1=l$ similarly. With the relationship $c=\sqrt[d]{\frac{n'}{n}}a$, $n'= m^d - \frac{d}{2}\cdot m^{d-1} $, and $A_2=d\cdot c a^{d-1}-(d-1)c^d$, we can get $A_2\approx\frac{l}{\sqrt[d]{n}}a^d$ when $l$, $n$, $m$ are large and the overlap of hyper-cuboids are neglected. In the G-skyline problem, $n\gg l$, hence $A_2\ll A_1$, we can see that our new algorithm can reduce the cost of time highly.

\subsection{Subspace Skyline Searching}
\label{subsec:subspace}

We have introduced a framework to construct the MSL in Subsection \ref{subsec:simul}, but the search technique for each point is still inadequate. Comparing with all the points visited already for dominance when processing each new point is not efficient. In this section, we illustrate some detailed searching strategies including utilizing skyline in the subspace and binary search to make searches more efficient.

A hyperplane mentioned in Subsection \ref{subsec:simul} is indeed a $d -1$ dimensional subspace, for instance, a line in planar space and a plane in $3$-dimensional space. When processing a point $p$, we can project it and the points that we need to compare against to the subspace. We utilize the properties of dominance in the subspace to construct the MSL. Of special note is that though points in certain $layer_i$, which is the skyline of $D(layer_{i - 1})$, cannot dominate each other in original space, but their projections can since one attribute is neglected in the subspace.

\iffalse
\begin{myDef}[Subspace Skyline]
For a set of points $S=\{p_1,\cdots,p_n\}$, $S'=\{p_1',\cdots,p_n'\}$ are their projects in the subspace. The skyline of $S'$ in the subspace is defined as the subspace skyline, denoted as $skyline_s(S')$
\end{myDef}
\fi

\vspace{-2mm}
\begin{figure}[ht!]
\centering
\includegraphics[scale = 0.25]{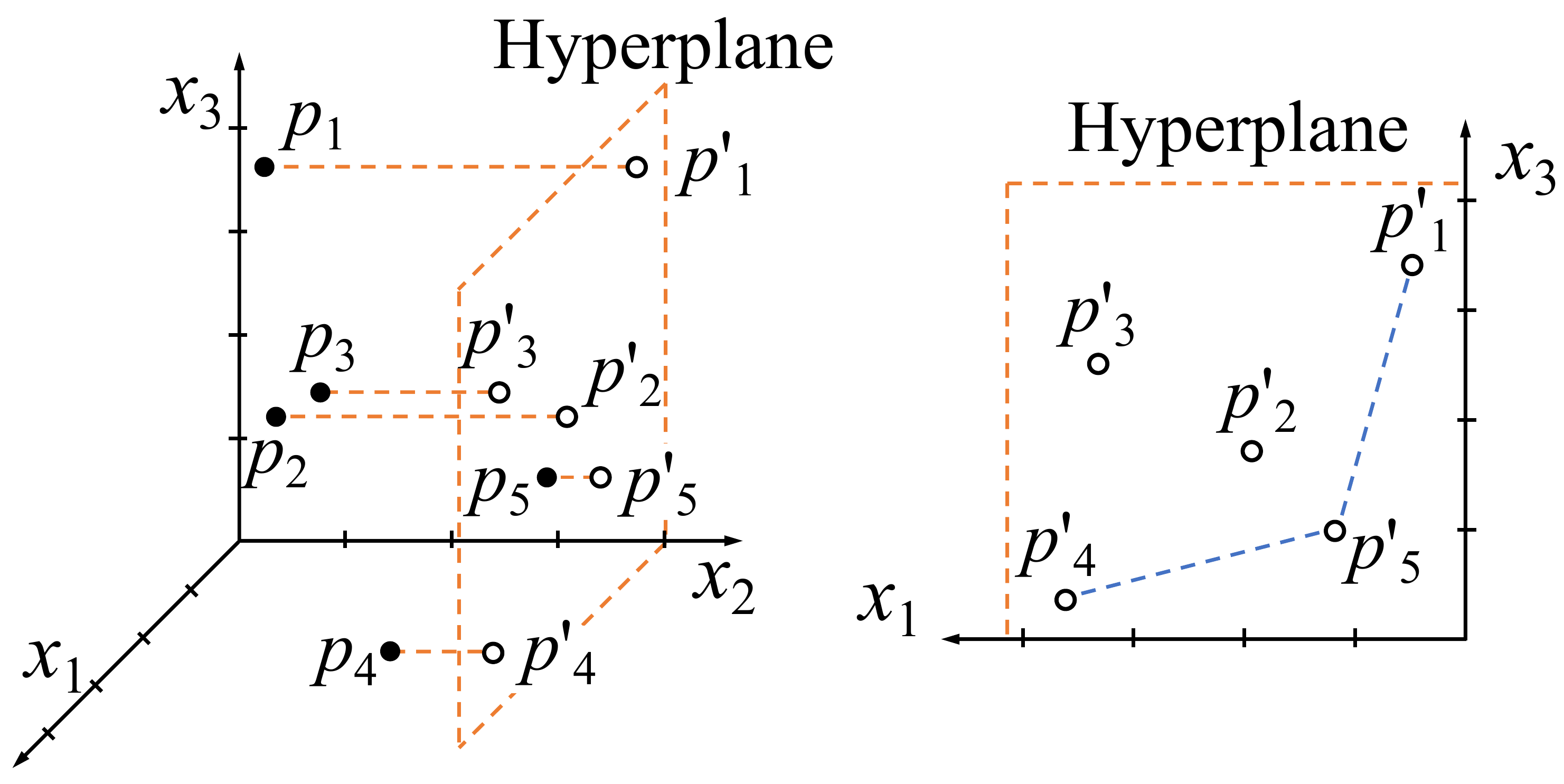}
%\vspace{-1em}
\caption{Points in original space and projected to the subspace.}
\label{fig:subspace}
\end{figure}

\begin{myPRO}
\label{pro:subspace}
When points are ordered increasingly by certain attribute $i$ and point $p_{2}$ is behind $p_{1}$, $p_{1}'$ and $p_{2}'$ are the projections of these two points in the $i$-th hyperplane. If $p_{1}'\prec p_{2}'$, then $p_{1}\prec p_{2}$. If $p_{1}'\nprec p_{2}'$, then $p_{1}\nprec p_{2}$. 
\end{myPRO}
\vspace{-1em}
\begin{myPRF}
Since point $p_{2}$ is behind $p_{1}$, we know that $p_{1}[i] \leq p_{2}[i]$. And since $p_{1}'\prec p_{2}'$, we know that $p_{1}[j] \leq p_{2}[j]$ ($j=1,\cdots, i-1, i+1,\cdots, m$), and $p_{1}[j] < p_{2}[j]$ for at least one j, so we know that $p_{1}\prec p_{2}$. Similarly, when $p_{1}'\nprec p_{2}'$, we can infer that $p_{1}\nprec p_{2}$.
\end{myPRF}

\newtheorem{myTHM}{Theorem}
\begin{myTHM}
When investigating if current point can be dominated by certain points in the current layer in our framework, we just need to compare its projection with the subspace skyline of this layer. 
\end{myTHM}
\vspace{-1em}
\begin{myPRF}
According to Property \ref{pro:subspace}, when searching if there is a point in certain layer can dominate current point, we just need to execute this procedure in the subspace. Notate current point as $p_0$ and the current layer as $layer_i$. According to the definition of skyline, for certain point $p_{1}'$ not in $skyline_{s}(layer_i')$, there must be some points $p_{2}',\cdots, p_m'$ in $layer_i'$ making $p_m'\preceq\cdots p_{2}'\preceq p_{1}'$. Since no point can dominate $p_m'$, $p_m'\in skyline_{s}(layer_i')$. So, if there is a point $p_{1}'$ that can dominate current point $p_0'$, there must be at least one corresponding skyline point $p_m'$ dominating $p_0'$, and vice versa. So, when investigating $p_0'$, only points in $skyline_{s}(layer_i')$ need to be compared.
\end{myPRF}

\begin{algorithm}[ht]
\label{alg:msl_2}
\caption{MSL with subspace skyline.}%算法名字
\LinesNumbered %要求显示行号
\small\KwIn{a set of $n$ points $S$ in $d$-dimensional space and the group size $l$.}%输入参数
\KwOut{$l$-layer $MSL$.}%输出

sort the $n$ points by the first dimension in ascending order $S = \{p_{o_1}, p_{o_2},\cdots, p_{o_n}\}$\;
$p_{o_1}.layer = 1$\;
$max\_layer = 1$\;
$sub\_skyline(1, \cdots, l) = \varnothing$\;
store $p_{o_1}$ to $sub\_skyline(1)$\;
$MSL = sub\_skyline$\;

\For{$i = 2$ to $n$}{
    \If{$p_{o_i}$ is in $D_{s}(sub\_skyline(max\_layer))$}{
        $p_{o_i}.layer = +\!+max\_layer$\;
        $max\_layer = \min\{max\_layer, l\}$\;
    }
    \ElseIf{$p_{o_i}$ is not in $D_{s}(sub\_skyline(1))$}{
        $p_{o_i}.layer = 1$\;
    }
    \Else{
        use binary search to find $j (2 \leq j \leq max\_layer)$ letting $p_{o_i}$ is in $D_{s}(sub\_skyline(j-1))$ but not in $D_{s}(sub\_skyline(j))$\;
        $p_{o_i}.layer = j$\;
    }
    \If{$p_{o_i}.layer\leq l$}{
        \If{there are points in $sub\_skyline(p_{o_i}.layer)$ dominated\footnotemark[2] by $p_{o_i}$}{
            delete these points from $sub\_skyline(p_{o_i}.layer)$;
        }
        store $p_{o_i}$ to $sub\_skyline(p_{o_i}.layer)$\;
        store $p_{o_i}$ to $MSL(p_{o_i}.layer)$\;
    }
}
\Return $MSL$\;
\end{algorithm}

\footnotetext[2]{All dominance relationships in Algorithm 2 are in the subspace.}

Consolidating theories above, we give the detailed algorithm in Algorithm \ref{alg:msl_2}. To make the introduction more concise, we use the framework of BS algorithm, where the points are ordered by only one dimension. In this algorithm, we use a binary search to label the current point (lines 7-20). When searching if the current point is in certain layer, compare its projection with the skyline of this layer's projection in the subspace (lines 8, 11, and 14, where $D_{s}(\,\,)$ is the dominance domain in the subspace). Renew the subspace skyline after each search (lines 16-19) and store the point to the corresponding layer (line 20). In fact, the tail point in BS algorithm is a special case in a $2$-dimensional dataset of our subspace skyline.

\vspace{-2mm}
%%%%%%%%%%%%%这是调整表格格式
\newcolumntype{L}[1]{>{\raggedright\arraybackslash}p{#1}}
\newcolumntype{C}[1]{>{\centering\arraybackslash}p{#1}}
\newcolumntype{R}[1]{>{\raggedleft\arraybackslash}p{#1}}

\begin{table}[ht!]
    \caption{Example steps of Algorithm 2.}
    \vspace{-1em}
    \centering
    \label{tab:exap_algor2}
    \scalebox{1}{
    \begin{tabular}{L{1cm}L{1.7cm}L{1.2cm}L{2.5cm}}
        \toprule
        Current point & Skyline of $layer_1'$ & Skyline of $layer_2'$ & MSL \\
        \midrule
        $p_1$ & $\{p_1'\}$ & $\{\}$ & $\{p_1\}, \{\}$\\
        $p_2$ & $\{p_1',p_2'\}$ & $\{\}$ & $\{p_1,p_2\}, \{\}$\\
        $p_3$ & $\{p_1',p_2'\}$ & $\{p_3'\}$ & $\{p_1,p_2\}, \{p_3\}$\\
        $p_4$ & $\{p_1',p_2',p_4'\}$ & $\{p_3'\}$ & $\{p_1,p_2,p_4\}, \{p_3\}$\\
        $p_5$ & $\{p_1',p_4',p_5'\}$ & $\{p_3'\}$ & $\{p_1,p_2,p_4,p_5\}, \{p_3\}$\\
        \bottomrule
    \end{tabular}}
\end{table}
%\vspace{-2mm}

\begin{myEXP}
Table \ref{tab:exap_algor2} shows the steps of Algorithm 2 with the example in Figure \ref{fig:subspace} ($n = 5$, $d = 3$, $l = 2$). Taking $p_{5}$ as an example, we compare $p_{5}'$ with the subspace skyline of each layer ($\{p_{1}', p_{2}', p_{4}'\}$ and $\{p_{3}'\}$). Once it is labeled and added to the first layer, we renew the subspace skyline of the first layer (adding $p_{5}'$ and removing $p_{2}'$, since it is dominated by $p_{5}'$).
\end{myEXP}

Recalling the Property \ref{pro:subspace}, we know that $p_{1}'\nprec p_{2}'$ then $p_{1}\nprec p_{2}$. However, $p_2$ may dominate $p_1$ (when $p_2'\prec p_1'$ and $p_2'[i] = p_1'[i]$). In this case, we need to make sure $p_2$ is ordered in front of $p_1$, or the skyline returned by Algorithm \ref{alg:msl_2} will be incorrect. So when ordering points by the $i$-th attribute and there are several points with the same value of this attribute, we need to find another attribute to order them increasingly. If there are still points with the same value, we keep this procedure till the order is determined.

\subsection{MSL Algorithm}
\label{subsec:MSL_alg}

We introduce a framework based on simultaneous search in each dimension in Subsection \ref{subsec:simul} and detailed searching strategies in Subsection \ref{subsec:subspace}. The complete algorithm for MSL is a combination of the two algorithms. We search the points concurrently in all dimensions. In each dimension, we use binary search and subspace skyline to investigate each point.

\partitle{Running Time} According to previous analysis in Subsection \ref{subsec:simul}, the number of points we need to investigate is $\frac{A_2}{a^d} n = n^{\frac{d-1}{d}}l$. For those points that are not in the first $l$ layers, we only need to compare it with the subspace skyline of the $l$-th layer. There are $\mathbb{T}_l$ points in it, so this part costs $O\left(\mathbb{T}_ln^{\frac{d-1}{d}}l\right)$. For each point $p_j$ in the first $l$ layers, we need to search and label it, the time complexity is $O\left(\sum_{i\in BS_j} \mathbb{T}_i\right)$, where $BS_j$ is the set of layer index in the binary search for $p_j$ and $\left|BS_j\right| = \log{l}$. When $n \gg \mathbb{S}_l$, we can ignore the influence of the first $i$ layers ($1\leq i\leq l$) and regard the complementary set $D(layer_i)$ as an even-distributed and independent dataset. In this situation, all $\mathbb{T}_i$, $1\leq i\leq l$, are approximately the same. Accordingly, the time complexity for investigating one points is $O\left(\mathbb{T}_l \log{l}\right)$ and the total cost of the second part is$\,O\left(\mathbb{T}_l \mathbb{S}_l\log{l}\right)$. Ignoring the preprocessing time, our algorithm requires $O\left(\mathbb{T}_l\left(n^{\frac{d-1}{d}}l+\mathbb{S}_l\log{l}\right)\right)$ time in total for constructing MSL. Noticing that $\mathbb{T}_l=1$ in the $2$-dimensional space, the time complexity becomes $O\left(\sqrt{n}l+\mathbb{S}_l\log{l}\right)$.

We note that the skyline layers are closely related to the $l$-skyband \cite{papadias2005progressive} (we use $l$ to represent the depth instead of $k$). The differences between $l$-skyband and MSL were discussed in \cite{liu2015finding}. In fact, $(l - 1)$-skyband is the subset of $l$-layer MSL. There are two reasons we use MSL instead of skyband: (1) We can design efficient algorithms to construct MSL, as we shown in this section. (2) The more important reason is that MSL is a multi-layer structure. Points in the same layer cannot dominate each other. Points in the low-layer can dominate points in the high-layer, but not vice versa. These properties are very important to construct G-skyline, as shown in next section.

\section{Finding G-skyline Groups}
\label{sec:gsky}

In this section, we devise a novel structure for finding G-skyline groups.

\begin{myDef}[G-Skyline]
\label{def:G-skyline}
Given a dataset $S$ of $n$ points in $d$-dimensional space. $G = \{p_{1}, p_{2},\cdots, p_{l}\}$ and $G' = \{p_{1}', p_{2}',\cdots,$ $ p_l'\}$ are two different groups with $l$ points. If we can find two permutations of the $l$ points for $G$ and $G'$, $G = \{p_{u_1}, p_{u_2},\cdots, p_{u_l}\}$ and $G' = \{p_{v_1}', p_{v_2}',\cdots, p_{v_l}'\}$, letting $p_{u_i}\preceq p_{v_i}'$ for all $i$ and $p_{u_i}\prec p_{v_i}'$ for at least one $i$ ($1 \leq i \leq l$), we say $G$ \textbf{g-dominates} $G'$, denoted by $G \prec_g G'$. All groups containing $l$ points that are not g-dominated by any other groups with the same size compose G-skyline, denoted by $GS$ \cite{liu2015finding}.
\end{myDef}

Obviously, the Brute-force method for G-skyline is extremely time-consuming and far from practical \cite{liu2015finding}. We demonstrate some important properties of G-skyline groups that can be utilized to construct the G-skyline.

\begin{myPRO}
\label{suf_nec}
For certain $l$-size group $G$, the necessary and sufficient conditions of $G\in GS$ is that for all $p_i \in G$ and all $p_j \prec p_i$ in dataset $S$, $p_j \in G$. 
\end{myPRO}
\vspace{-1em}
\begin{myPRF}\partitle{Necessity} By contradiction, for certain group $G\in GS$ and point $p_i \in G$, assume there is a point $p_j$ satisfying $p_j\notin G$ and $p_j\prec p_i$. We construct a new group $G'$ by using $p_j$ to replace $p_i$, then $G' \prec_g G$ since $p_j$ dominates $p_i$ and all the other points are the same, which contradicts the G-Skyline definition. We come to the conclusion, for all points in a G-skyline group, all its parents are in this group.

\partitle{Sufficiency} By contradiction, we assume that there exits a group $G$ satisfying for all $p_i \in G$ and all parents of $p_i$ are in $G$, and $G$ is not a G-skyline group. According to Definition \ref{def:G-skyline}, we have $G\notin GS$, so there is a group $G'$ g-dominating $G$ and two permutations of these groups, $G' = \{p_{u_1}', p_{u_2}',\cdots, p_{u_l}'\}$ and $G = \{p_{v_1}, p_{v_2},\cdots, p_{v_l}\}$, letting $p_{u_i}'\preceq p_{v_i}$ and $p_{u_i}'\prec p_{v_i}$ for at least one $i$ ($1\leq i\leq l$). We know that $p_{u_i}'$ is a parent node of $p_{v_i}$, so $p_{u_i}'\in G$, and assume it is notated as $p_{v_j}$ in $G$. Considering $G' \prec_g G$, there is a point $p_{u_j}'\in G'$ making $p_{u_j}'\preceq p_{v_j}$. Since $p_{u_i}'=p_{v_j}$, we can get $p_{u_j}'\neq p_{v_j}$ (there are no coincident points in the dataset $S$ after the preprocessing mentioned in Section \ref{sec:MSL}) and $p_{u_j}'\prec p_{v_j}$ strictly. For the same reason, $p_{u_j}'\in G$, which can be denoted as $p_{v_m}$, and there is a point $p_{u_m}'\in G'$ that can dominate it, so $p_{v_i}\succ p_{v_j}\succ p_{v_m}\cdots$. We can continue this procedure until there is a point $p_{v_n}\in G$, no points in $G$ can dominate it, which is the end of the dominance chain $p_{v_i}\succ p_{v_j}\succ p_{v_m}\cdots\succ p_{v_n}$. Since all parents of $p_{v_i}$ are in $G$, $p_{v_n}$ is a parent of it and no points in $G$ can dominate $p_{v_n}$, $p_{v_n}$ is a skyline point. However, we can get a point $p_{u_n}'\in G'$ making $p_{u_n}'\prec p_{v_n}$ according to the relationship $G'\prec_g G$, which is a contradiction. We come to the conclusion that if a group $G$ satisfies for all $p_i \in G$ and all parents of $p_i$ are in $G$, $G$ is a G-skyline group.
\end{myPRF}

Property \ref{suf_nec} is a very important property of the G-skyline. The sufficient condition is the theoretical foundation of all algorithms for Pareto optimal groups even though it has not been explicitly studied. We can demonstrate other interesting properties with Property \ref{suf_nec}. Next, we demonstrate the G-skyline is a complete candidate set of top-$l$ solutions. To do so, we demonstrate these two following properties.

\begin{myPRO}
For all monotonically decreasing criterion function $f(\,\,)$, the group $G$ consists of the top-$l$ solutions belongs to G-skyline. 
\end{myPRO}
\vspace{-1em}
\begin{myPRF}
For all monotonically decreasing function $f(\,\,)$, $G$ is the set of top-$l$ solutions. For any $p_i\in G$, assuming there is a point $p_j\prec p_i$. It is obvious that $f(p_j)>f(p_i)$ according to the definition of dominance and monotonicity, so $p_j\in G$. Finally, we can get $G\in GS$ according to Property \ref{suf_nec}.
\end{myPRF}

\begin{myPRO}
For all $G\in GS$, there exists at least one monotonically decreasing function $f(\,\,)$ that lets $G$ be the set of top-$l$ solutions. 
\end{myPRO}
\vspace{-1em}
\begin{myPRF}
We prove the existence by giving a specific example. For a group $G=\{p_1,p_2,\cdots,p_l\}$ in G-skyline, the criterion function could be in the form of 
\begin{equation*}
\label{cri_fuc}
 f({\bm x}) = \left\{
\begin{array}{ll}
\sum\limits_{i=1}^d\left(\max\limits_j\{p_j[i]\}-x_i\right), &\bigvee\limits_{j=1}^l\left(\bigwedge\limits_{i=1}^d x_i\leq p_j[i]\right)\\
-\sum\limits_{i=1}^d x_i, &otherwise
\end{array},
\right.
\end{equation*}
where ${\bm x} = (x_1,x_2,\cdots,x_d)$ are all attributes. It is a piecewise function and decreases monotonically in all domain. All points in $G$ can get non-negative scores while points in $S\backslash G$ get negative scores, so $G$ is the top-$l$ solutions.
\end{myPRF}

We come to a conclusion that all G-skyline groups can be the top-$l$ solution, while all non-skyline groups can not. So G-skyline is the exact candidate set of optimal solution. This is the main superiority over other definitions of group-based skyline \cite{im2012group,zhang2014on}.

G-skyline is established by the MSL, however we observed that not all layers in MSL contribute equally to G-skyline. The first layer is much more important. Groups that consist of only the first layer points take a great proportion. 

\begin{myDef}[Primary Group]
$l$-point groups consisting of only $layer_1$ points in MSL are primary groups. According to the definition of G-skyline, all primary groups are G-skyline groups because every individual point cannot be dominated.
\end{myDef}

\begin{myDef}[Secondary Group]
G-skyline groups that are not primary groups are defined as secondary groups.
\end{myDef}

\begin{myEXP}
\label{exa:observe}
Taking the synthetic INDE dataset ($n = 1000$, $d = 4$, $l = 3$) as an example, there are 76 points in $layer_1$, 152 points in $layer_2$, 189 points in $layer_3$ in MSL and 75034 groups in G-skyline. The number of primary groups and secondary groups is $\binom{76}{3}= 68450$ and 6584 respectively. We can see that the number of primary groups dominates with an overwhelming advantage.
\end{myEXP}

\vspace{-2mm}
\begin{table}[ht!]
\caption{Percentage of primary groups.}  
\label{sample_tab} 
\vspace{-2em}
\begin{center}  
\scalebox{0.7}{
\begin{tabular}{|c|c|c|c|c|c|c|c|c|c|c|c|c|c|}  
\hline
%\cline{1,6,10,14}
\multicolumn{2}{|c|}{dataset} & \multicolumn{4}{c|}{CORR} & \multicolumn{4}{c|}{INDE} & \multicolumn{4}{c|}{ANTI} \\ 
\hline
\multicolumn{2}{|c|}{$d$} & 2 & 4 & 6 & 8 & 2 & 4 & 6 & 8 & 2 & 4 & 6 & 8\\
\hline
\multirow{3}{*}{$l$} & 2 & 62.5 & 97.6 & 98.8 & 99.5 & 80.3 & 97.9 & 99.6 & 99.9 & 98.1 & 99.9 & 99.8 & 99.9\\
\cline{2-14}  %%\cline and \hline are all for horizontal line !!!
& 3 & 18.6 & 87.8 & 97.8 & 98.9 & 38.2 & 91.2 & 98.0 & 99.3 & 91.0 & 98.9 & $-$ & $-$\\
\cline{2-14}
& 4 & 2.1 & 84.3 & $-$ & $-$ & 16.9 & 87.4 & $-$ & $-$ & 88.2 & $-$ & $-$ & $-$\\
\hline
\end{tabular}} 
\end{center}  
\end{table}

Table \ref{sample_tab} shows the percentage of primary groups in different datasets, with various parameters. We can see that the primary groups occupy a large proportion in most cases, especially with a small $l$ and a large $d$. This phenomenon motivates the following. 

(1) Primary groups, which are combinations of $l$ points in $layer_1$, contribute to a majority number of G-skyline groups. It is important to give priority to enumerate all these combinations in the most direct and efficient way. (2) Though there are significantly more points in later layers, they contribute very little to the G-skyline groups. So, an efficient pruning strategy is desired.

Considering these motivations, we introduce two new approaches, fast PWise algorithm (F\_PWise) and fast UWise algorithm (F\_UWise), based on the PWise and UWise+ algorithms proposed in \cite{liu2015finding}. Here we give a very brief summary of them: Authors first constructed the directed skyline graph (DSG), which is a graph saves all points in MSL and the dominance relationship among them. DSG is then pruned with preprocessing (remove all points with more than $l-1$ parents). In DSG, points with larger index than current point/a group of points are stored as a \textit{tail set} of this point/group, denoted as $T_{DSG}(p)$/$T_{DSG}(G)$. Point $p$ and all its parents form the \textit{unit group}. In PWise and UWise+ algorithm, points and unit groups from the tail set of the current group are added into it to construct a new group based on Property \ref{suf_nec}. The new group is removed if it is not a G-skyline group and is outputted if the size is $l$.

\subsection{Fast PWise Algorithm}
\label{subsec:PWise}
In this subsection, we propose the fast PWise algorithm (F\_PWise) to construct G-skyline based on the combination queue. Besides the previous pruning strategies in PWise algorithm, we give the edge pruning for DSG to make the algorithm more effective. We first give the definition of DSG and then illustrate the pruning strategy.

\begin{myDef}[Directed Skyline Graph (DSG)]
A directed skyline graph is a graph where a node represents a point and an edge represents a dominance relationship. Each node has a structure as $[layer\;index; point\;index; parents; children]$
\end{myDef}

\partitle{Edge Pruning} For any two points $p_i$ and $p_j$ in DSG ($p_j$ is in the tail set of $p_i$), if $p_j.layer - p_i.layer > 1$, delete the edge between them. Also, taking Figure \ref{fig:MSL} as an example, we can delete ($p_{7}, p_{8})$ because when adding $p_{8}$ to a group containing $p_{7}$, it must contain $p_{5}$ and $p_{4}$ too or the new group will be a non-skyline group and removed by subtree pruning \cite{liu2015finding}. We can find $p_{8}$ in the child set of $p_{5}$ hence the connection between $p_{7}$ and $p_{8}$ is not necessary.

\begin{myDef}[Combination Queue]
A queue to enumerate combinations by adding elements in the tail set of current set.
\end{myDef}

We enumerate all combinations of no more than $l$ $layer_1$ points with a combination queue. It can output all the primary groups and can also be used to find secondary groups in the next procedure. The detailed algorithm is shown in Algorithm 3. The algorithm prunes the DSG by preprocessing \cite{liu2015finding} and edge pruning (line 1). A combination queue is used to find all primary groups (line 3). Lines 4-15 show the procedure to find the secondary group with the intermediate results of the combination queue. Lines 10-14 show the subtree pruning since a group which is not an $(i-1)$-size G-skyline group cannot be an $i$-size G-skyline group by adding a point from its tail set \cite{liu2015finding}.

\begin{algorithm}[ht]
\label{alg:g_sl}
\caption{Fast PWise algorithm}%算法名字
\LinesNumbered %要求显示行号
\small\KwIn{a DSG and group size $l$.}%输入参数
\KwOut{G-skyline $GS$.}%输出

prune DSG with preprocessing and edge pruning\;
$GS = \varnothing$\;
construct combination queue $Q$ to find out all primary groups and store them to $GS$\;

$queue\_pointer = 1$\;
\While{$queue\_pointer \leq length(Q)$}{
    \For{each $p_i$ in $Q(queue\_pointer)$}{
        \For{each $p_j$ in $p_i$'s children set in DSG}{
            \If{$p_j$ is in $T_{DSG}(Q(queue\_pointer))$}{
                combine $Q(queue\_pointer)$ and $p_i$ as $G$\;
                \If{$G$ is a skyline group}{
                    \If{there are $l$ points in $G$}{
                        add $G$ to $GS$\;
                    }
                    \Else{add $G$ to $Q$\;}                    
                }
                $queue\_pointer++$\;
            }
            
        }
    }
}
\Return $GS$\;
\end{algorithm}

\begin{figure}[ht!]\vspace{-1mm}
\setlength{\abovecaptionskip}{2mm}
  \centering
  \subfigure[DSG]{
    \label{fig:ESA3:a} %% label for first subfigure
    \includegraphics[scale = 0.28]{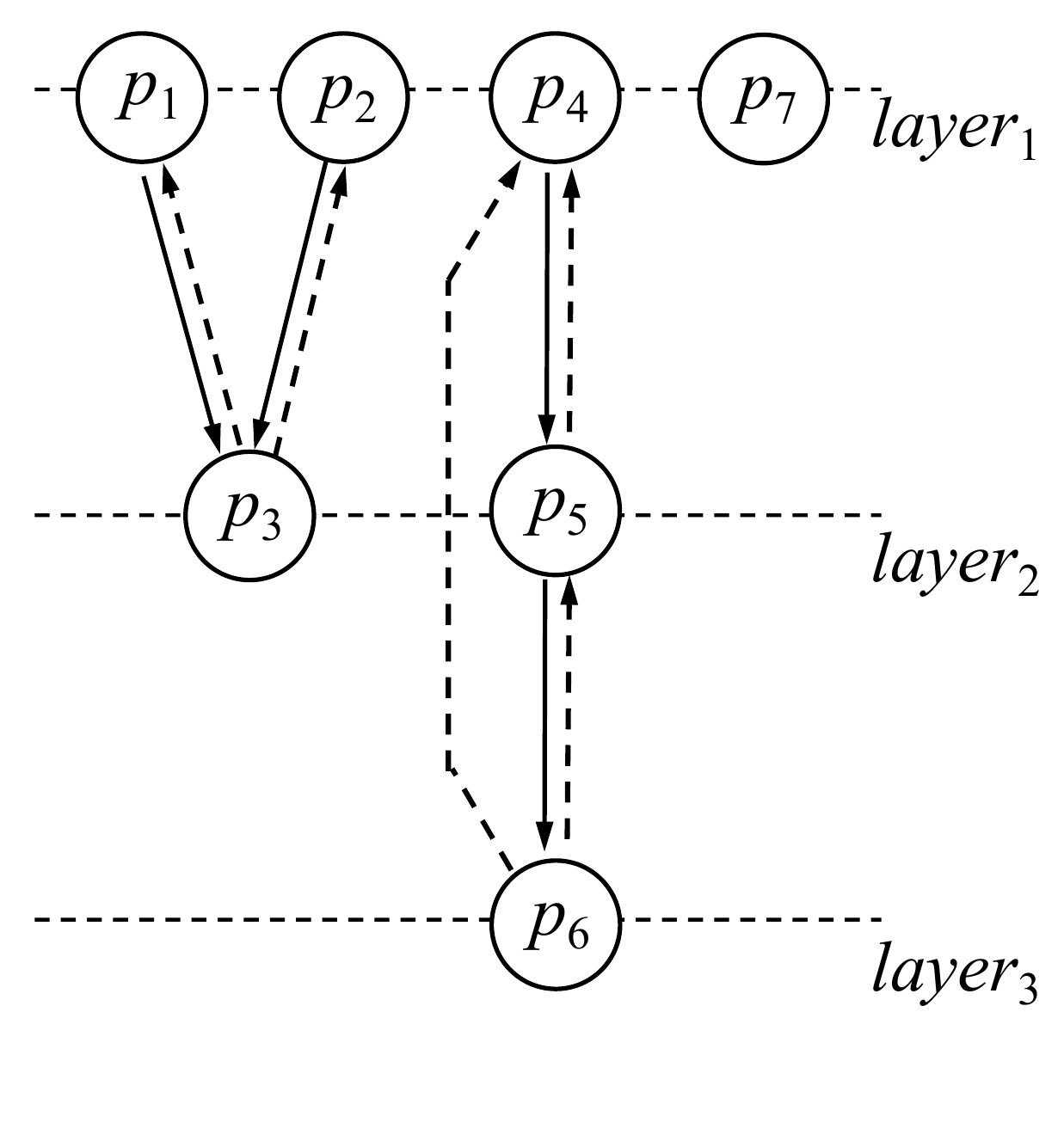}}
  \hspace{0.1in}
  \subfigure[Example steps]{
    \label{fig:ESA3:b} %% label for second subfigure
    \includegraphics[scale = 0.48]{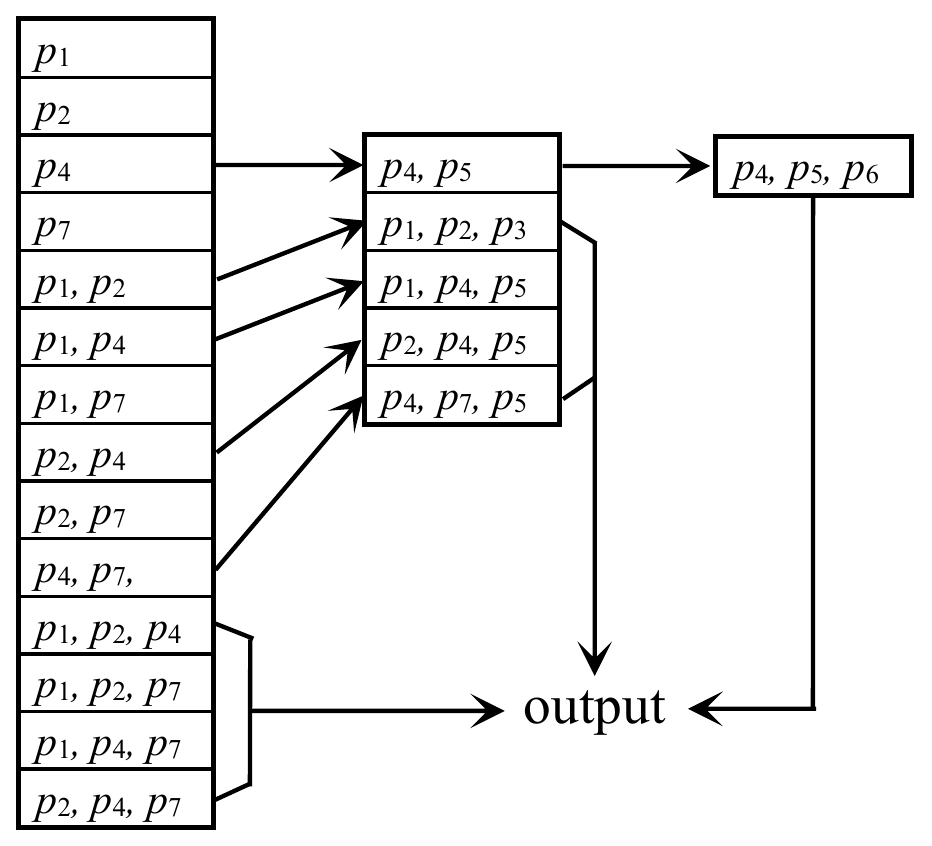}}
  \caption{Example steps of Algorithm 3.}
  \label{fig:ESA3} %% label for entire figure
  \vspace{-2mm}
\end{figure}

\begin{myEXP}
Figure \ref{fig:ESA3} uses MSL in Figure \ref{fig:MSL} as an example to show the steps of Algorithm \ref{alg:g_sl}. Figure \ref{fig:ESA3:a} shows a DSG, where solid line arrows point to child nodes and dotted arrows point to parent node. Node $p_{8}$ and edge ($p_{4}$, $p_{6}$) are all removed (only the solid line arrow is removed, the dotted arrows is reserved for subtree pruning). In Figure \ref{fig:ESA3:b}, the first column is a combination queue to enumerate all primary groups and the later two columns find secondary groups by adding points in $layer_2$ and $layer_3$ to groups in the first column.
\end{myEXP}

\subsection{Fast UWise Algorithm}
\label{subsec:UWise}

In this subsection, we propose the fast UWise algorithm (F\_UWise) by adding combination queue to UWise+ algorithm. UWise+ method constructs G-skyline by adding unit groups (groups composed of certain point and all its parents in DSG) \cite{liu2015finding}. We find all primary groups by the combination queue and secondary groups by UWise+ method. Unlike F$\_$PWise, F$\_$UWise gives these two kinds of groups independently by different methods. It is a simple combination but also performs well.

\section{Finding RG-skyline Groups}
\label{sec:repre}

We demonstrated the completeness of G-skyline in Section \ref{sec:gsky}. However, it can be a rather costly process returning a huge number of groups, especially when $l$ is large, or in high-dimensional space. We introduced several efficient algorithms in Sections \ref{sec:MSL} and \ref{sec:gsky} to deal with the problem of high time consumption. In this section, we focus on the problem of huge output size by defining RG-skyline. In real-world applications, reasonable output size is very important for user experience, so we just show $k$ RG-skyline groups that represent the ``contour'' of the whole G-skyline to users rather than the original G-skyline. \cite{group_repre} computed $k$-SGQ by finding G-skyline groups that can dominate the most points. However, it only returns the groups with similar attribute patterns hence is not representative enough. 

\begin{myEXP}
Recalling the hotel example shown in Figure \ref{fig:MSL}, the $k$-SGQ groups are $\{p_1,p_2,p_4\}$, $\{p_1,p_4,p_7\}$, and $\{p_2,p_4,p_7\}$ ($l=3$, $k=3$).
\end{myEXP}

We can see that a max-dominance method tends to return the groups that all their points are scattered. In fact $\{p_1,p_2,p_4\}$, $\{p_1,p_4,p_7\}$, $\{p_2,p_4,p_7\}$ are quite similar, they are all for agencies who want to provide extensive service. The agencies targeting low-end travelers (may collaborate with hotels $\{p_{4}, p_{7}, p_{5}\}$ or \{$p_{4}$, $p_{5}$, $p_{6}$\}) and high-end travelers (may prefer \{$p_{1}$, $p_{2}$, $p_{3}$\}) cannot find their choice in $k$-SGQ. To bridge this gap, we define a novel distance-based representative skyline in the group level, RG-skyline, to show all main patterns of groups and propose a novel G-clustering method to construct it.

\begin{myDef}[RG-Skyline]
Given a G-skyline $GS$ and an integer $k$, cluster all $GS$ groups into $k$ clusters, the RG-skyline consists of $k$ cluster centers \footnote[3]{The clustering is on group-level thus the cluster centers returned are groups.}. 
\end{myDef}

To construct RG-skyline, all G-skyline groups are clustered. We regard a cluster, which is considered as a category in unsupervised learning, as a pattern and use the cluster center to represent it. However, the conventional clustering algorithms are not competent in the group situation. Next, we introduce a $k$-means G-clustering algorithm. There are two steps in $k$-means algorithm, assign each observation to the cluster whose mean has the least distance and update the new means to be the centroids of the observations in the new clusters. So, in the G-clustering algorithm, we need to define how to calculate the distance between two groups (introduced in Subsetion \ref{subsec:dis_group}) and how to update the cluster centers (introduced in Subsetion \ref{subsec:upd_pro}).

\vspace{-2mm}
\subsection{Distance Between Groups} 
\label{subsec:dis_group}

There are several ways to measure the distance between two groups, such as aggregating a group into one point with MAX, MIN, or SUM. However, aggregate functions capture fragmentary attributes (strongest/weakest/average attributes with MAX/MIN/SUM). To address this gap, we give a new definition of distance in the group level. Considering groups with the same pattern have similar spacial distribution of points, we define the group distance by calculating the distance among their points. 

\begin{myDef}[Group Distance]
For two groups $G=\{p_{1},p_{2},$ $\cdots,p_{l}\}$ and $G'=\{p'_{1},p'_{2},\cdots,p'_{l}\}$, the distance between them is defined as:
\begin{equation}
\label{equ:distance}
    d_g(G,G')=\sqrt{\min\limits_{u,v}\sum\limits_{i=1}^l d^2(p_{u_i},p'_{v_i})},
\end{equation}
where $u$ and $v$ are two permutations that minimize Equation \ref{equ:distance} and $d(\,\,)$ is the Euclidean distance between two points.
\end{myDef}

Calculating the group distance is an assignment problem \cite{AP}. We use a distance matrix $\bm{D}\in \mathbb{R}^{l\times l}$ to denote the distances between points in two groups and a Boolean matrix $\bm{X}\in \{0,1\}^{l\times l}$ to denote the matching strategy. $\bm{D}_{i,j}$ is the distance between $p_i$ and $p'_j$; $\bm{X}_{i,j}=1$ means $p_i$ is matched to $p'_j$ and $\bm{X}_{i,j}=0$ otherwise. An assignment problem can be formulated as:
\begin{align*}
&\min\quad \sum\limits_{i=1}^l\sum\limits_{j=1}^l \bm{X}_{i,j} \bm{D}_{i,j}\\
& \begin{array}{r@{\quad}r@{}l@{\quad}l}
%\texttt{s}.\texttt{t}.&\sum\limits_{i=1}^l \bm{X}_{i,j}=1\\
{\rm s. t.}&\sum\limits_{i=1}^l \bm{X}_{i,j}=1\\
&\sum\limits_{j=1}^l \bm{X}_{i,j}=1 
\end{array}
\end{align*}

There are several existing methods to solve the assignment problem, such as Hungarian algorithm \cite{Hungarian} and heuristic algorithms \cite{Particle,Ant,ant2}. Nevertheless, in our RG-skyline group problem, $l$ is usually small, and these existing algorithms are inefficient (even slower than Brute-force, shown in Figure \ref{fig:greed_time}). Compared with methods for global optimization, the greedy method is much more efficient while it usually returns an inaccurate solution. As the basic operation of G-clustering, a very fast and accurate algorithm for small scale matching is greatly desired. To bridge these gaps, we modify the greedy method and propose a Greedy+ algorithm (represented in Algorithm \ref{alg:greedy}). In the naive greedy method, all elements in $\bm D$ are selected in an increasing order to try to establish the matching strategy. Since only one element can be chosen in each row and column, once an element is chosen, the elements in the same row or column are out of consideration, even though they are involved in the optimal matching strategy. So greedy method easily gets trapped in a local optimum. In our Greedy+ algorithm, more strategies are investigated (lines 1-11). We define the elements searched in all previous iterations as candidate matchings (lines 2, 9-11), they are better than the worst matching in the local optimization, hence with considerable probability to be involved in the global optimal solution. In Greedy+ algorithm, we investigate matching strategies covering all candidate matchings (lines 3-11) and return the best strategy (lines 12, 13).

\begin{algorithm}[ht]
\label{alg:greedy}
\caption{Greedy+ algorithm}%算法名字
\LinesNumbered %要求显示行号
\small\KwIn{a distance matrix $\bm D$.}%输入参数
\KwOut{distance between two groups $d$ and corresponding matching strategy $u$ and $v$.}%输出
sort all elements of $\bm D$ increasingly and record in $List$\;
$candidate\_matchings=List$\;
\While{$candidate\_matchings$ are not empty}{
    $strategy=\varnothing$\;
    add $candidate\_matchings(1)$ to $strategy$\;
    \For{each ${\bm D}_{i,j}$ in $List$}{
        \If{${\bm D}_{i,j}$ are not in the same row or column with elements in $strategy$}{
            add ${\bm D}_{i,j}$ to $strategy$\;
        }
        \If{$length(strategy)==l$}{
            remove the elements appearing in $strategy$ or larger than ${\bm D}_{i,j}$ from $candidate\_matchings$\;
            \textbf{break}\;
        }
    }
    record the minimum group distance $d$ and corresponding matching strategy $u$, $v$ during iterations\;
}
\Return $d$, $u$, $v$\;
\end{algorithm}

\subsection{Updating Strategy} 
\label{subsec:upd_pro}

In conventional $k$-means, we update the cluster centers by calculating the mean of each cluster. However, in G-skyline, a group is a set and all points in it are unordered, so we cannot sum groups to calculate the mean. To deal with this problem, we need to determine the sequence of points in each group. To do this, we utilize the permutations ($u$ and $v$) we get in Equation \ref{equ:distance}. However, Equation \ref{equ:distance} only gives the permutations between each pair of groups hence cannot be used to sum several groups up. To deal with the problem, we revise the updating procedure. We first formulate the procedure in the point level and then extend the form to the group level.

In the point level, for certain cluster $C$, we have the mean $\bar p$: $$\bar p=\frac{\sum\limits_{p_i\in C}p_i}{|C|}=\frac{\sum\limits_{i\in \{j|p_j\in C\}}\left(p_0+\Delta_i\right)}{|C|}=p_0+\frac{\sum\limits_{i\in \{j|p_j\in C\}}\Delta_i}{|C|},$$ where $p_0$ is the previous center and $\Delta_i$ is the vectorial difference between $p_i$ and $p_0$. 

Similarly, in the group level, the new centroid of cluster $C$ is:
\begin{equation}
\label{equ:upd_pro}
    \bar G=G_0+\frac{\sum\limits_{G_i\in C}\Delta(G_i,G_0)}{|C|},
\end{equation}
where the previous cluster centroid $G_0$ is not a set ($ \left\{ p^{(0)}_1,p^{(0)}_2,\cdots,p^{(0)}_l \right\} $) but a concatenated vector:
$$G_0=\left(\left(p^{(0)}_{v_1}[1],\cdots p^{(0)}_{v_1}[d]\right),\cdots,\left(p^{(0)}_{v_l}[1],\cdots p^{(0)}_{v_l}[d]\right)\right),$$
and $\Delta(\,\,)$ in Equation \ref{equ:upd_pro} is also defined as a concatenated vector: 
\begin{flalign}
\label{equ:delta1}
\Delta(G_i,G_0)=\Big(&\Big(p_{u_1}^{(i)}[1]-p_{v_1}^{(0)}[1],\cdots p_{u_1}^{(i)}[d]-p_{v_1}^{(0)}[d]\Big),\cdots, \nonumber\\
&\Big(p_{u_l}^{(i)}[1]-p_{v_l}^{(0)}[1],\cdots p_{u_l}^{(i)}[d]-p_{v_l}^{(0)}[d]\Big)\Big), 
\end{flalign}
where $u$ and $v$ are determined in Equation \ref{equ:distance}. We fix $v$ as $\{1,2,\cdots,l\}$ to give the unique solution and the vectorial difference in Equation \ref{equ:delta1} is denoted as $\Delta_u(G_i,G_0)$. We can see that the distance in Equation \ref{equ:distance} $d_g(G_i,G_0)= \left\Arrowvert \Delta(G_i,G_0) \right\Arrowvert_2$. All additions in Equation \ref{equ:upd_pro} are for vectors.

\begin{myEXP}
For the hotel example illustrated in Figure \ref{fig:MSL}, assuming there are three groups in current cluster $C = \{G_1,G_2,G_3\}$, where $G_1=\{p_2,p_4,p_7\}$, $G_2=\{p_4,p_5,p_7\}$, and $G_3=\{p_2,p_4,p_5\}$. The cluster center is initialized to $G_1$ and the new cluster center is:
\begin{equation*}
    \bar G=G_0+\frac{\sum\limits_{i=1}^3\Delta(G_i,G_0)}{3},
\end{equation*}
where
\iffalse
\begin{flalign}
    &\,\,\,\,\,G_0=(6,250,13,50,23,25),\nonumber\\
    \Delta(G_1,G_0)=&(6,250,13,50,23,25)-(6,250,13,50,23,25) \nonumber\\
    =&(0,0,0,0,0,0),\nonumber\\
    \Delta(G_2,G_0)=&(16,75,13,50,23,25)-(6,250,13,50,23,25) \nonumber\\
    =&(10,-175,0,0,0,0),\nonumber\\
    \Delta(G_3,G_0)=&(6,250,13,50,16,75)-(6,250,13,50,23,25) \nonumber\\
    =&(0,0,0,0,-7,50),\nonumber\\
    \bar G=\!\!\!&\,\,\,\,(9.33,191.67,13,50,20.67,41.67).\nonumber
\end{flalign}
\fi
$$G_0=(6,250,13,50,23,25),$$
\vspace{-6mm}
\begin{flalign}
\Delta(G_1,G_0)=&(6,250,13,50,23,25)-(6,250,13,50,23,25) \nonumber\\
    =&(0,0,0,0,0,0),\nonumber
\end{flalign}
\vspace{-6mm}
\begin{flalign}
\Delta(G_2,G_0)=&(16,75,13,50,23,25)-(6,250,13,50,23,25) \nonumber\\
    =&(10,-175,0,0,0,0),\nonumber
\end{flalign}
\vspace{-6mm}
\begin{flalign}
\Delta(G_3,G_0)=&(6,250,13,50,16,75)-(6,250,13,50,23,25) \nonumber\\
    =&(0,0,0,0,-7,50),\nonumber
\end{flalign}
\vspace{-4mm}
$$\bar G=(9.33,191.67,13,50,20.67,41.67).$$
The new cluster center is $$\{(9.33,191.67),(13,50),(20.67,41.67)\}.$$

\begin{figure}[ht!]\vspace{-1mm}
  \setlength{\abovecaptionskip}{1mm}
  \centering
    \includegraphics[scale = 0.4]{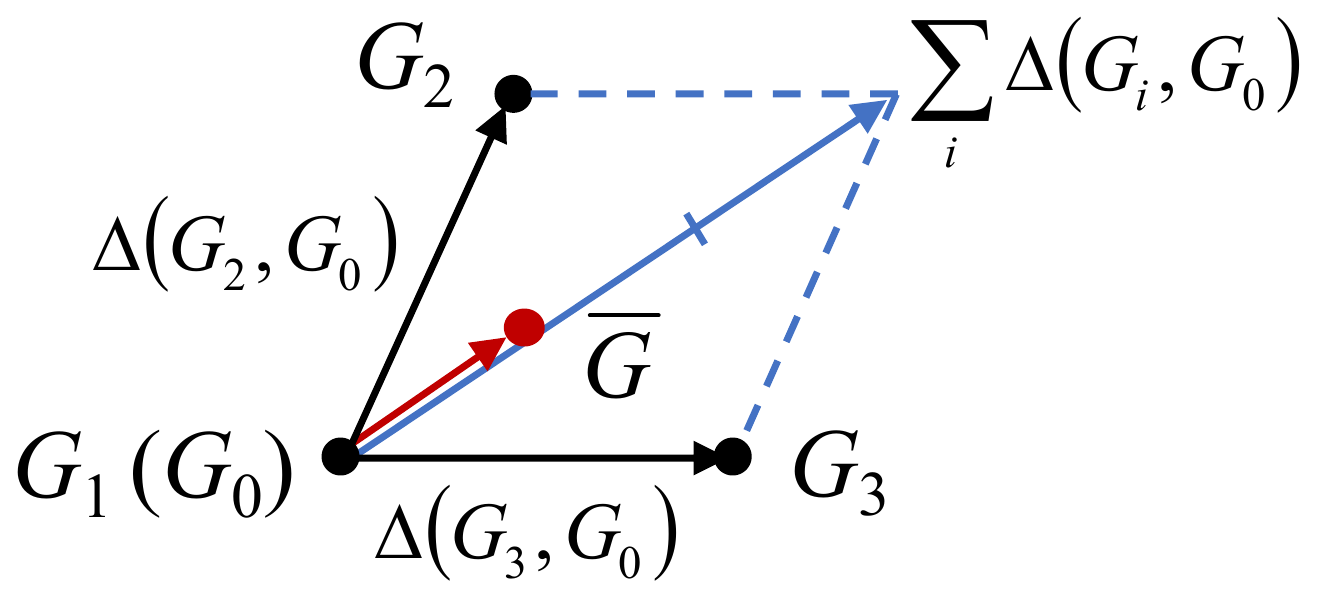}
  \caption{An example for cluster center updating. Each point in this figure indicates a group, the red point represents the new cluster center and the red arrow is the shift vector $\frac{\sum_{i}\Delta(G_i,G_0)}{|C|}$. $G_1$, $G_2$, and $G_3$ cannot be summed up directly since they are sets and the order of their elements are not determined. However, $\Delta(G_i,G_0), i=1,2,3$ are vectors so we can sum them together.}
\vspace{-2mm}
\end{figure}
\end{myEXP}

We can see that after updating, the cluster centers may not correspond to any G-skyline groups. When the iteration stops, for each cluster center, we find a G-skyline group with the shortest distance to replace it (please see Algorithm \ref{alg:g_clu} for details).

\begin{algorithm}[ht]
\label{alg:g_clu}
\caption{G-clustering algorithm}%算法名字
\LinesNumbered %要求显示行号
\small\KwIn{a G-skyline $GS$, the size of RG-skyline $k$.}%输入参数
\KwOut{a set of RG-skyline groups $RS$.}%输出
initialize the RG-skyline: $RS=\{G_{c_j}\}, j=1,\cdots,k$\;
\While{not converge}{
    \For{$j=1$ to $k$}{
        $C_j=\varnothing$\;
        initialize $\Delta_j$ to a zero vector of length $d\times l$\;
    }
    \For{each $G_i$ in $GS$}{
        \For{each $G_{c_j}$ in $RS$}{
            calculate the distance matrix $D$\;
            calculate $d_g(G_i,G_{c_j})$ and $u$ by Greedy+ algorithm\;
        }
        find the $G_{c_j}$ with the minimum $d_g(G_i,G_{c_j})$, record $G_{c_j}$ as $G_{cen}$, record $u$ as $u_{cen}$\;
        add $G_i$ to $C_j$\;
        $\Delta_j+\!\!=\Delta_{u_{cen}}(G_i,G_{cen})$\;
    }
    \For{$j=1$ to $k$}{
        $G_{c_j}+\!\!=\Delta_j/length(C_j)$\;
    }
}

\For{each $G_{c_j}$ in $RS$}{
    \For{each $G_i$ in $GS$}{
        calculate $d_g(G_i,G_{c_j})$\;
    }
    record the nearest group $G_i$ and replace $G_{c_j}$ with it in $RS$\;
}
\Return $RS$
\end{algorithm}

Our final G-clustering algorithm is presented in Algorithm \ref{alg:g_clu}. To initialize the clustering centers (line 1), we first choose an arbitrary group and then add a group to $RS$ $k-1$ times. For each addition, we select a group in $GS$ with the largest distance to those in $RS$. In fact, our initialized clustering centers are sufficient to represent the whole G-skyline \cite{distance_based}. In each iteration, we assigning each group to the nearest cluster (lines 6-12) and then updating the centers with new means (lines 13-14). The iteration is terminated when the composition of each cluster remains stable (lines 2-14). After being updated many times, the cluster centers we get may not be skyline groups, so we return groups with the shortest distance as RG-skyline groups (lines 15-18).

\begin{myEXP}
RG-skyline groups returned by G-clustering are $\{p_1,p_2,p_3\}$, $\{p_4,p_5,p_7\}$, and $\{p_2,p_4,p_5\}$ ($l=3$, $k=3$). We can see that agencies for high-end travelers, low-end travelers, or extensive service can all find their choices.
\end{myEXP}

\vspace{-1mm}
\section{Experiments}
\label{sec:experiments}

In this section, we demonstrate the effectiveness and the efficiency of our proposed methods in the synthetic and real-world datasets. All algorithms were implemented in python\footnote[4]{All the codes and datasets used in this paper are available on \url{https://github.com/Wenhui-Yu/Gskyline}.}. We conducted the experiments on a PC with Intel Core i7 2.8GHz processors, 512K L2 cash, 3M L3 cash, and 8G RAM.

\partitle{Datasets} We performed each algorithm in synthetic datasets and a real NBA dataset. To examine the scalability of our methods, we generated independent (INDE), correlated (CORR), and anti-correlated (ANTI) datasets. These three types of data distribution are first proposed by \cite{borzsony2001the} to simulate different actual application scenarios.

CORR is for the objects whose attributes are all correlated. As an example, an outstanding student may perform well in all subjects. ANTI is for the objects whose attributes are all anti-correlated, like the hotel example shown in Figure \ref{fig:examp}, the long distance one may have a lower price. 

We crawl the data of NBA players from the official website\footnote[5]{\url{http://stats.nba.com.}} to construct the NBA dataset. There are five attributes to investigate players' performance, points (PTS), rebounds (REB), assists (AST), steals (STL), and blocks (BLK).

\vspace{-1mm}
\subsection{MSL on the Synthetic Data}
\label{subsec:MSL_syn}
\begin{figure}[ht!]
  \setlength{\abovecaptionskip}{1mm}
  \centering
    \includegraphics[scale = 0.37]{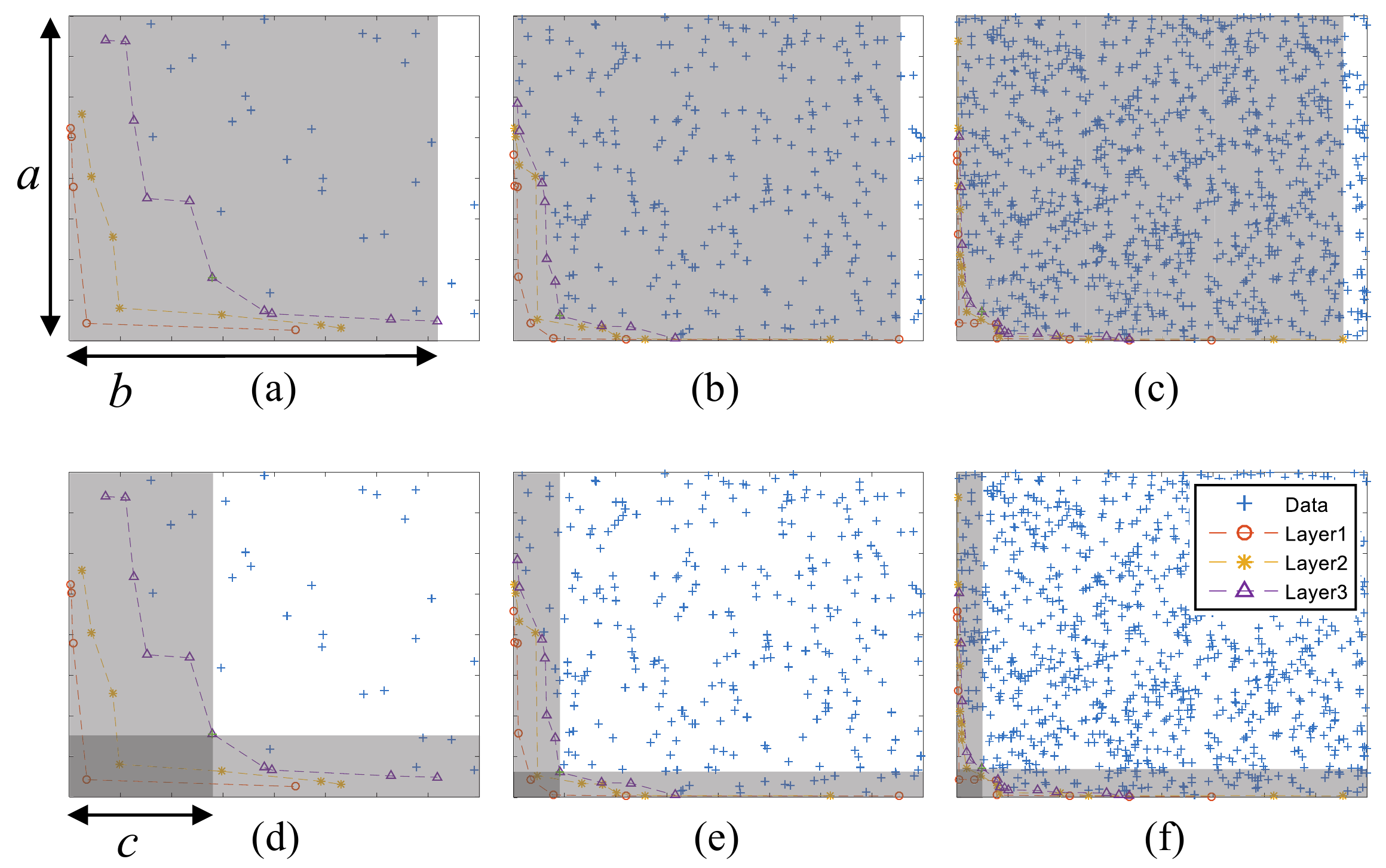}
  \caption{Comparison between previous methods and our method.}
  \label{fig:comp}
\end{figure}

\begin{figure}[ht!]
\setlength{\abovecaptionskip}{1mm}
  \centering
  \subfigure[CORR]{
    \label{fig:MSL_l_corr} %% label for first subfigure
    \includegraphics[scale = 0.15]{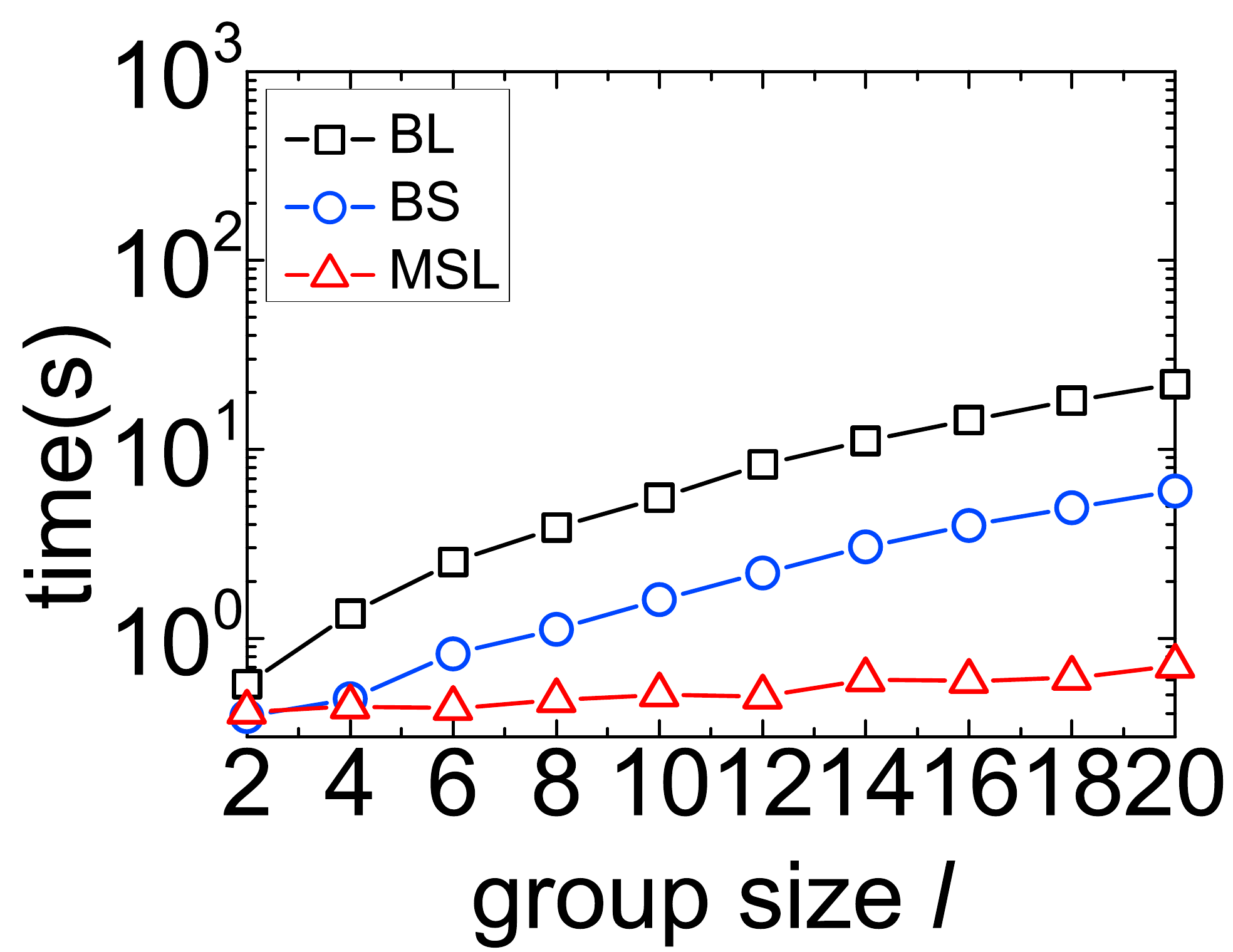}}
  \hspace{-0.15in}
  \subfigure[INDE]{
    \label{fig:MSL_l_inde} %% label for second subfigure
    \includegraphics[scale = 0.15]{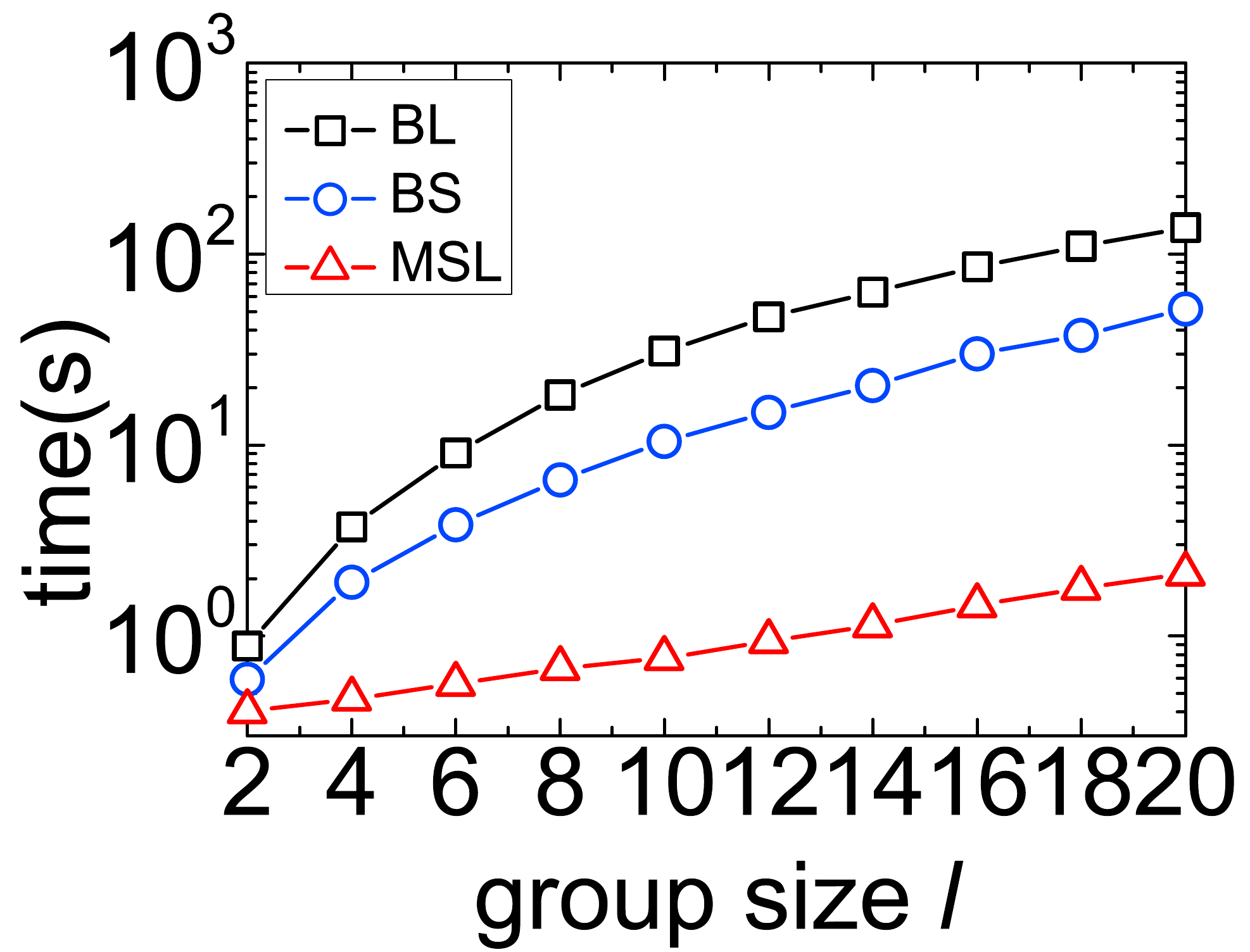}}
  \hspace{-0.15in}
  \subfigure[ANTI]{
    \label{fig:MSL_l_anti} %% label for second subfigure
    \includegraphics[scale = 0.15]{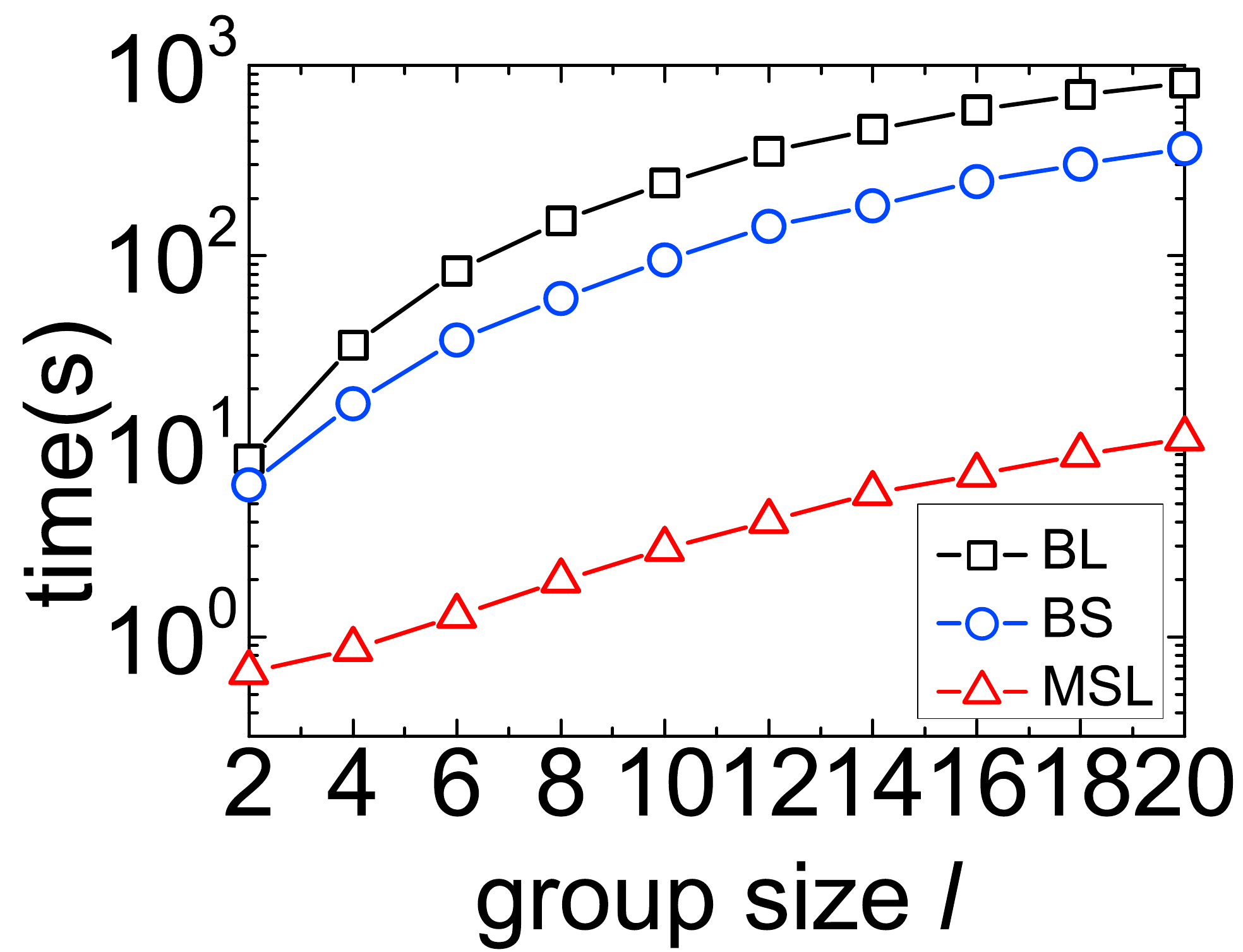}}
  \caption{MSL in synthetic datasets of varying $l$.}
  \label{fig:MSL_l} %% label for entire figure
  \vspace{-2mm}
\end{figure}

\begin{figure}[ht!]
\setlength{\abovecaptionskip}{1mm}
  \centering
  \subfigure[CORR]{
    \label{fig:MSL_n_corr} %% label for first subfigure
    \includegraphics[scale = 0.147]{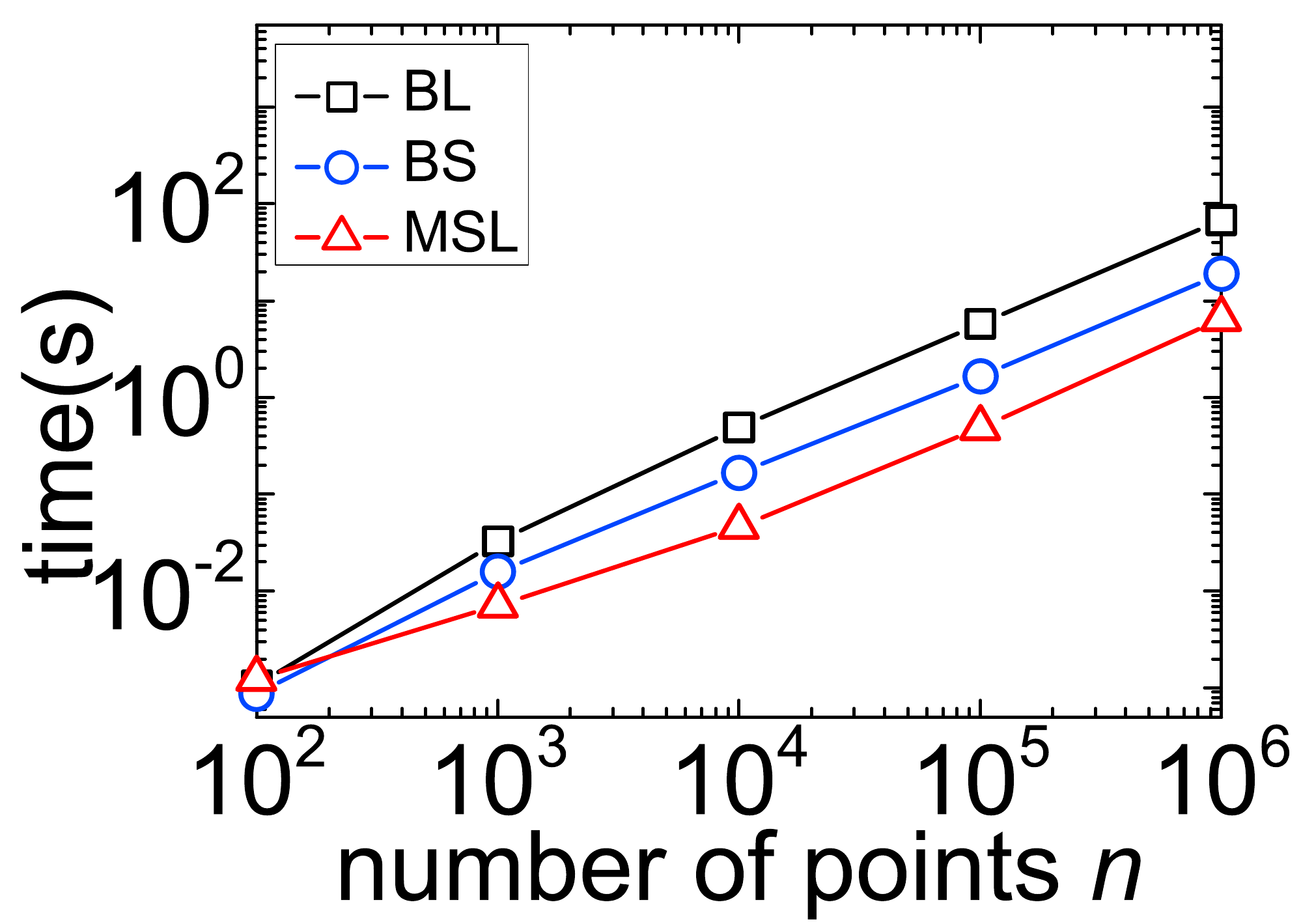}}
  \hspace{-0.15in}
  \subfigure[INDE]{
    \label{fig:MSL_n_inde} %% label for second subfigure
    \includegraphics[scale = 0.147]{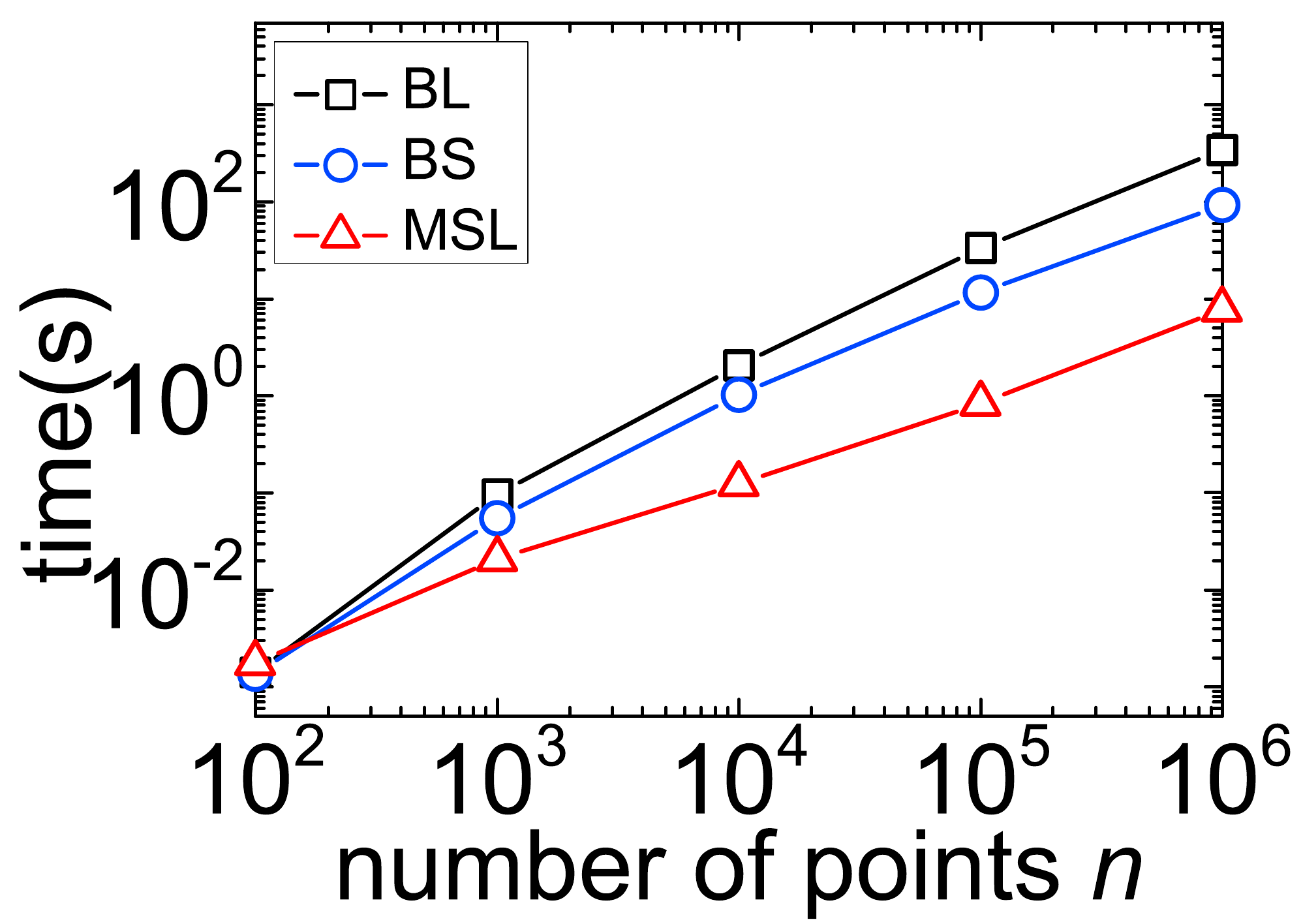}}
  \hspace{-0.15in}
  \subfigure[ANTI]{
    \label{fig:MSL_n_anti} %% label for second subfigure
    \includegraphics[scale = 0.147]{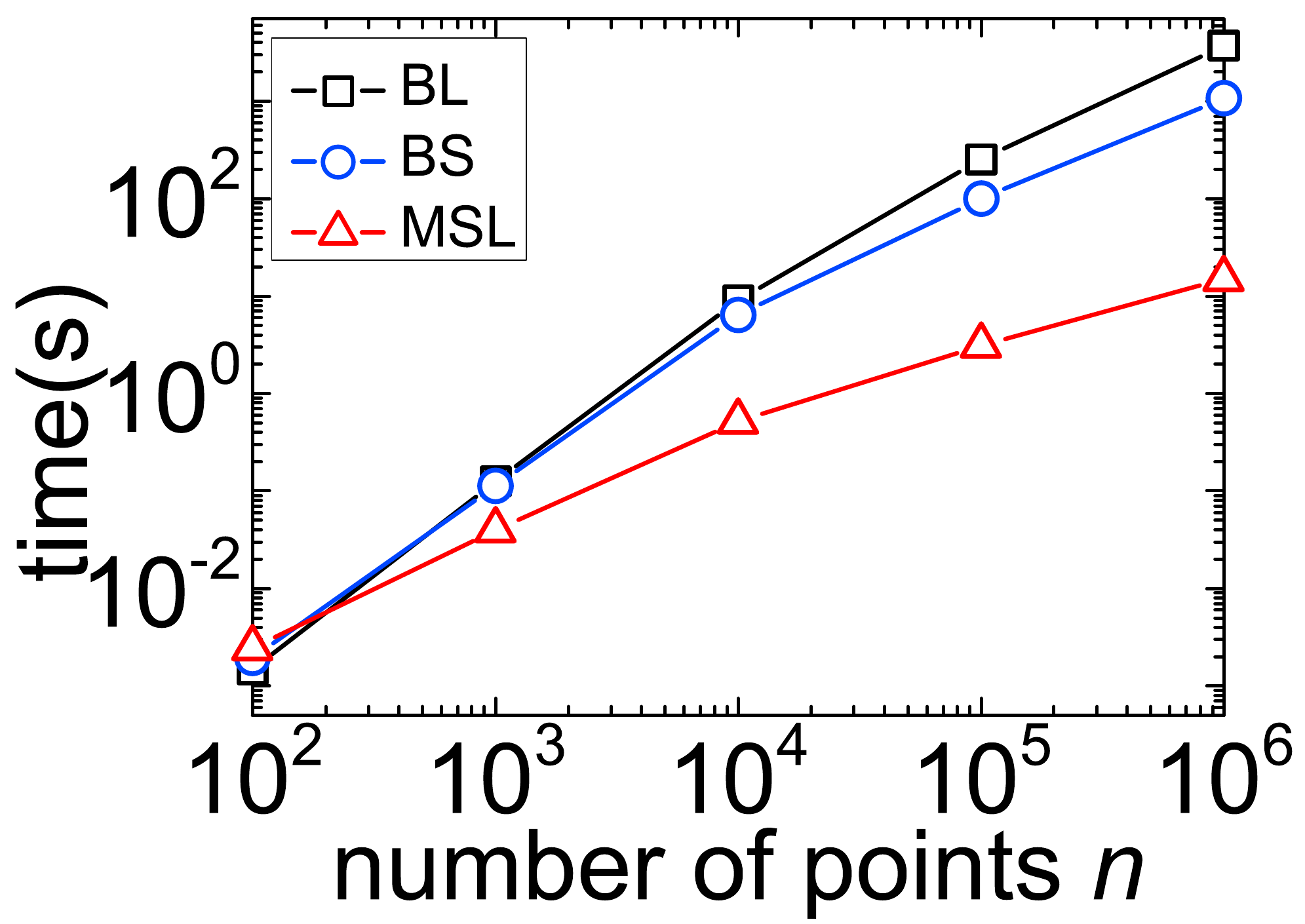}}
  \caption{MSL in synthetic datasets of varying $n$.}
  \label{fig:MSL_n} %% label for entire figure
\end{figure}

\begin{figure}[ht!]
\setlength{\abovecaptionskip}{1mm}
  \centering
  \subfigure[CORR]{
    \label{fig:MSL_d_corr} %% label for first subfigure
    \includegraphics[scale = 0.15]{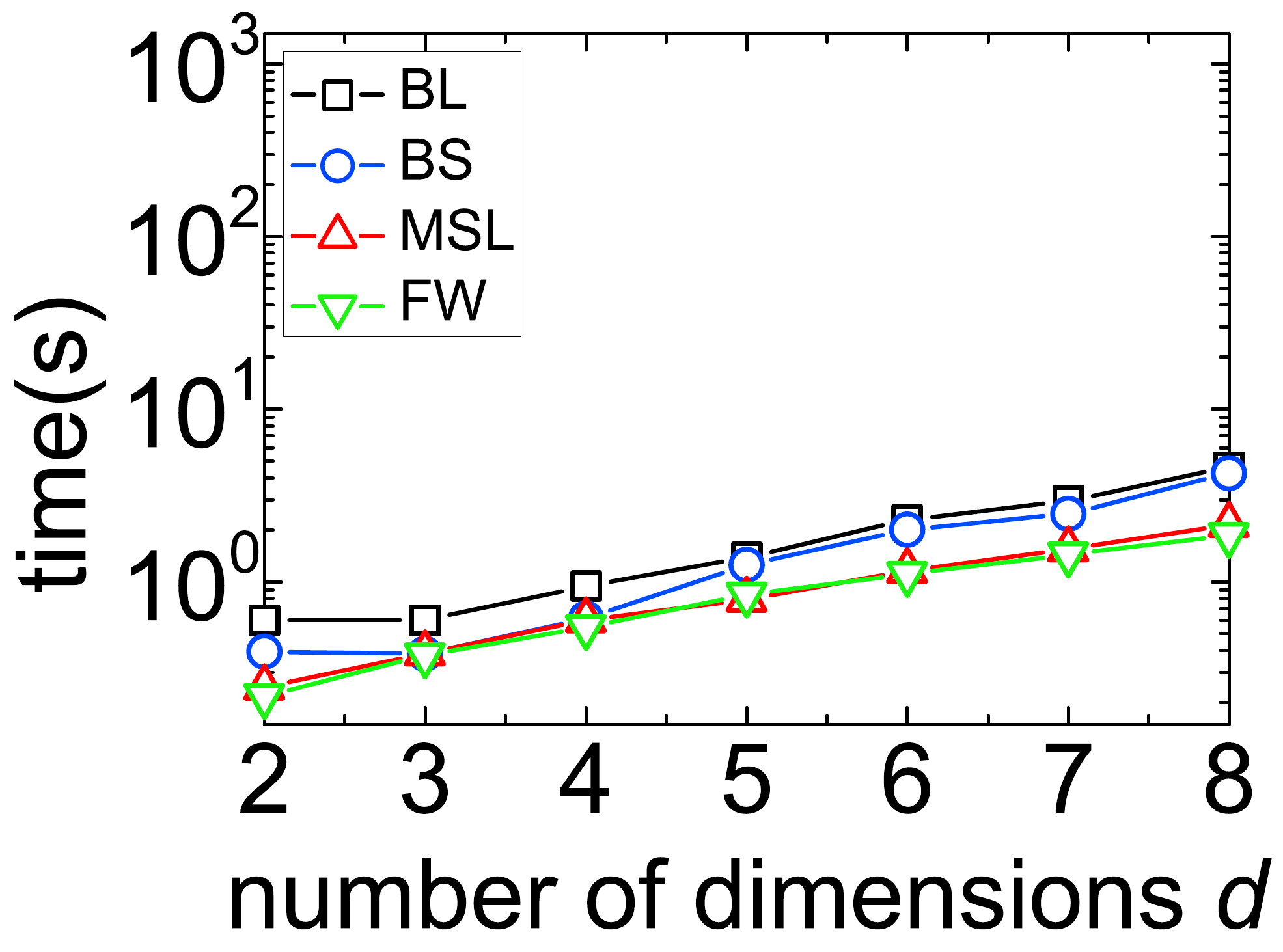}}
  \hspace{-0.15in}
  \subfigure[INDE]{
    \label{fig:MSL_d_inde} %% label for second subfigure
    \includegraphics[scale = 0.15]{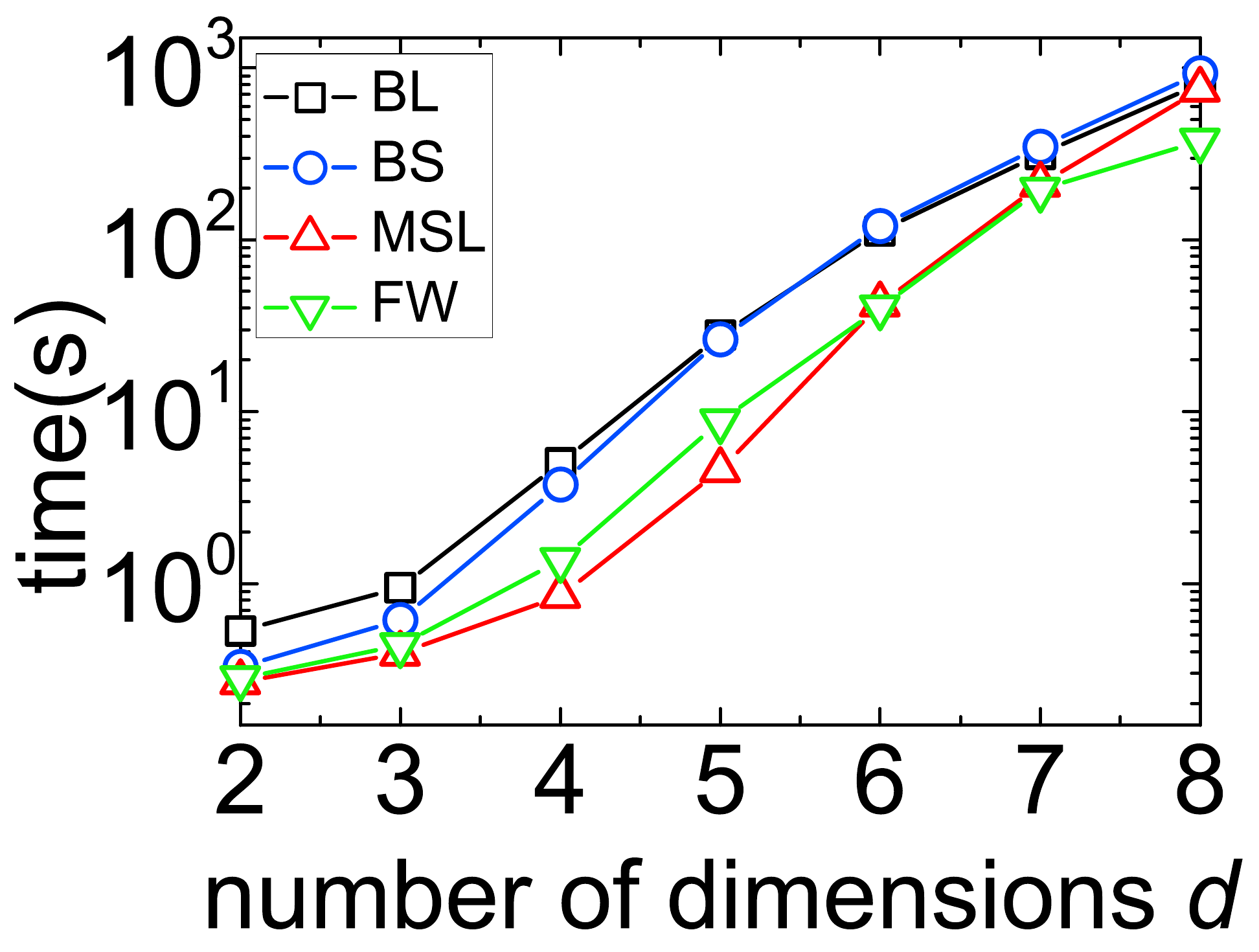}}
    \hspace{-0.15in}
  \subfigure[ANTI]{
    \label{fig:MSL_d_anti} %% label for second subfigure
    \includegraphics[scale = 0.15]{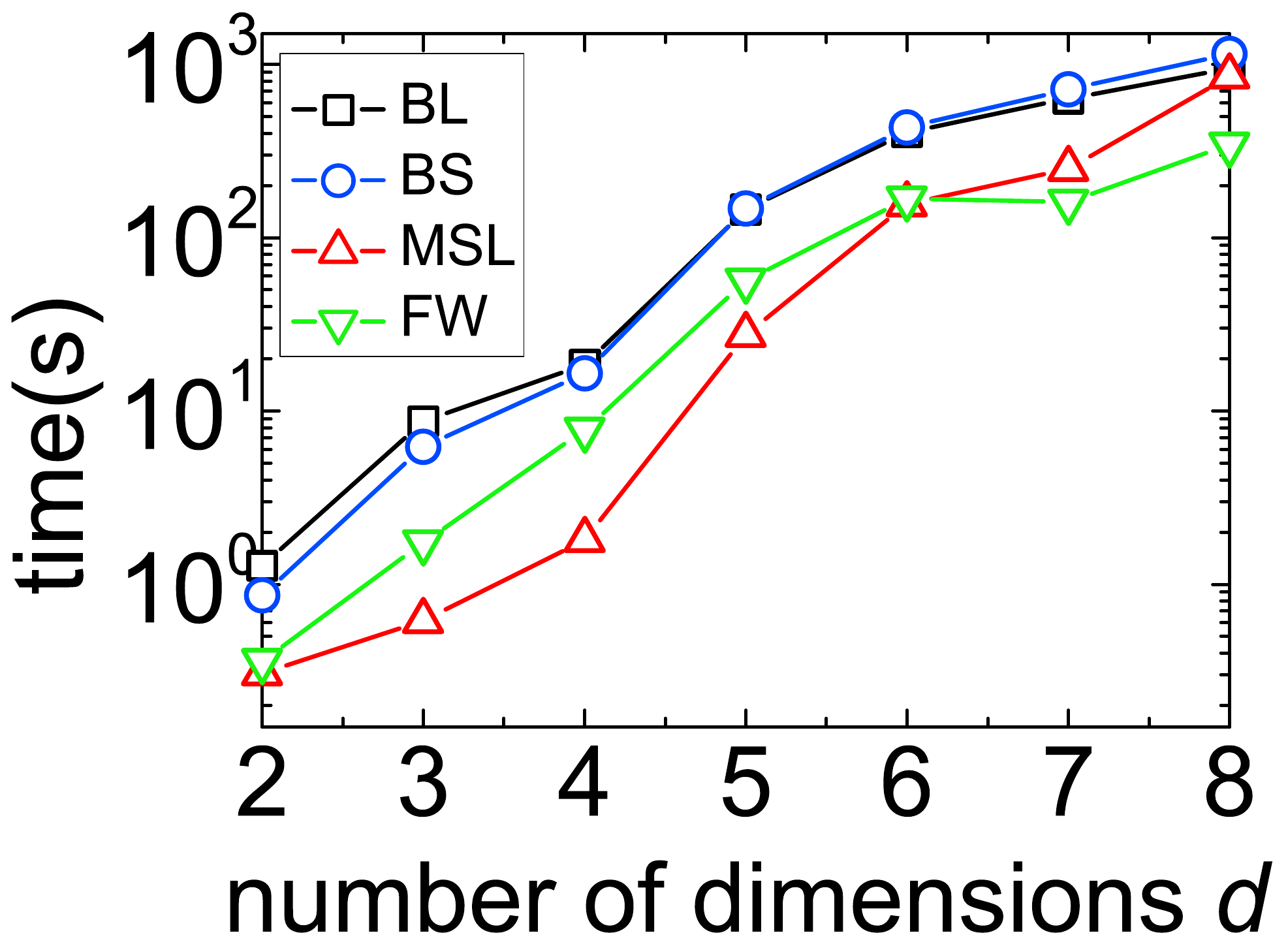}}
  \caption{MSL in synthetic datasets of varying $d$.}
  \label{fig:MSL_d} %% label for entire figure
  \vspace{-4mm}
\end{figure}

In this subsection, we run the approaches for MSL in synthetic datasets. Figure \ref{fig:comp} gives examples of MSL and compares the number of points need to be investigated by different methods. Figures \ref{fig:MSL_l}-\ref{fig:MSL_d} show the running time of MSL algorithm (presented in Subsection \ref{subsec:MSL_alg}), the binary search algorithm (BS) \cite{liu2015finding}, and the baseline approach (BL, iteratively compute and remove each skyline layer).

Figure \ref{fig:comp} compares different investigate ranges between previous methods and our method (INDE, $d = 2, l = 3, n = 50$ in Figures \ref{fig:comp}(a)(d), $n = 300$ in Figures \ref{fig:comp}(b)(e) and $n = 1,000$ in Figures \ref{fig:comp}(c)(f)). Shadow areas in Figures \ref{fig:comp}(a)(b)(c) mean the points that are investigated in binary search (BS) algorithm. Shadow areas in Figures \ref{fig:comp}(d)(e)(f) mean the points need to be investigated in our approach. It is evident that there is less redundant computation in our approach. And the larger $n$ is, the more efficient our work is.

Figure \ref{fig:MSL_l} shows the time cost of each approach with varying group size $l$ in three different datasets respectively ($n = 100,000, d = 3$). It indicates that our approach outperforms others especially when the output size is large.  When $l$ is very small, there are very few points in MSL (Figure \ref{fig:MSL_l_corr}, $l = 2$, as an example), ordering before building MSL spends much time so that our approach is not the fastest. But with the increasing of $l$, the advantage of our approach becomes more and more significant. Another interesting observation is that running the skyline algorithm 20 times (BL, $l=20$) costs 100 times more than running it twice (BL, $l=2$). This is because when we construct and remove skylines layer by layer, there are more and more points in later layers (it is obvious in Example \ref{exa:observe} and Figure \ref{fig:comp}(a)). It costs more time to construct later layers so the increasing of the total cost is not linear with the increasing of $l$. 

Figure \ref{fig:MSL_n} presents the time cost of each approach with varying number of points $n$ ($d = 3, l = 10$). For each approach, the running time grows almost linearly with the increasing of the point number $n$ since the amount of computation is almost proportional to the scale of the dataset. Also, when output size is very small, our approach does not perform the best due to the overhead cost of ordering.

Figure \ref{fig:MSL_d} illustrates the running time of each approach with different number of dimensions $d$ ($n = 100,000, l = 2$). The result shows that the efficiency advantage of our approach decreases with the increasing of $d$. It is because in our search approach (shown in Algorithm \ref{alg:msl_2}), we save the subspace skyline of each MSL layer instead of the layer itself and compare with them when investigating a new point. We reduce the number of points to store and compare to save time but we waste some to update the subspace skyline each time. However, both theoretical analysis and experiment show that the scale of skyline will grow intensively with the increasing of $d$ since a large number of attributes means difficulty for points to dominate each other (and the same situation in the subspace). In this case, the time wasted for skyline updating is more than the time gained in search strategy. When $d$ is large, the framework (FW) only, shown in Algorithm 1, performs better than the whole MSL algorithm.

\subsection{MSL on the NBA Data}
\label{subsec:MSL_NBA}
In this section, we implemented all algorithms in a real NBA dataset. We gathered $5,000$ records of players after filtering out some inferior ones (we added 5 attributes to rank players and removed the bottom ones).

\begin{figure}[ht!]
\setlength{\abovecaptionskip}{1mm}
  \centering
  \subfigure[Varying $l$]{
    \label{fig:MSL_l_NBA} %% label for first subfigure
    \includegraphics[scale = 0.15]{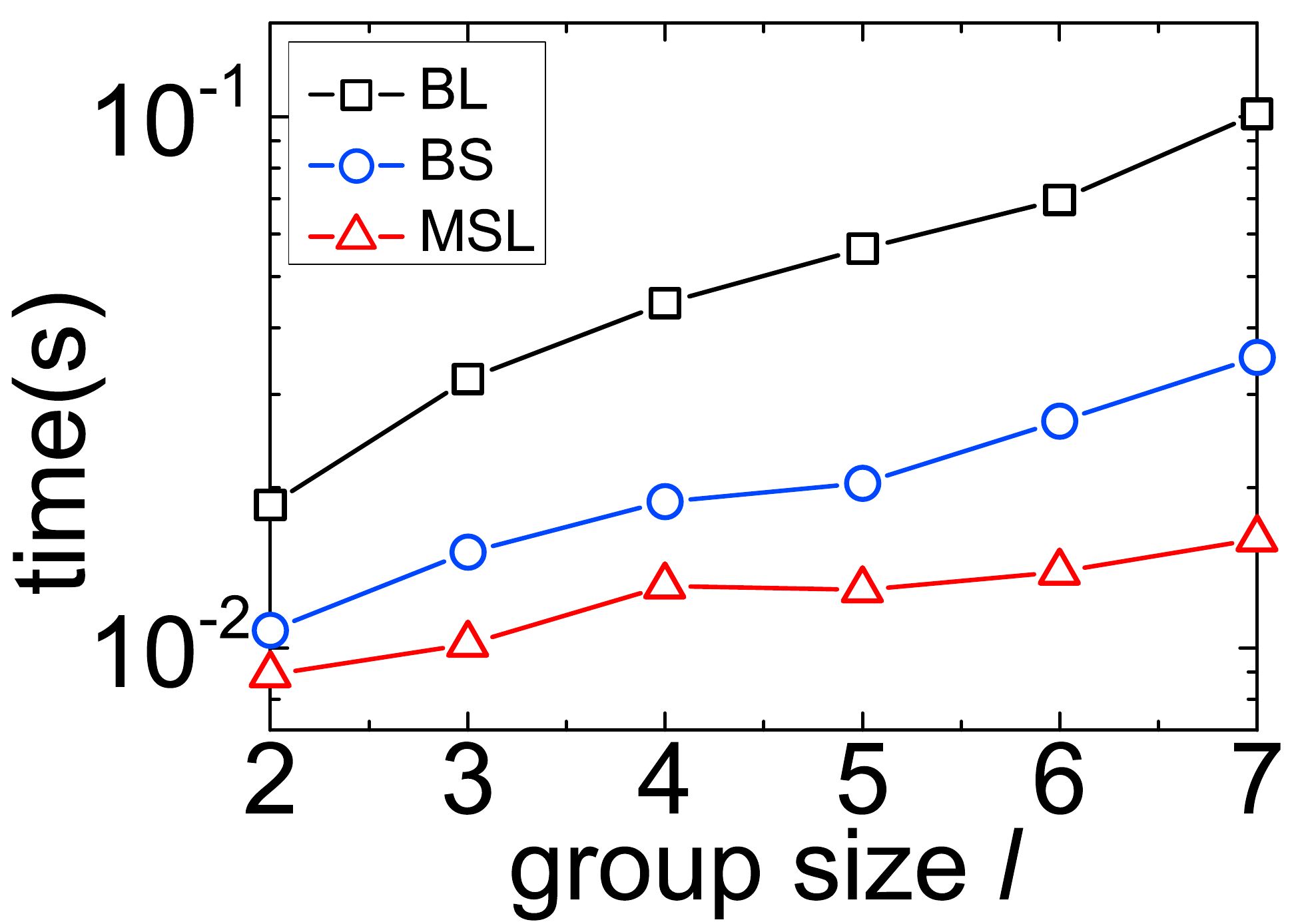}}
  \hspace{-0.1in}
  \subfigure[Varying $n$]{
    \label{fig:MSL_n_NBA} %% label for second subfigure
    \includegraphics[scale = 0.15]{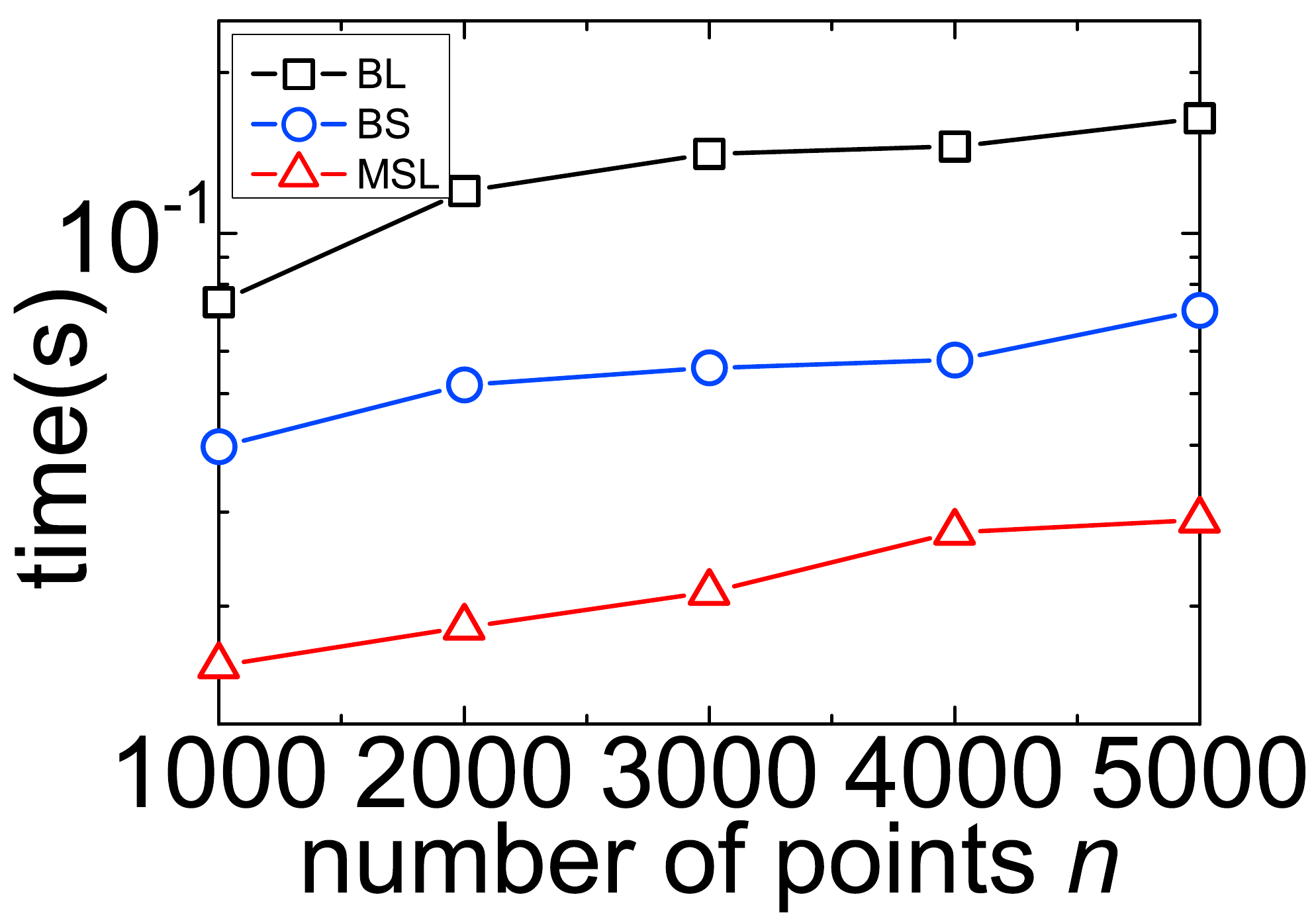}}
    \hspace{-0.2in}
  \subfigure[Varying $d$]{
    \label{fig:MSL_d_NBA} %% label for second subfigure
    \includegraphics[scale = 0.15]{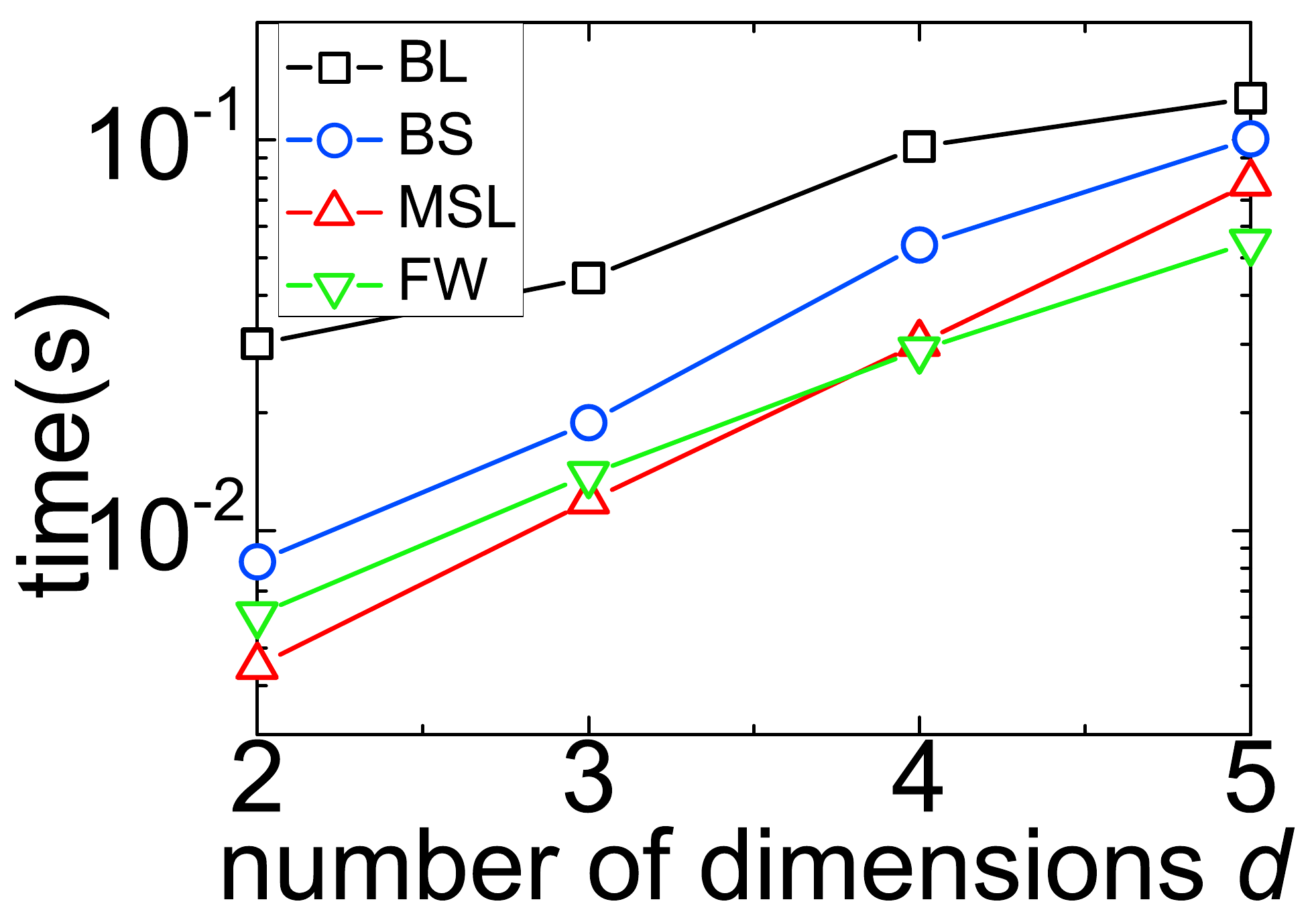}}
  \caption{MSL in NBA datasets of varying parameters.}
  \label{fig:MSL_NBA} %% label for entire figure
\end{figure}

Figure \ref{fig:MSL_NBA} shows the influence of different parameters on the time cost of MSL in the NBA dataset. Figure \ref{fig:MSL_l_NBA} represents the variation of the running time with the impact of group size $l$ ($n = 5,000, d = 3$). We can see that our approach performs better than previous approaches. The variation of the time cost with the impact of dataset size $n$ is presented in Figure \ref{fig:MSL_n_NBA} with $d = 3$ and $l = 10$. The time cost does not vary intensively with the varying $n$ due to the ``saturation'' of each MSL layer. And the variation of the running time with the impact of dimension size $d$ is represented in Figure \ref{fig:MSL_d_NBA} when $n = 5,000$ and $l = 5$. As we have analyzed in Subsection \ref{subsec:MSL_syn}, the efficiency advantage of our approach decreases distinctly with the increase of $d$ because of the dramatic increase of $\mathbb{T}_l$. Our approach costs $O\left(\mathbb{T}_l\left(n^{\frac{d-1}{d}}l+\mathbb{S}_l\log{l}\right)\right)$ time and becomes inefficient in high-dimensional space. So, in this case, we can implement framework (FW, shown in Algorithm 1) only instead of the whole MSL algorithm.

\subsection{G-skyline on the Synthetic Data}
\label{subsec:gsky_syn}

In this section, we show the experimental result of the proposed approaches for computing G-skyline. F$\_$PWise and F$\_$UWise are executed and the baseline approach (BL) is the UWise+ algorithm, which is the best performing algorithm in \cite{liu2015finding}. To show the effectiveness of our G-skyline algorithms, we only compare the time after building MSL in the experiments, to eliminate the effect of our proposed MSL algorithm.

All existing approaches for G-skyline return a candidate set that is too large to be useful. In fact, if G-skyline is used for data pruning, primary groups are not necessary to return as result. This is because when investigating if a group is a primary group, we can search if all its points are in the skyline ($l\mathbb S_1$ times for the worst case) rather than searching if it is in G-skyline (more than $\binom{\mathbb{S}_1}{l}$ times for the worst case). Only when investigating if it is a secondary group, it needs to be searched in G-skyline. So, we just output secondary groups in our methods and the result shows the time cost of each approach.

\begin{figure}[ht!]
\setlength{\abovecaptionskip}{1mm}
  \centering
  \subfigure[CORR]{
    \label{fig:gsky_l_corr} %% label for first subfigure
    \includegraphics[scale = 0.137]{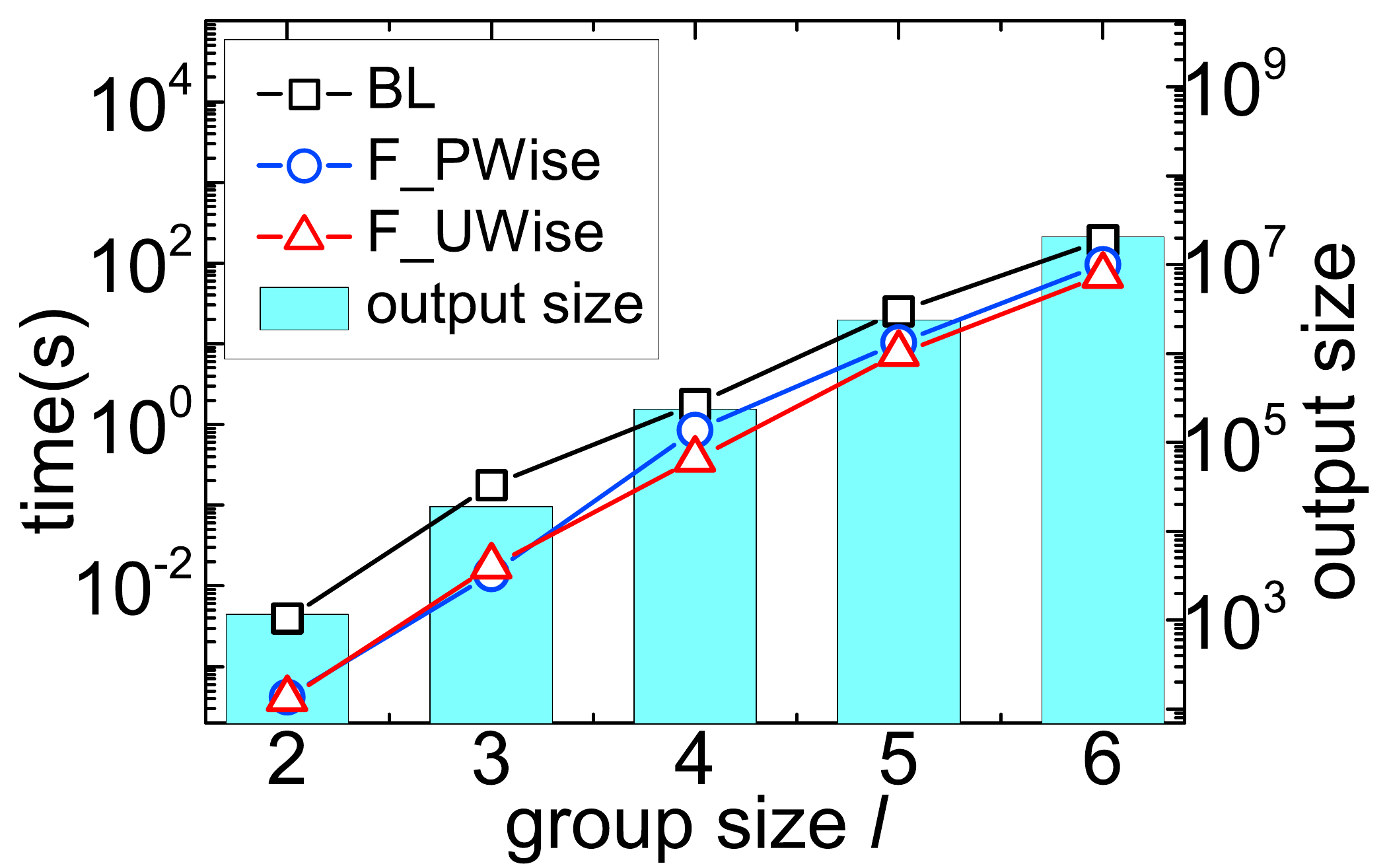}}
    \hspace{-0.1in}
  \subfigure[INDE]{
    \label{fig:gsky_l_inde} %% label for second subfigure
    \includegraphics[scale = 0.137]{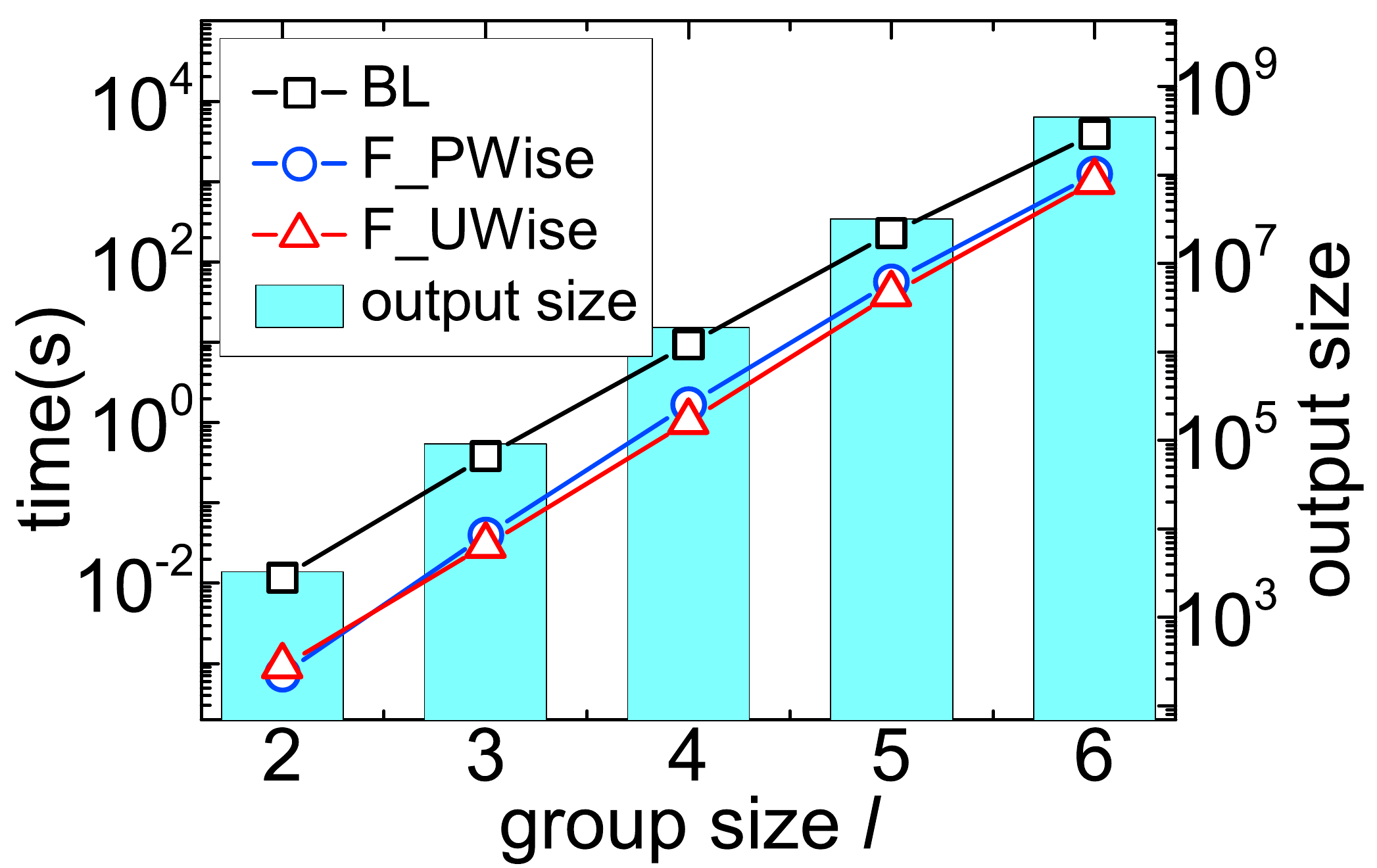}}
    \hspace{-0.1in}
  \subfigure[ANTI]{
    \label{fig:gsky_l_anti} %% label for second subfigure
    \includegraphics[scale = 0.137]{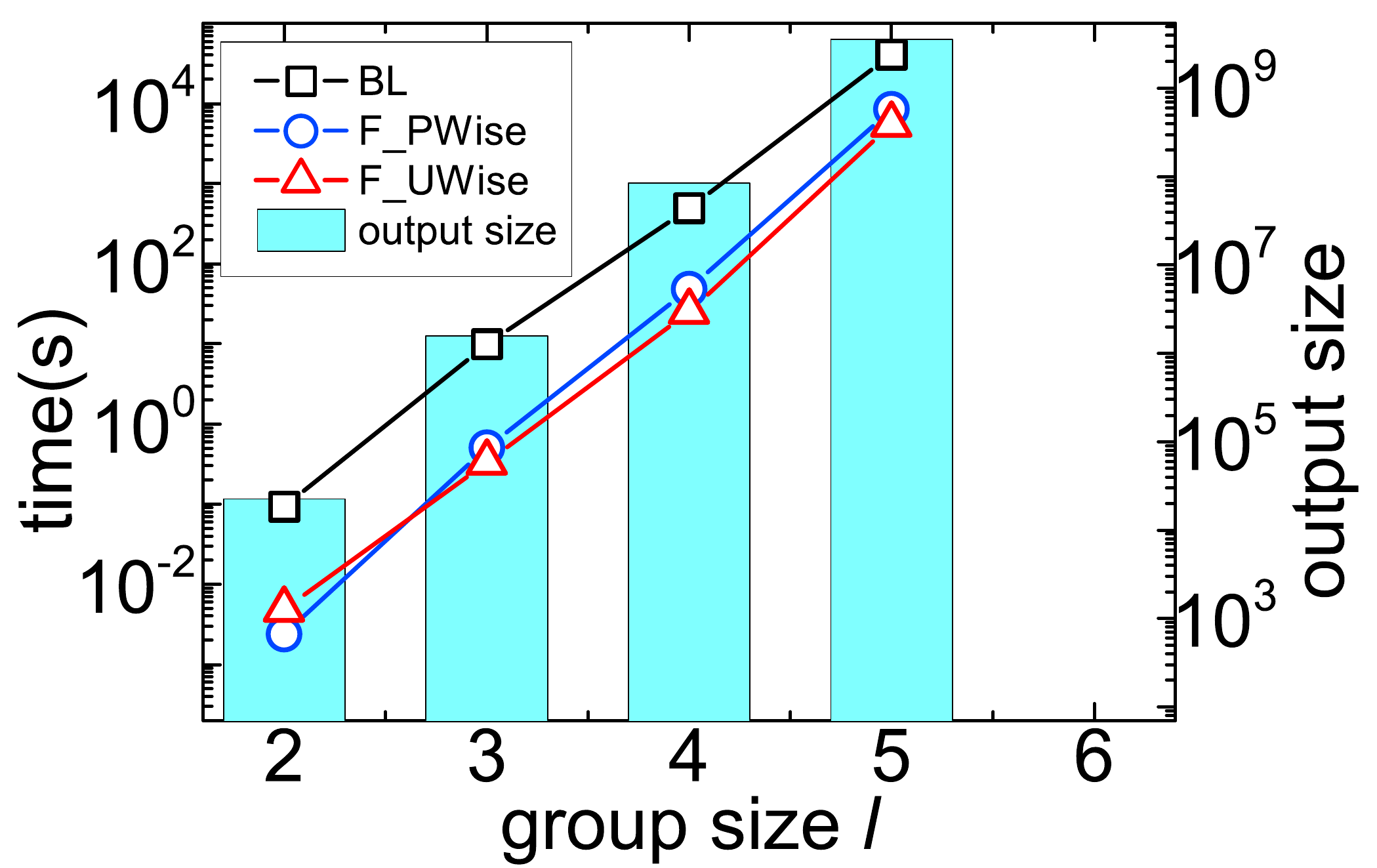}}
  \caption{G-skyline in synthetic datasets of varying $l$.}
  \label{fig:gsky_l} %% label for entire figure
\end{figure}

\begin{figure}[ht!]
\setlength{\abovecaptionskip}{1mm}
  \centering
  \subfigure[CORR]{
    \label{fig:gsky_n_corr} %% label for first subfigure
    \includegraphics[scale = 0.137]{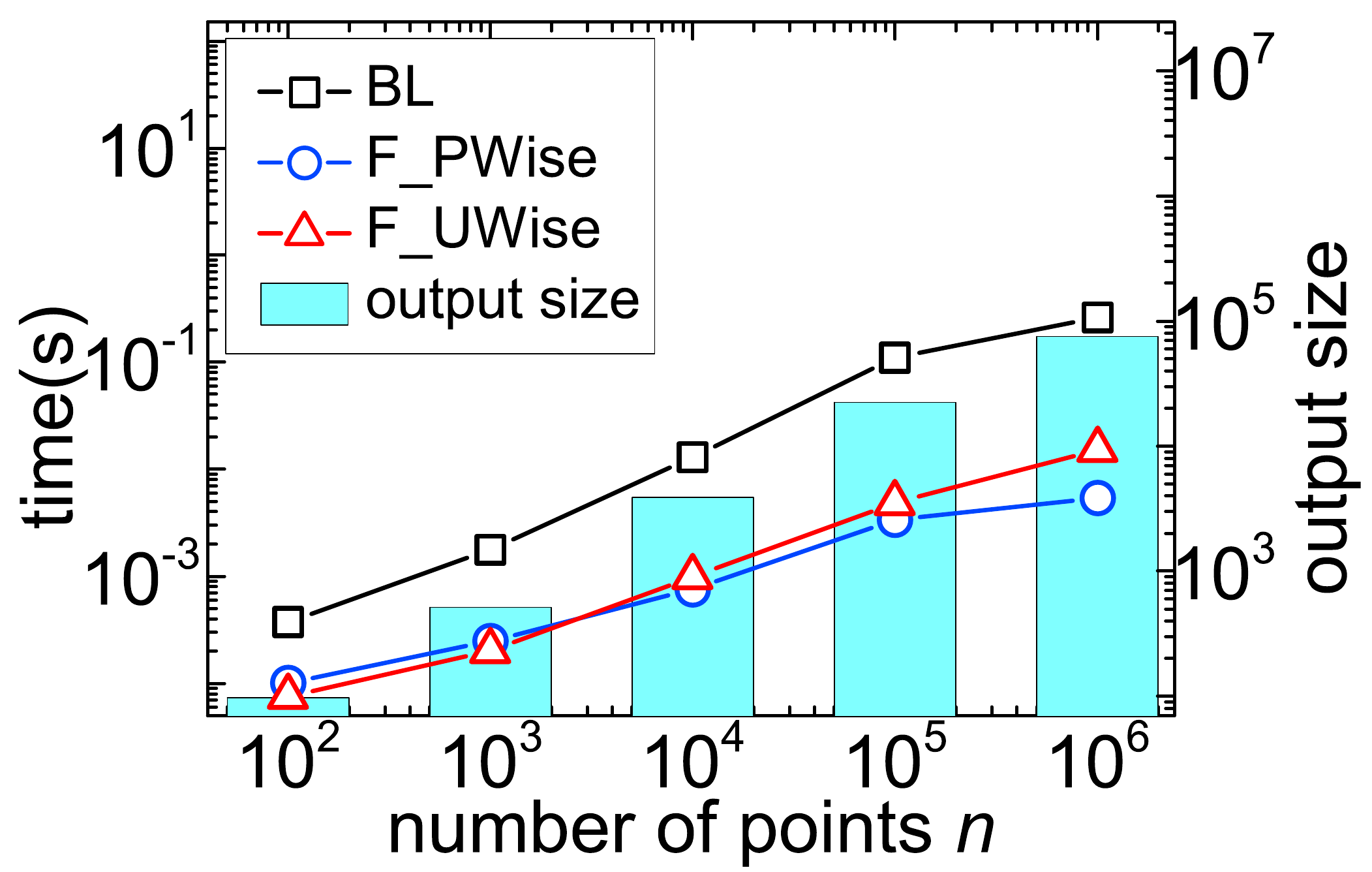}}
    \hspace{-0.12in}
  \subfigure[INDE]{
    \label{fig:gsky_n_inde} %% label for second subfigure
    \includegraphics[scale = 0.137]{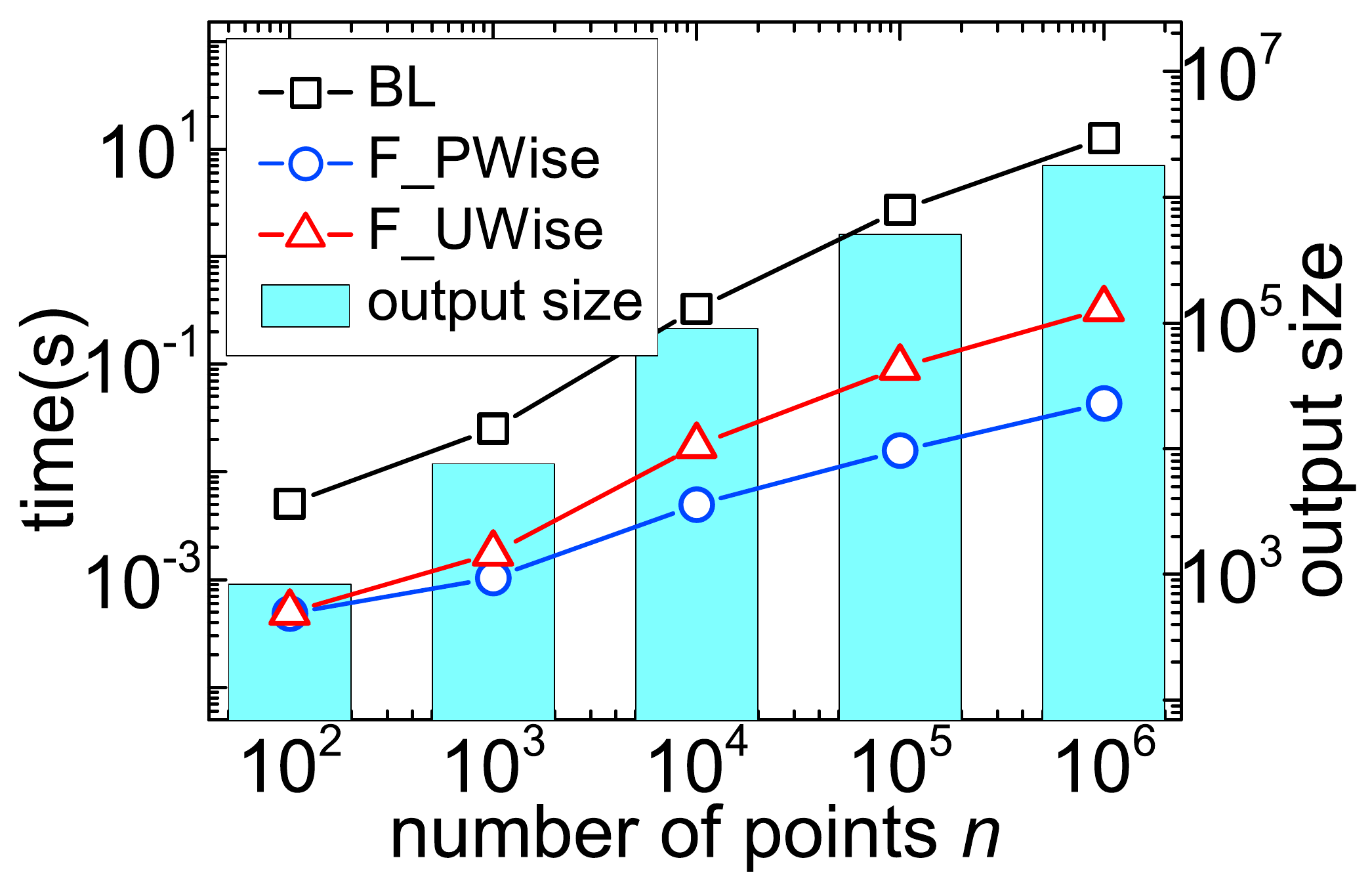}}
    \hspace{-0.12in}
  \subfigure[ANTI]{
    \label{fig:gsky_n_anti} %% label for second subfigure
    \includegraphics[scale = 0.137]{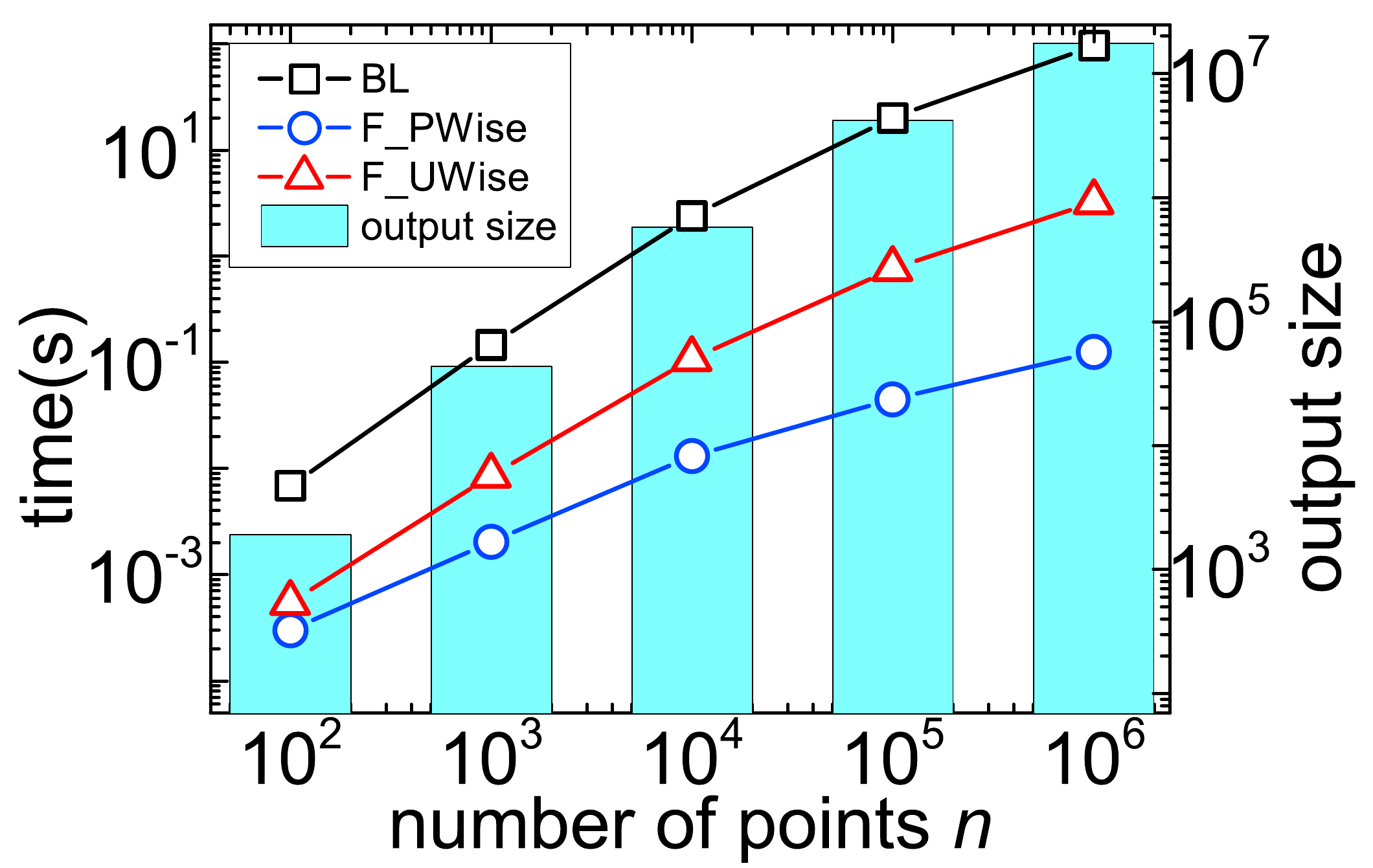}}
  \caption{G-skyline in synthetic datasets of varying $n$.}
  \label{fig:gsky_n} %% label for entire figure
\end{figure}

\begin{figure}[ht!]
\setlength{\abovecaptionskip}{1mm}
  \centering
  \subfigure[CORR]{
    \label{fig:gsky_d_corr} %% label for first subfigure
    \includegraphics[scale = 0.137]{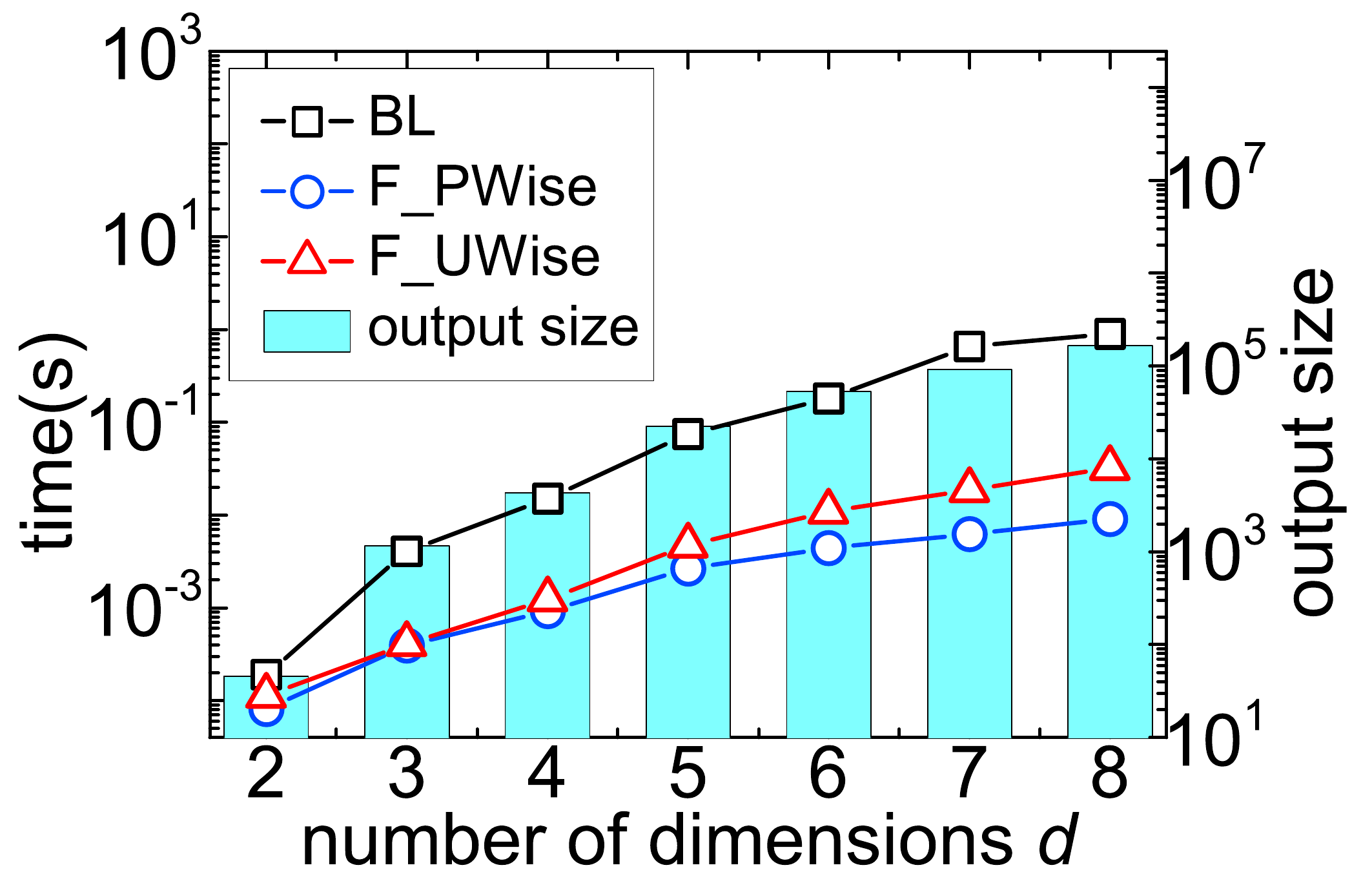}}
    \hspace{-0.15in}
  \subfigure[INDE]{
    \label{fig:gsky_d_inde} %% label for second subfigure
    \includegraphics[scale = 0.137]{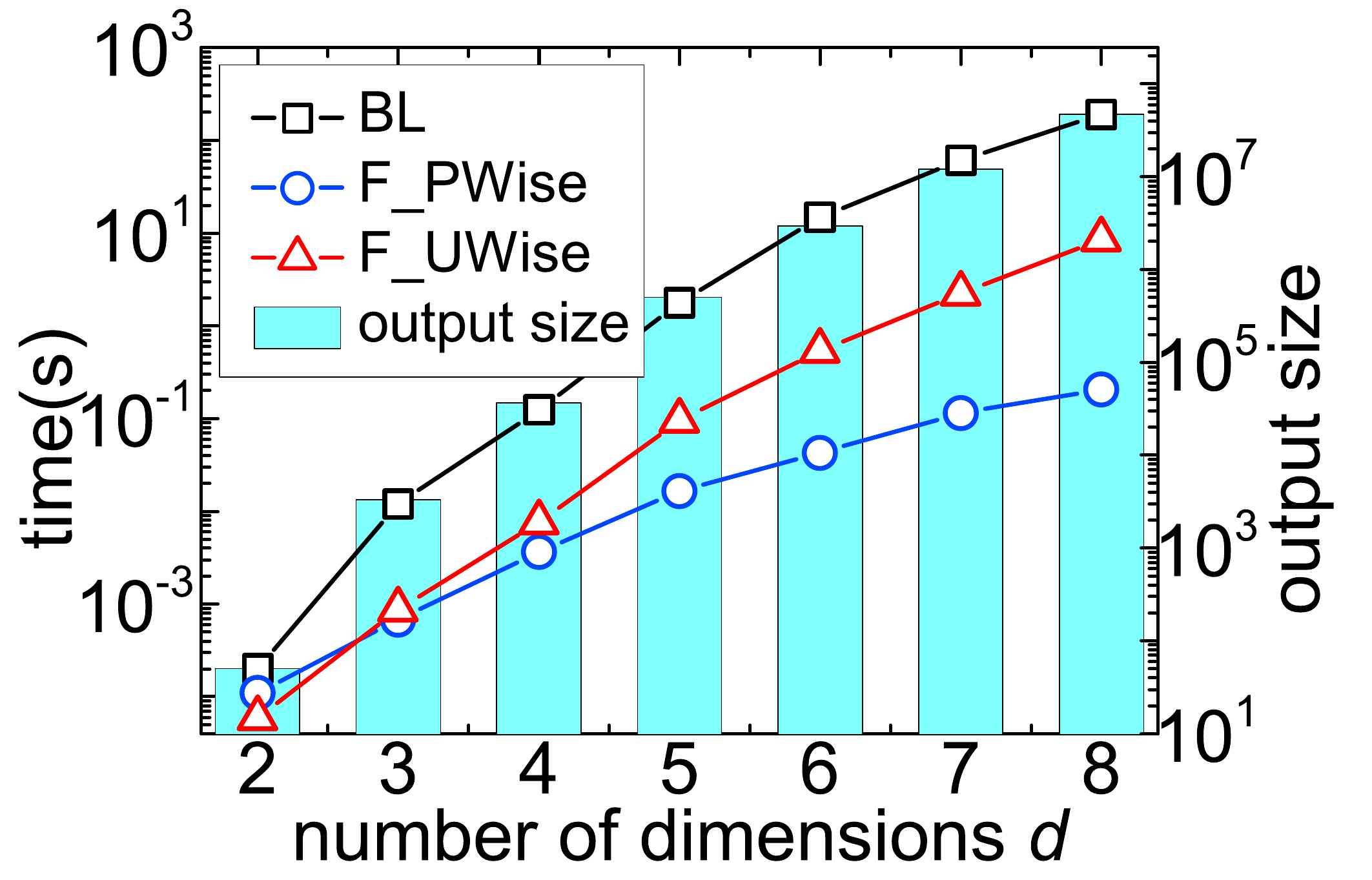}}
    \hspace{-0.15in}
  \subfigure[ANTI]{
    \label{fig:gsky_d_anti} %% label for second subfigure
    \includegraphics[scale = 0.137]{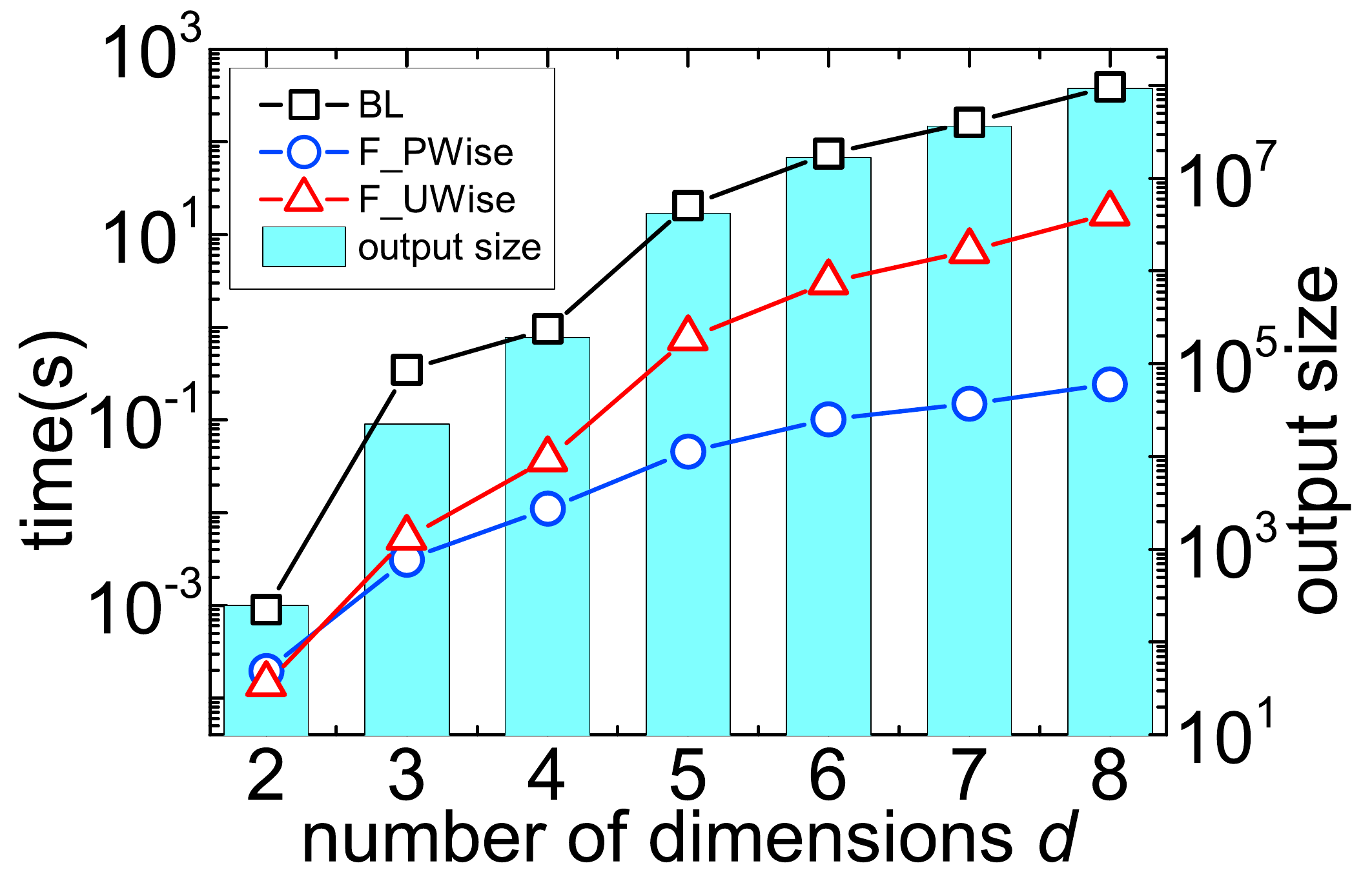}}
  \caption{G-skyline in synthetic datasets of varying $d$.}
  \label{fig:gsky_d} %% label for entire figure
\end{figure}

Line charts in Figures \ref{fig:gsky_l} to \ref{fig:gsky_d} show the time cost of BL, F$\_$PWise and F$\_$UWise and histograms in Figures \ref{fig:gsky_l} to \ref{fig:gsky_d} show the size of G-skyline with certain varying parameter (group size $l$, number of points $n$, and number of dimensions $d$) in CORR, INDE, and ANTI dataset respectively.

Figure \ref{fig:gsky_l} shows the time cost and the G-skyline size with varying group size $l$ ($n = 100,000, d = 3$). We can see that the increasing of output size is almost exponential with the increasing of the group size $l$, accordingly the running time of each approach also increases exponentially with it. 

Figure \ref{fig:gsky_n} shows the time cost and the output size with varying number of points $n$ ($d = 5, l = 2$). Distinctly, they both increase approximately linearly with the increasing of $n$.

Figure \ref{fig:gsky_d} shows the time cost and the size of G-skyline with varying number of dimensions $d$ ($n = 100,000, l = 2$), which increase approximately exponentially with the increasing of $d$.

Unit group based approach is a pretty artful way for the G-skyline problem. It fits the property of G-skyline group well and UWise+ usually performs better than PWise \cite{liu2015finding}. But in our new approaches, F$\_$PWise performs much better than F$\_$UWise in most situations, especially in a large output size. We can see that the new pruning strategy promotes the efficiency of F$\_$PWise dramatically. In addition, there are many overlaps between each unit groups. When the output size is large, there are a large number of unit groups to search and the overlap issue becomes serious whereas our point-based method is enhanced dramatically by the edge pruning.

\subsection{G-skyline on the NBA Data}
\label{subsec:gsky_NBA}

\begin{figure}[ht!]
\setlength{\abovecaptionskip}{1mm}
  \centering
  \subfigure[Varying $l$]{
    \label{fig:gsky_l_NBA} %% label for first subfigure
    \includegraphics[scale = 0.137]{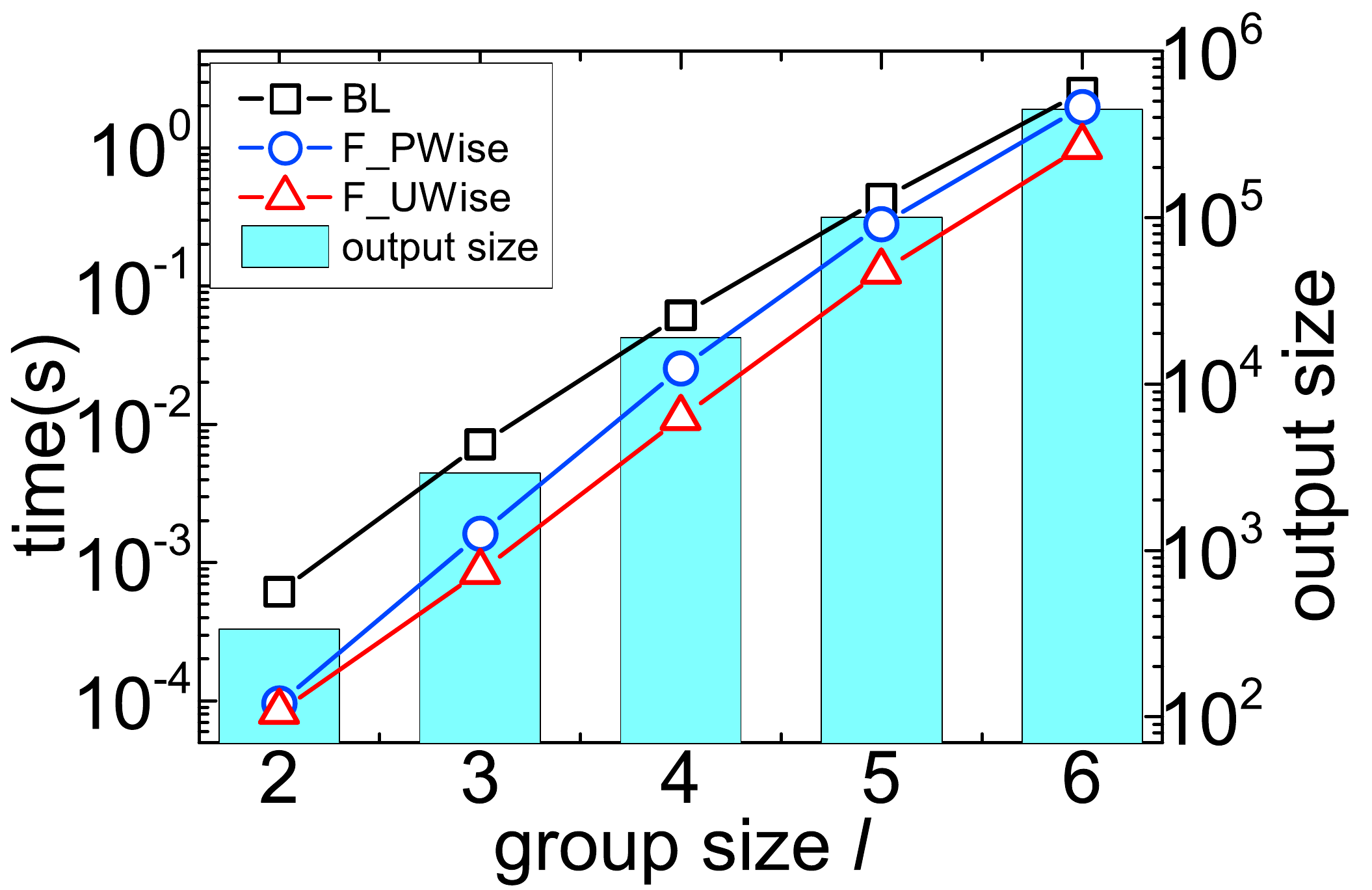}}
  \hspace{-0.13in}
  \subfigure[Varying $n$]{
    \label{fig:gsky_n_NBA} %% label for second subfigure
    \includegraphics[scale = 0.137]{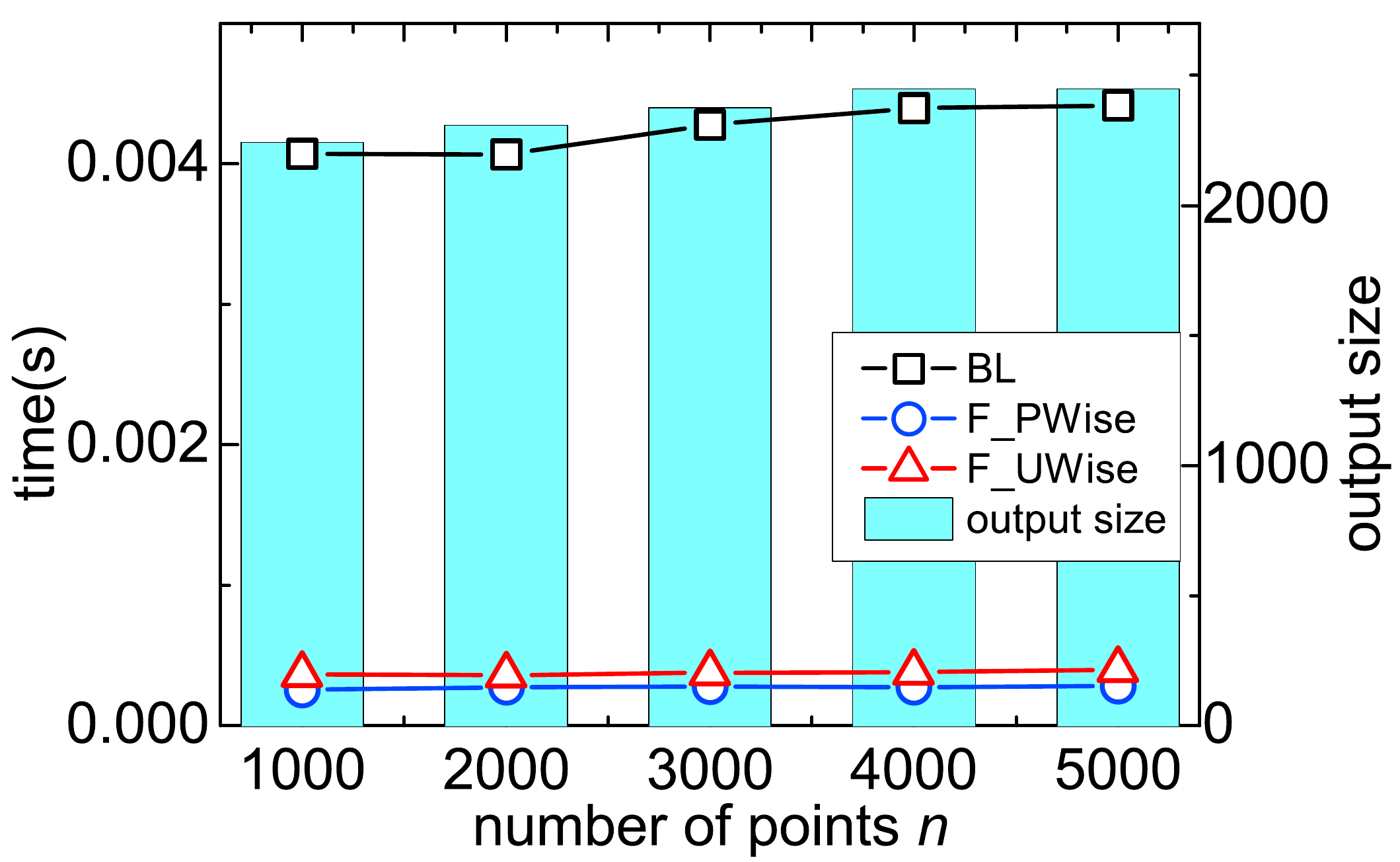}}
    \hspace{-0.12in}
  \subfigure[Varying $d$]{
    \label{fig:gsky_d_NBA} %% label for second subfigure
    \includegraphics[scale = 0.137]{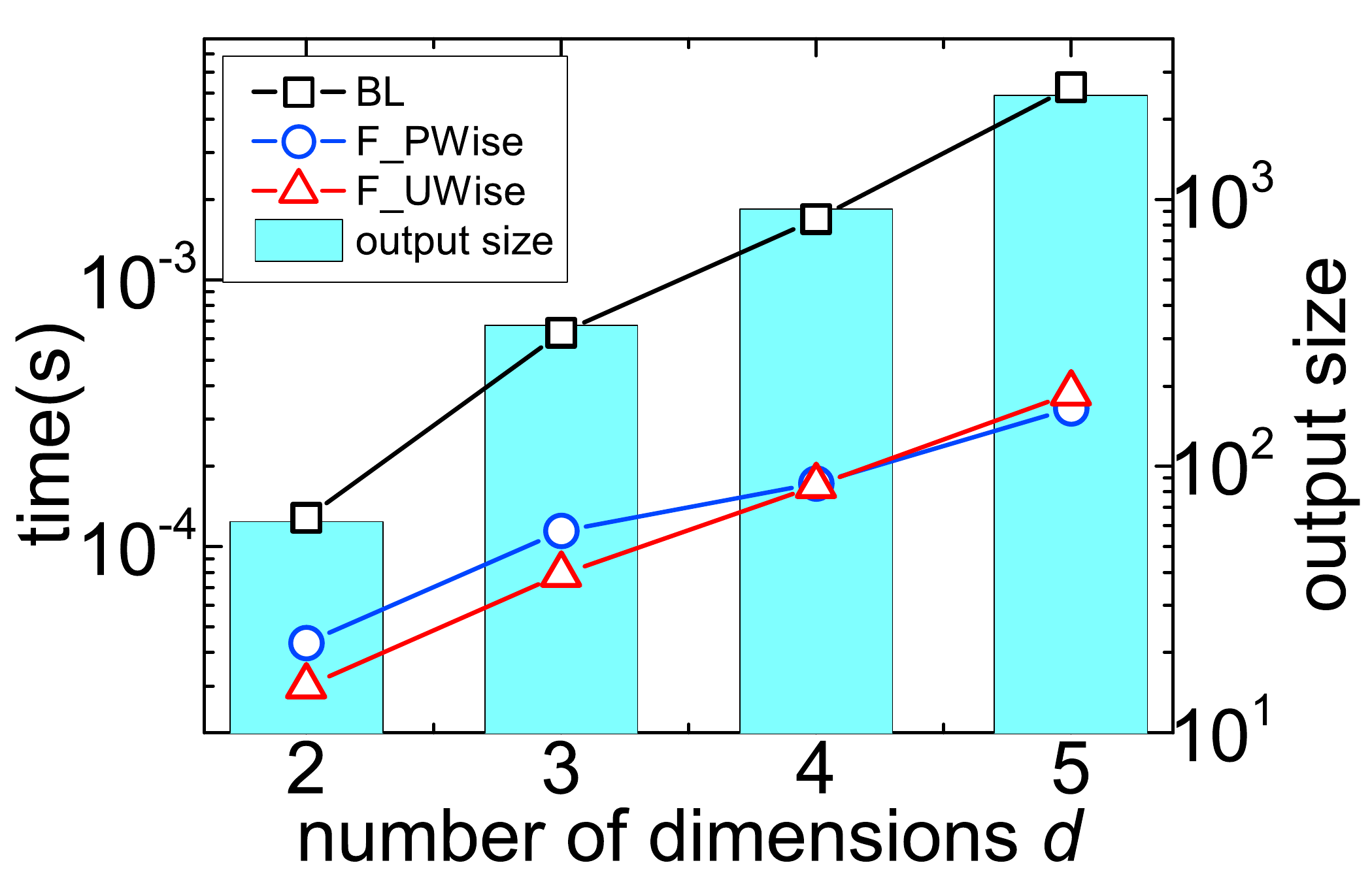}}
  \caption{G-skyline in NBA dataset with varying parameters.}
  \label{fig:gsky_NBA} %% label for entire figure
\end{figure}

In this subsection, we implemented BL (UWise+), F$\_$UWise, and F$\_$PWise in NBA dataset and report the result. Figure \ref{fig:gsky_NBA} shows the output size and the time cost of G-skyline in NBA dataset with different parameters.

The line chart in Figure \ref{fig:gsky_l_NBA} shows the variation of the running time for G-skyline with the impact of group size $l$ ($n = 5,000, d = 3$). The improvement of our approaches is not so significant compared with that in other cases shown in Figure \ref{fig:gsky_n_NBA} or Figure \ref{fig:gsky_d_NBA}, it may be because that our approaches omit primary groups, which account for a lower proportion when there are many layers in MSL (shown in Table \ref{sample_tab}). As a result, the time complexity is the same order of magnitude with the baseline. Another interesting phenomenon is that in most situation, F$\_$PWise performs better, especially with a large $n$ or $d$, however F$\_$UWise generally performs better in this situation. The reason may be that with a large $l$, there are too many children of each node in DSG and enumerating them takes significant time in F$\_$PWise. And in F$\_$UWise, the overlap issue, mentioned in Subsection \ref{subsec:gsky_syn}, is not so serious. The histogram in Figure \ref{fig:gsky_l_NBA} is the illustration of output size with different group size $l$.

The line chart in Figure \ref{fig:gsky_n_NBA} illustrates the variation of the time cost with the impact of dataset size $n$ with $d = 5$ and $l = 2$ and the histogram in Figure \ref{fig:gsky_n_NBA} illustrates the variation of the output size. We can observe that the output size does not vary prominently. The reason may be that the NBA dataset is correlated, so the skyline becomes ``saturate'' even with the increase of the dataset size, hence the running time and the output size of G-skyline keep constant.

The line chart in Figure \ref{fig:gsky_d_NBA} presents the variation of the running time with the impact of dimension size $d$ when $n = 5,000$ and $l = 2$. Line chart in Figure \ref{fig:gsky_d} and Figure \ref{fig:gsky_d_NBA} show that with the increase of $d$, F$\_$Pwise becomes more and more efficient and surpasses F$\_$UWise eventually. The reason may be that when there are more attributes, it is harder for a point to dominate another and there are fewer child nodes of each node in DSG, so the enumeration complexity reduces and F$\_$PWise becomes more efficient.

\subsection{Representative skyline on the Synthetic Data}
\label{subsec:repre_syn}

In this subsection, we devise experiments to validate the effectiveness of all algorithms for the assignment problem and the representative skyline (including $k$-SGQ and RG-skyline). Finally, we report the performance.

\begin{figure}[ht!]
\setlength{\abovecaptionskip}{1mm}
  \centering
  \subfigure[Time consumption]{
    \label{fig:greed_time} %% label for first subfigure
    \includegraphics[scale = 0.15]{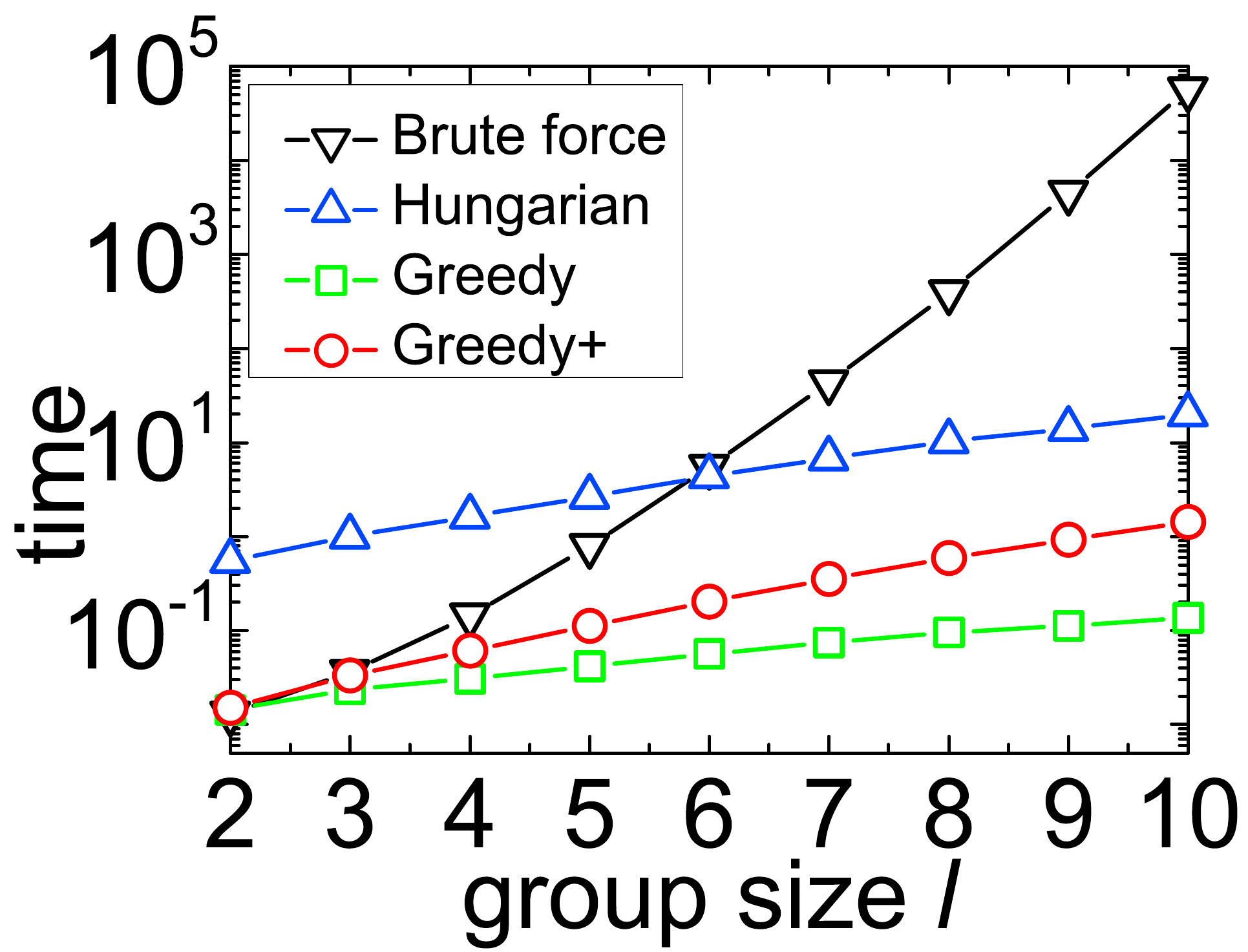}}
  \hspace{0.1in}
  \subfigure[Error]{
    \label{fig:greed_error} %% label for second subfigure
    \includegraphics[scale = 0.15]{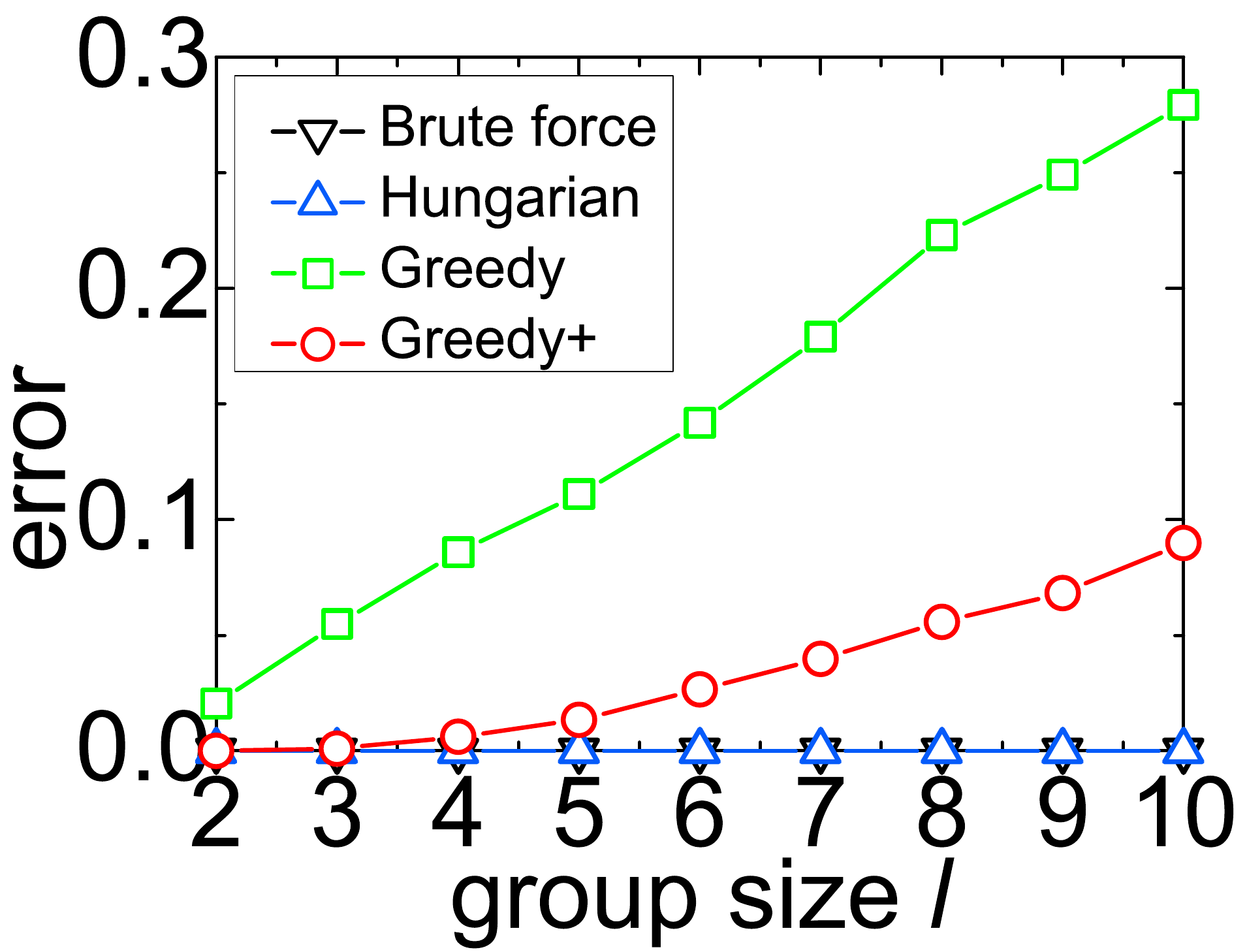}}
  \caption{Comparison of the methods for the assignment problem.}
  \label{fig:greed} %% label for entire figure
\end{figure}

In Figure \ref{fig:greed}, the performance of four methods for the assignment problem (Brute-force, Hungarian, Greedy, and Greedy+) are reported. Figure \ref{fig:greed_time} shows the cost of time of each method and Figure \ref{fig:greed_error} shows the error. Brute-force calculates the distance by enumerating all possible matchings. We can see that though small at the beginning, the cost of Brute-force increases with the explosive exponential growth when $l$ grows. With much additional computation, Hungarian algorithm is far from satisfactory in the situation with small $l$. Greedy is the most efficient method though very inaccurate. Our proposed Greedy+ shows a good balance between cost of time and accuracy.

\begin{figure}[ht!]
\setlength{\abovecaptionskip}{1mm}
  \centering
  \subfigure[CORR]{
    \label{fig:repre_k_corr} %% label for first subfigure
    \includegraphics[scale = 0.15]{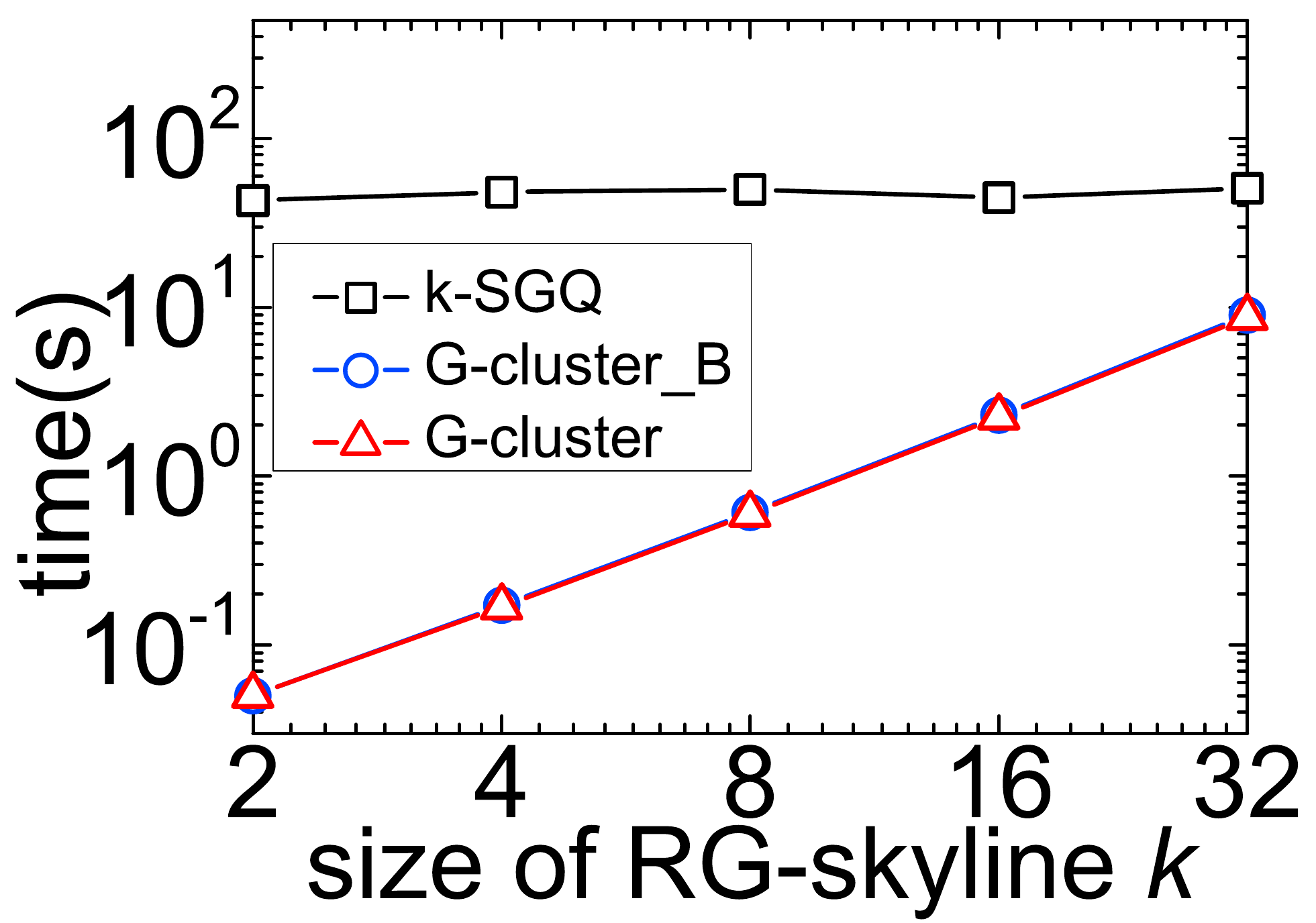}}
  \hspace{-0.15in}
  \subfigure[INDE]{
    \label{fig:repre_k_inde} %% label for second subfigure
    \includegraphics[scale = 0.15]{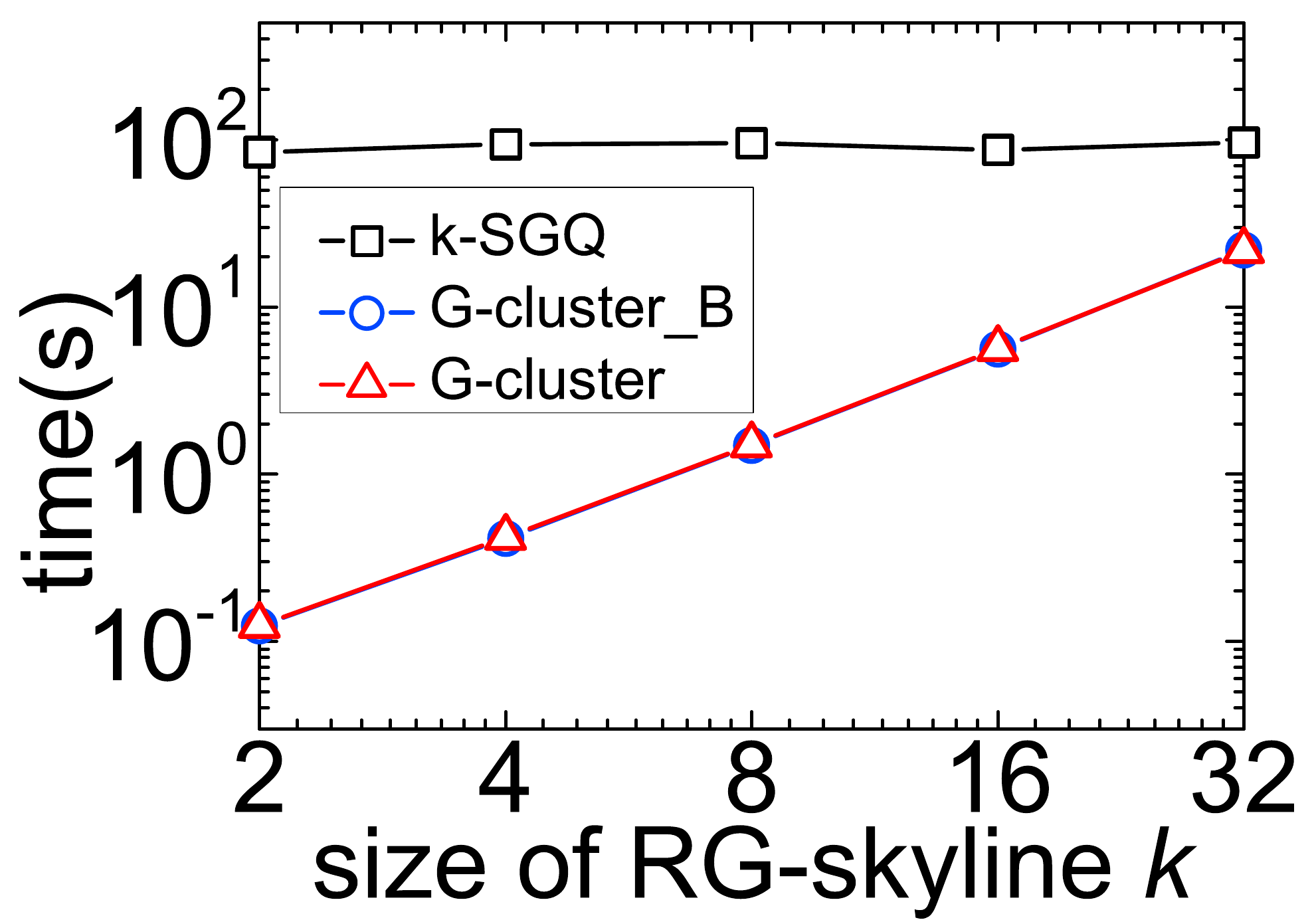}}
    \hspace{-0.15in}
  \subfigure[ANTI]{
    \label{fig:repre_k_anti} %% label for second subfigure
    \includegraphics[scale = 0.15]{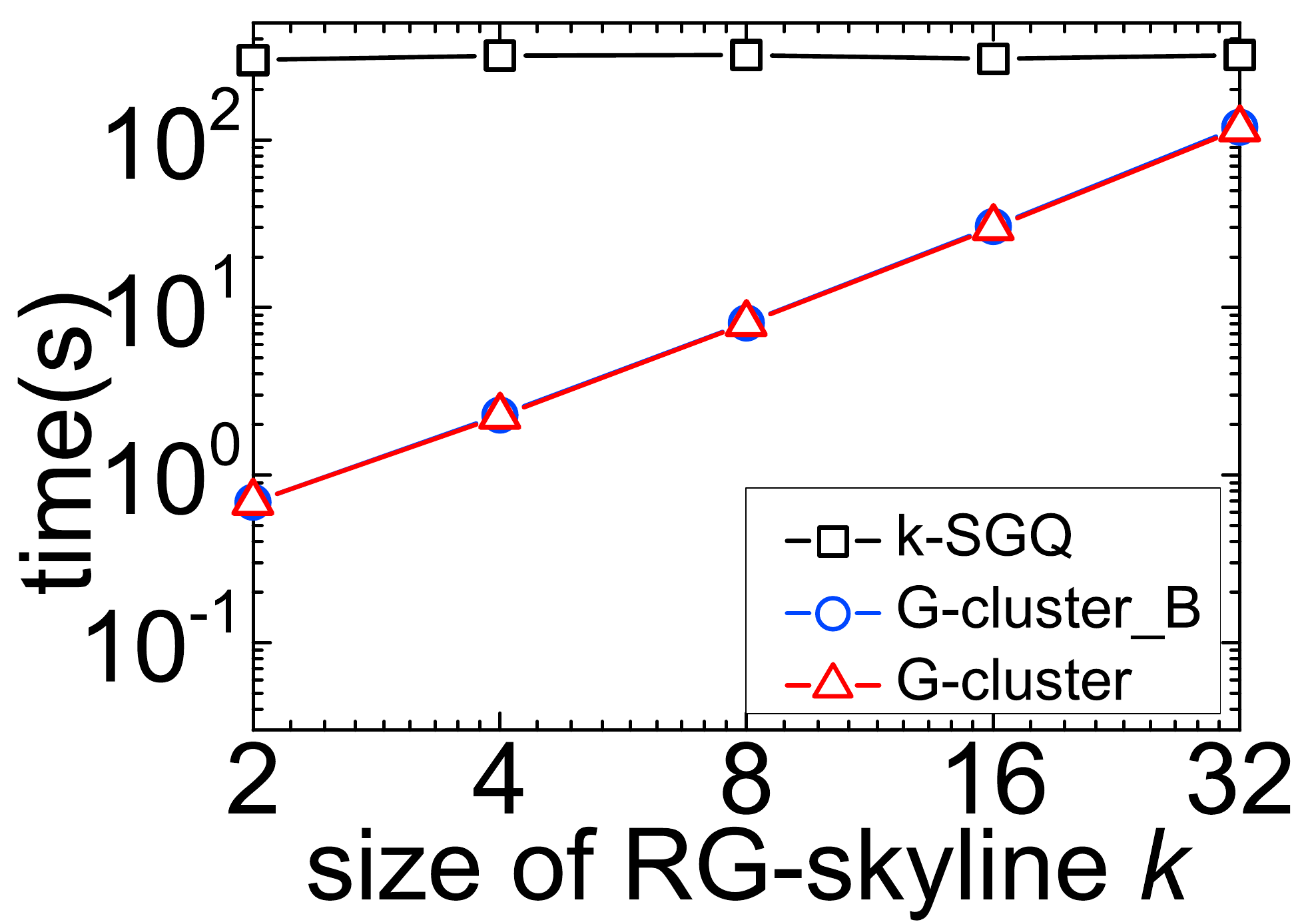}}
  \caption{Representative skyline in synthetic datasets of varying $k$.}
  \label{fig:repre_k} %% label for entire figure
\end{figure}

\begin{figure}[ht!]
\setlength{\abovecaptionskip}{1mm}
  \centering
  \subfigure[CORR]{
    \label{fig:repre_l_corr} %% label for first subfigure
    \includegraphics[scale = 0.15]{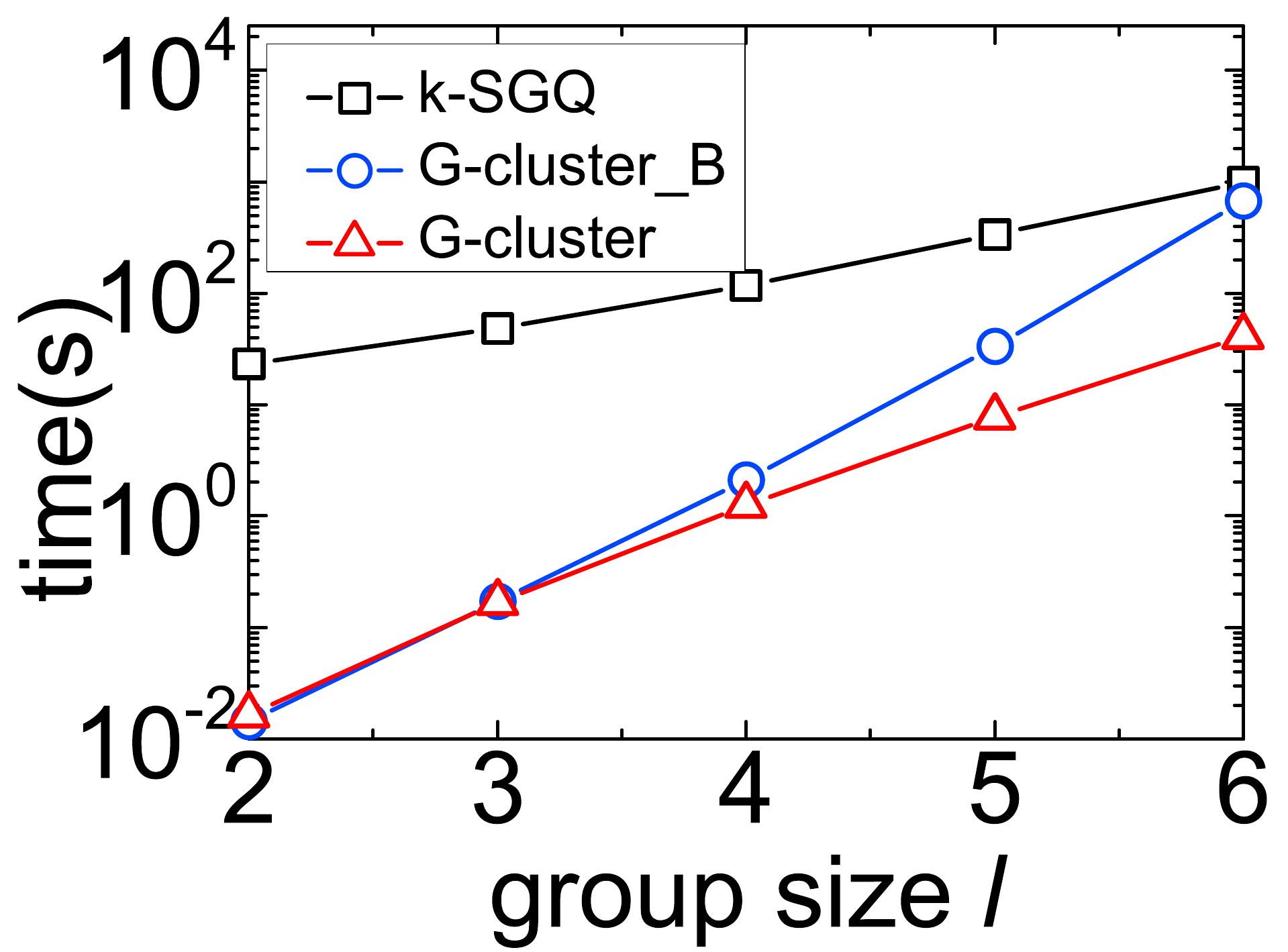}}
  \hspace{-0.15in}
  \subfigure[INDE]{
    \label{fig:repre_l_inde} %% label for second subfigure
    \includegraphics[scale = 0.15]{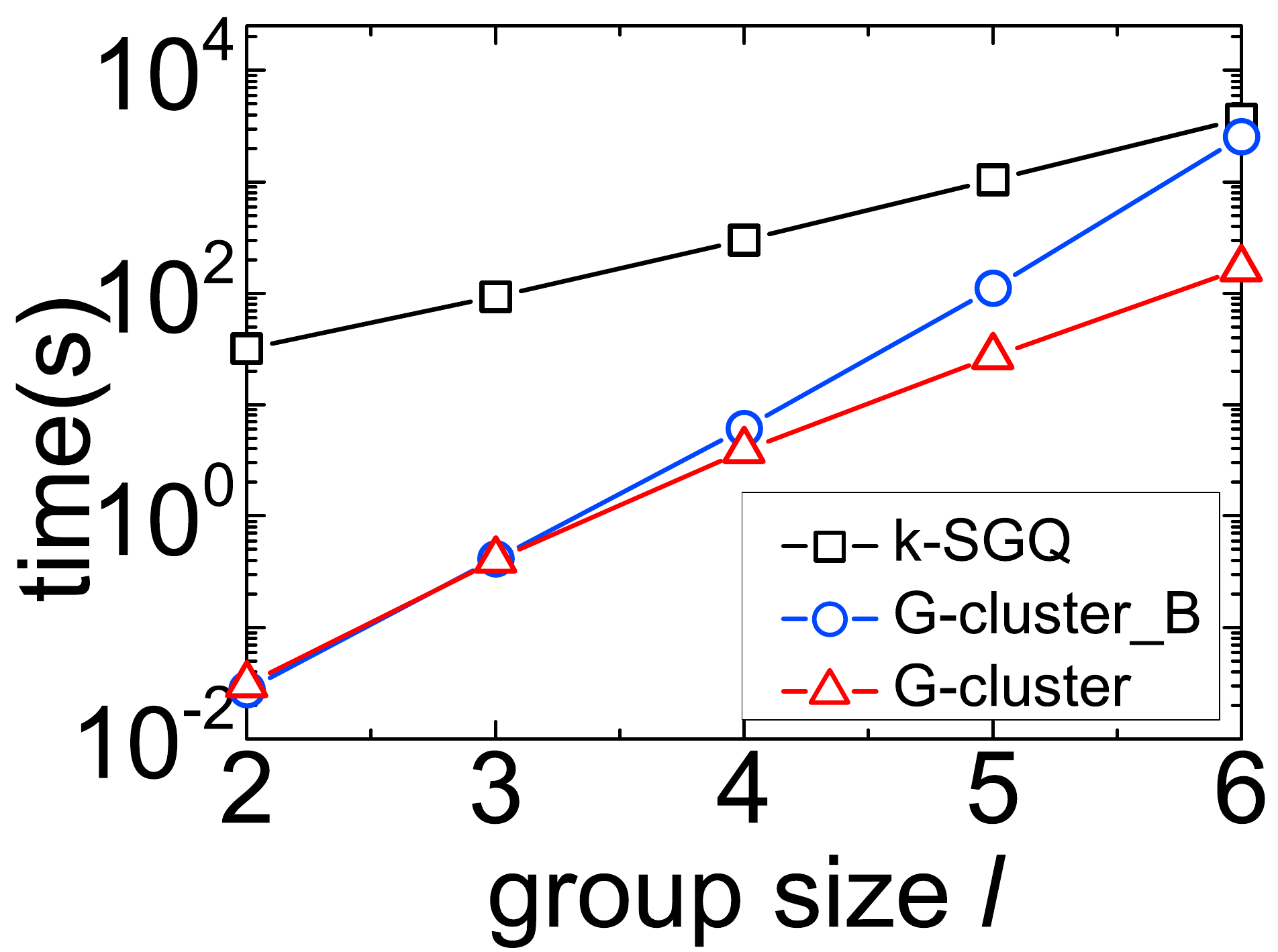}}
    \hspace{-0.15in}
  \subfigure[ANTI]{
    \label{fig:repre_l_anti} %% label for second subfigure
    \includegraphics[scale = 0.15]{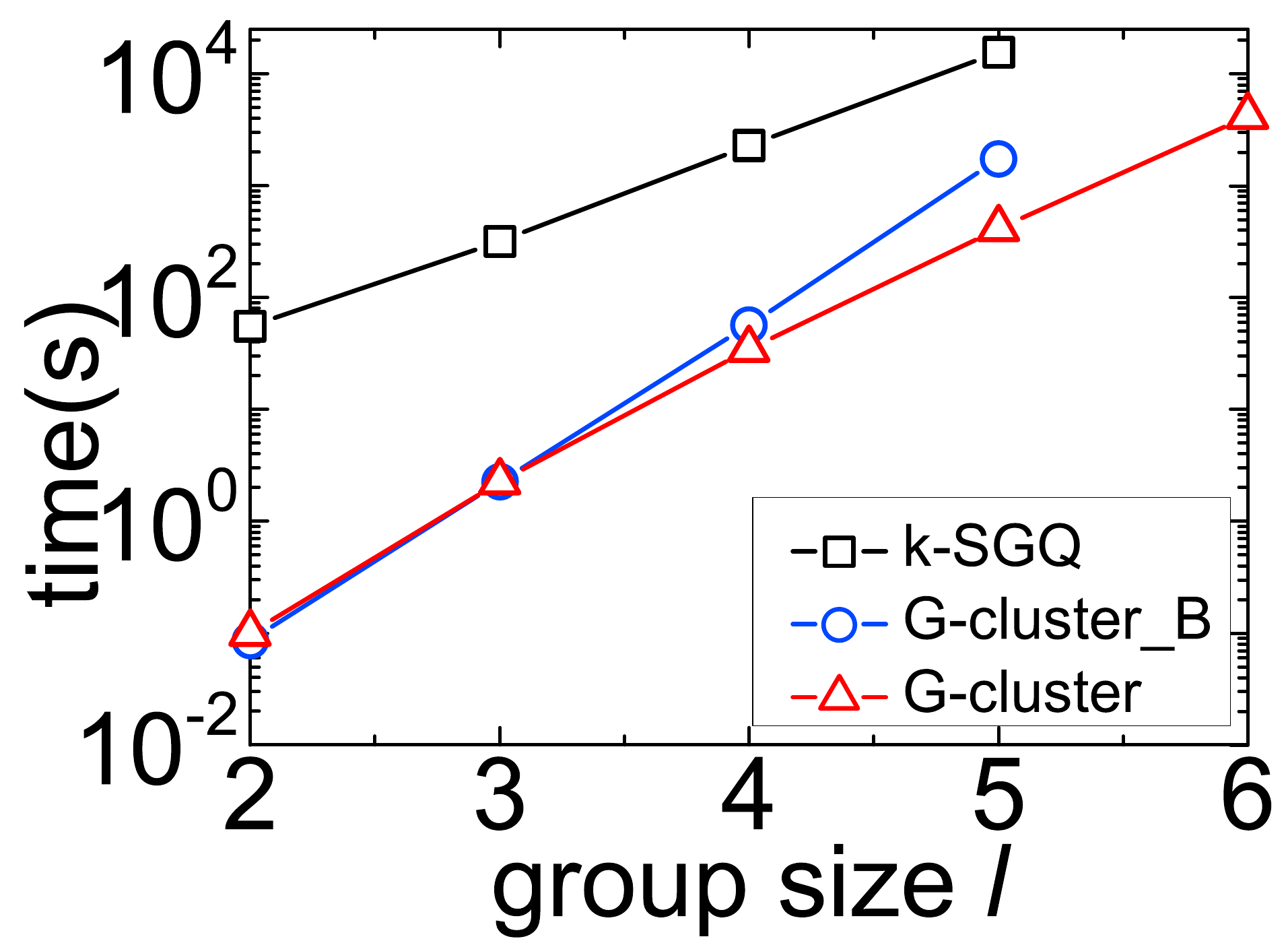}}
  \caption{Representative skyline in synthetic datasets of varying $l$.}
  \label{fig:repre_l} %% label for entire figure
\end{figure}

\begin{figure}[ht!]
\setlength{\abovecaptionskip}{1mm}
  \centering
  \subfigure[CORR]{
    \label{fig:repre_n_corr} %% label for first subfigure
    \includegraphics[scale = 0.15]{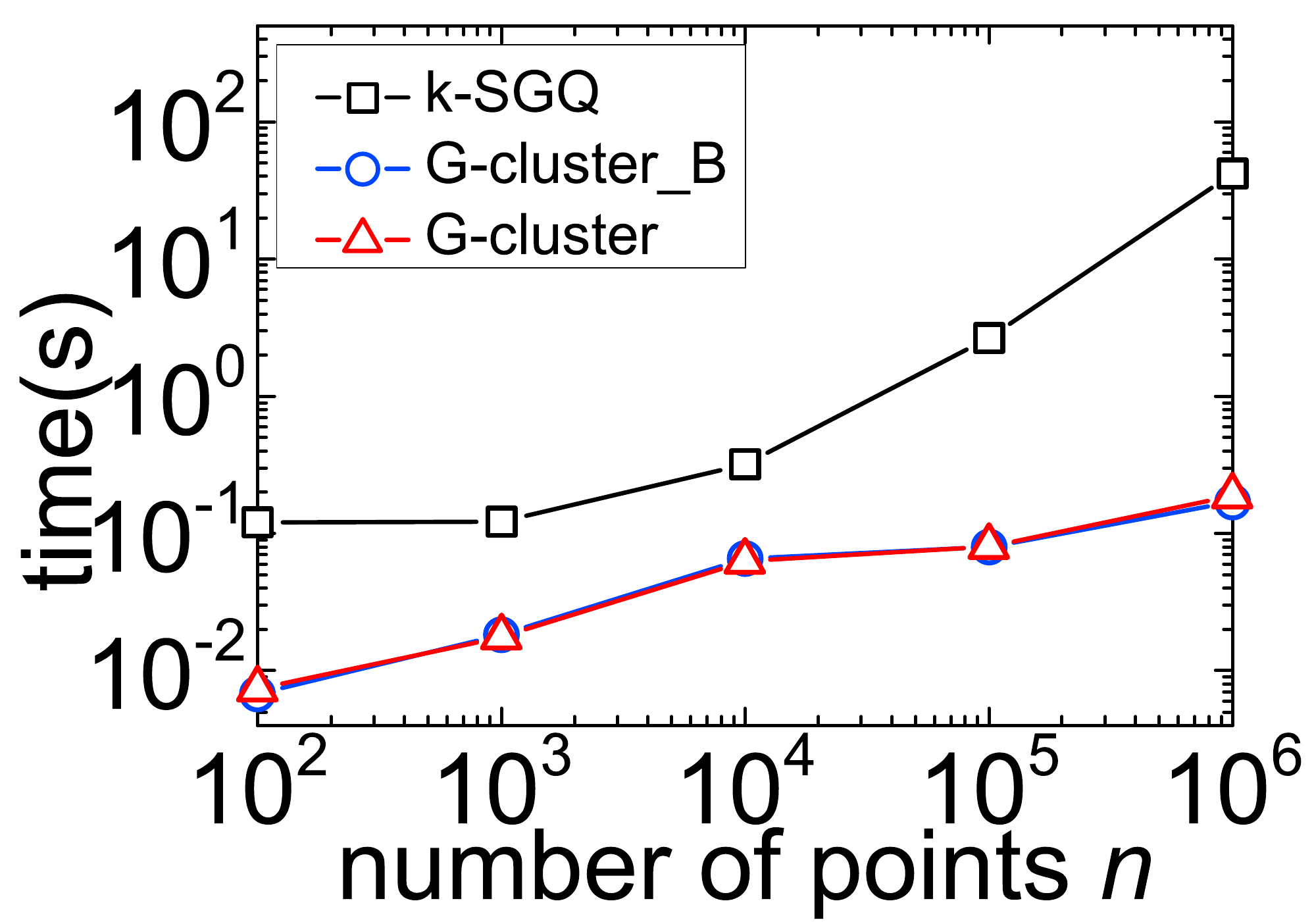}}
  \hspace{-0.18in}
  \subfigure[INDE]{
    \label{fig:repre_n_inde} %% label for second subfigure
    \includegraphics[scale = 0.15]{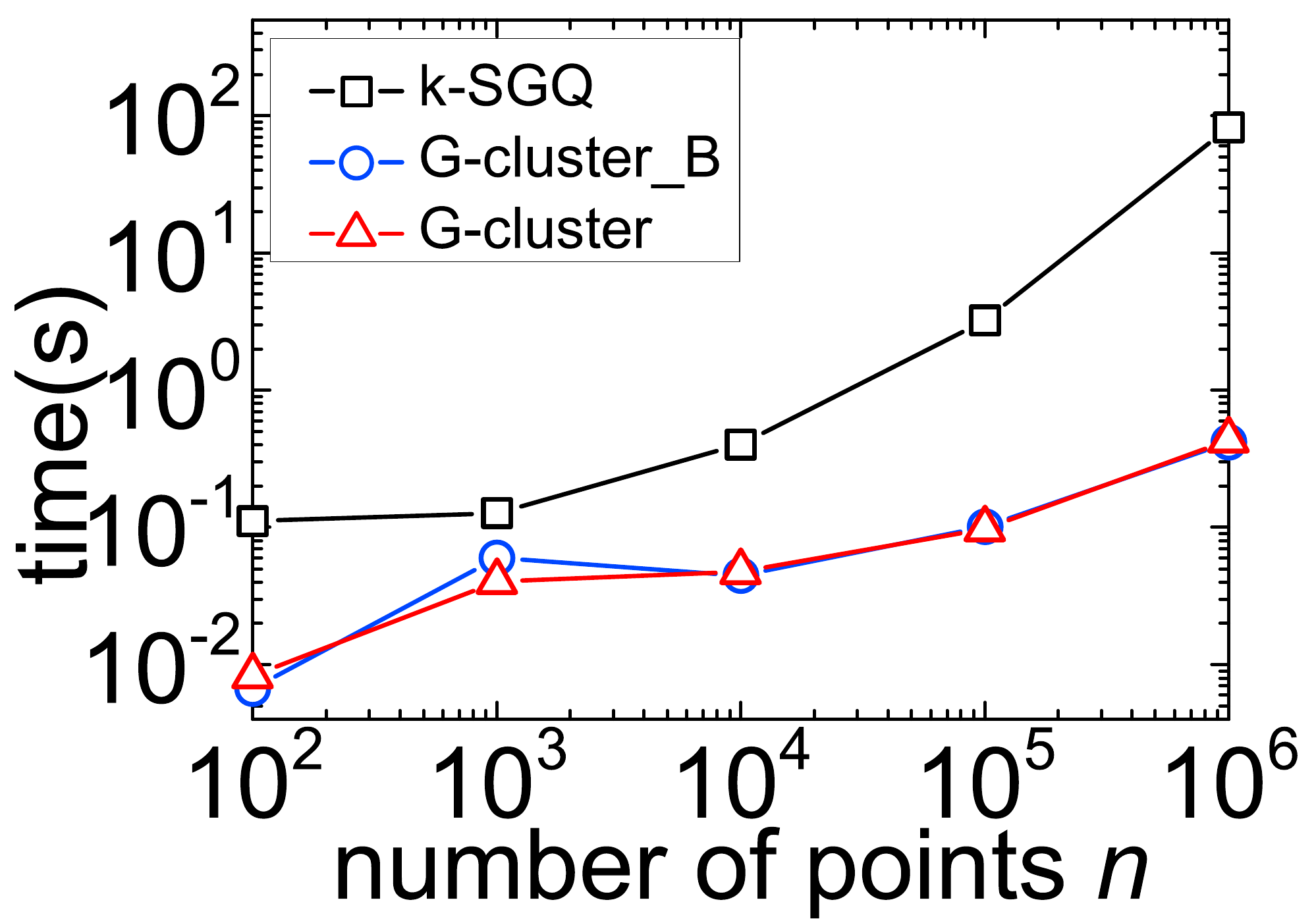}}
    \hspace{-0.18in}
  \subfigure[ANTI]{
    \label{fig:repre_n_anti} %% label for second subfigure
    \includegraphics[scale = 0.15]{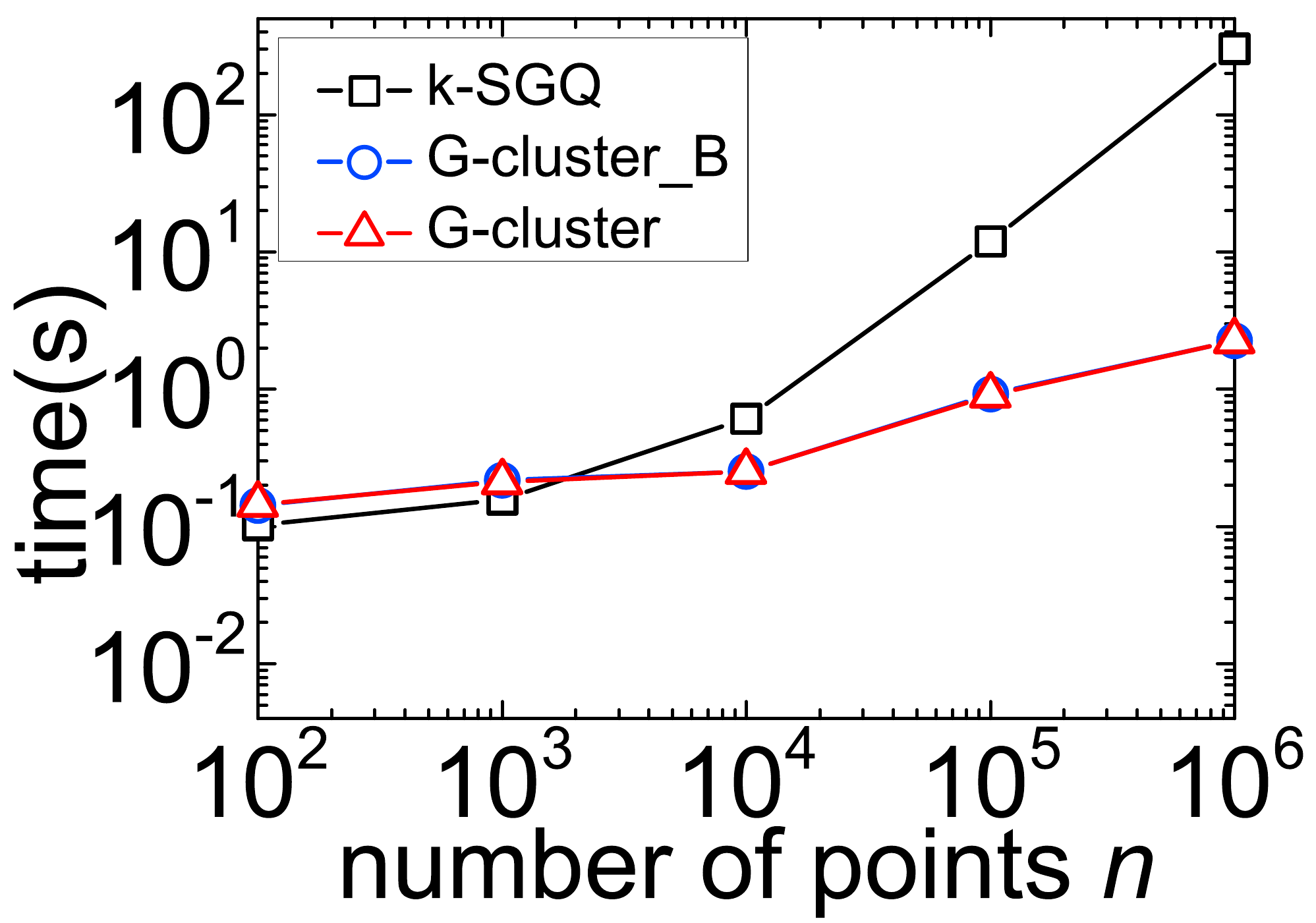}}
  \caption{Representative skyline in synthetic datasets of varying $n$.}
  \label{fig:repre_n} %% label for entire figure
\end{figure}

We then show the performance of algorithms for representative skyline: $k$-SGQ \cite{max-dominance}, 
G-cluster\_B (G-clustering with Brute-force to match the points), and G-clustering. $k$-SGQ algorithm constructs the $k$-SGQ while G-cluster\_B and G-clustering construct the RG-skyline. Figures \ref{fig:repre_k} to \ref{fig:repre_n} show the cost of time to calculate representative skyline in three synthetic datasets.

Figure \ref{fig:repre_k} shows performance with varying $k$ ($l=3$, $n=1,000,000$, $d=2$). As shown in the figure, the time cost of $k$-SGQ keeps constant with the increasing of $k$, since we only need to iterate one time to construct the RG-skyline. G-cluster\_B and G-clustering has a linear time cost with respect to $k$, since we need to iterate $k$ times. Since the time cost of Brute-force and Greedy+ are similar (represented in Figure \ref{fig:greed_time}), their lines almost coincide.

Figure \ref{fig:repre_l} illustrates performance of three algorithms with varying group size $l$ ($k=4$, $n=1,000,000$, $d=2$). Benefiting from our novel Greedy+ method, G-clustering outperforms G-cluster\_B as $l$ grows.

Figure \ref{fig:repre_n} presents the time cost with a varying number of points $n$ ($k=4$, $l=3$, $d=2$). The time cost of $k$-SGQ increases obviously with the increase of $n$. This is because when there are a large number of points in the dataset, it will cost more time to search how many points a group can dominate. 

\subsection{Representative skyline on the NBA Data}
\label{subsec:repre_NBA}

In this subsection, we report the performance of all algorithms for representative skyline in the NBA real-world dataset.

\begin{figure}[ht!]
\setlength{\abovecaptionskip}{1mm}
  \centering
  \subfigure[Varying $k$]{
    \label{fig:repre_k_NBA} %% label for first subfigure
    \includegraphics[scale = 0.15]{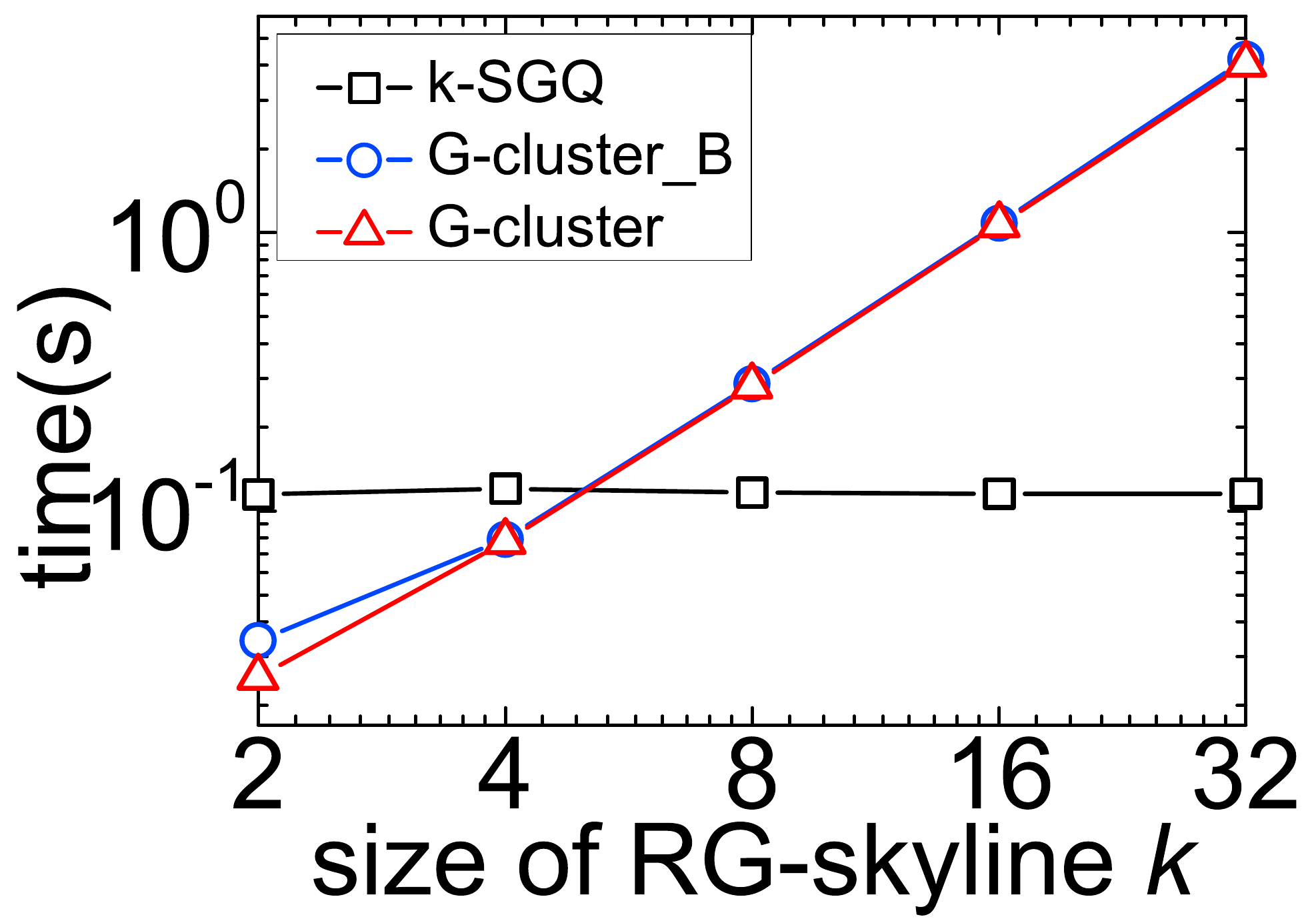}}
  \hspace{-0.15in}
  \subfigure[Varying $l$]{
    \label{fig:repre_l_NBA} %% label for second subfigure
    \includegraphics[scale = 0.15]{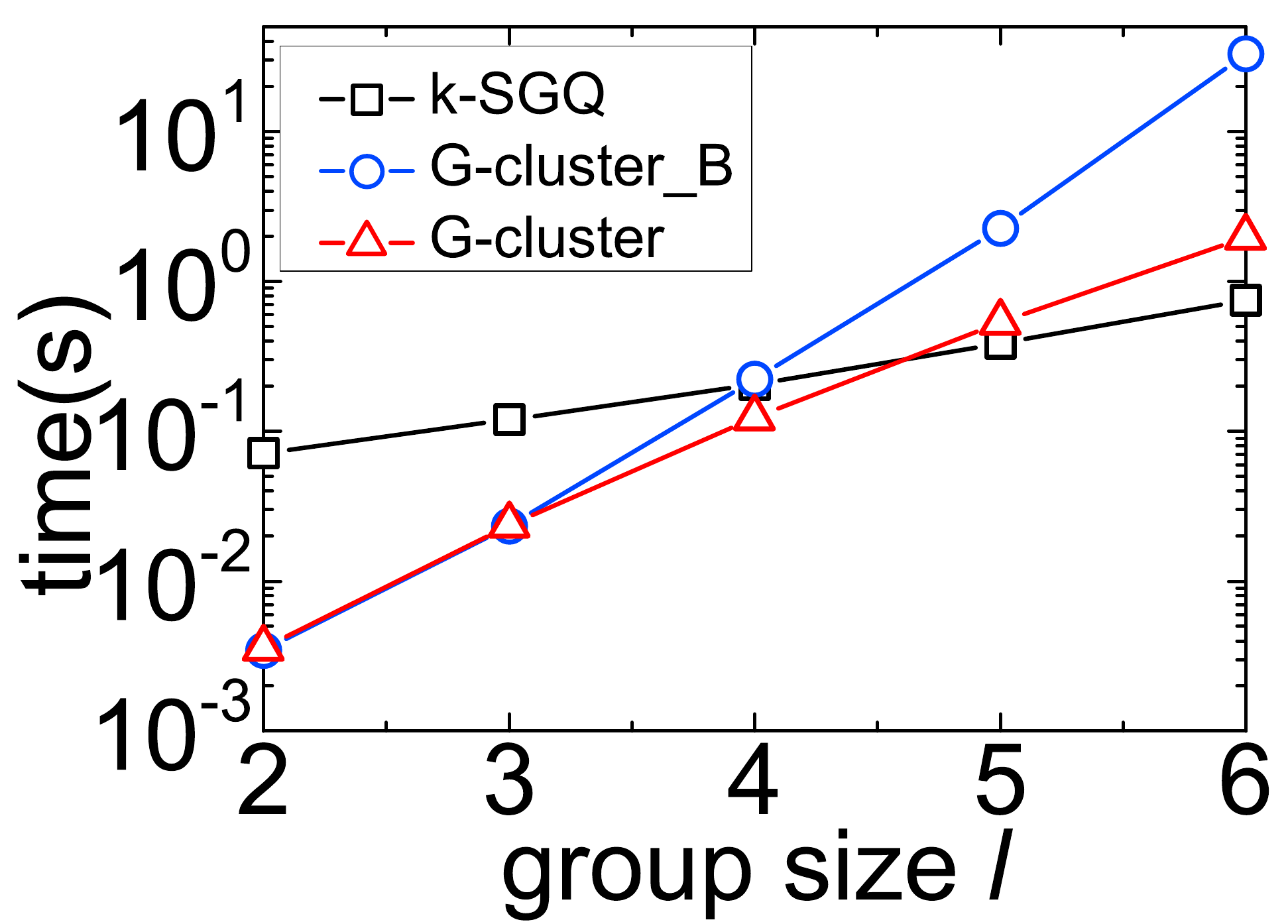}}
    \hspace{-0.15in}
  \subfigure[Varying $n$]{
    \label{fig:repre_n_NBA} %% label for second subfigure
    \includegraphics[scale = 0.15]{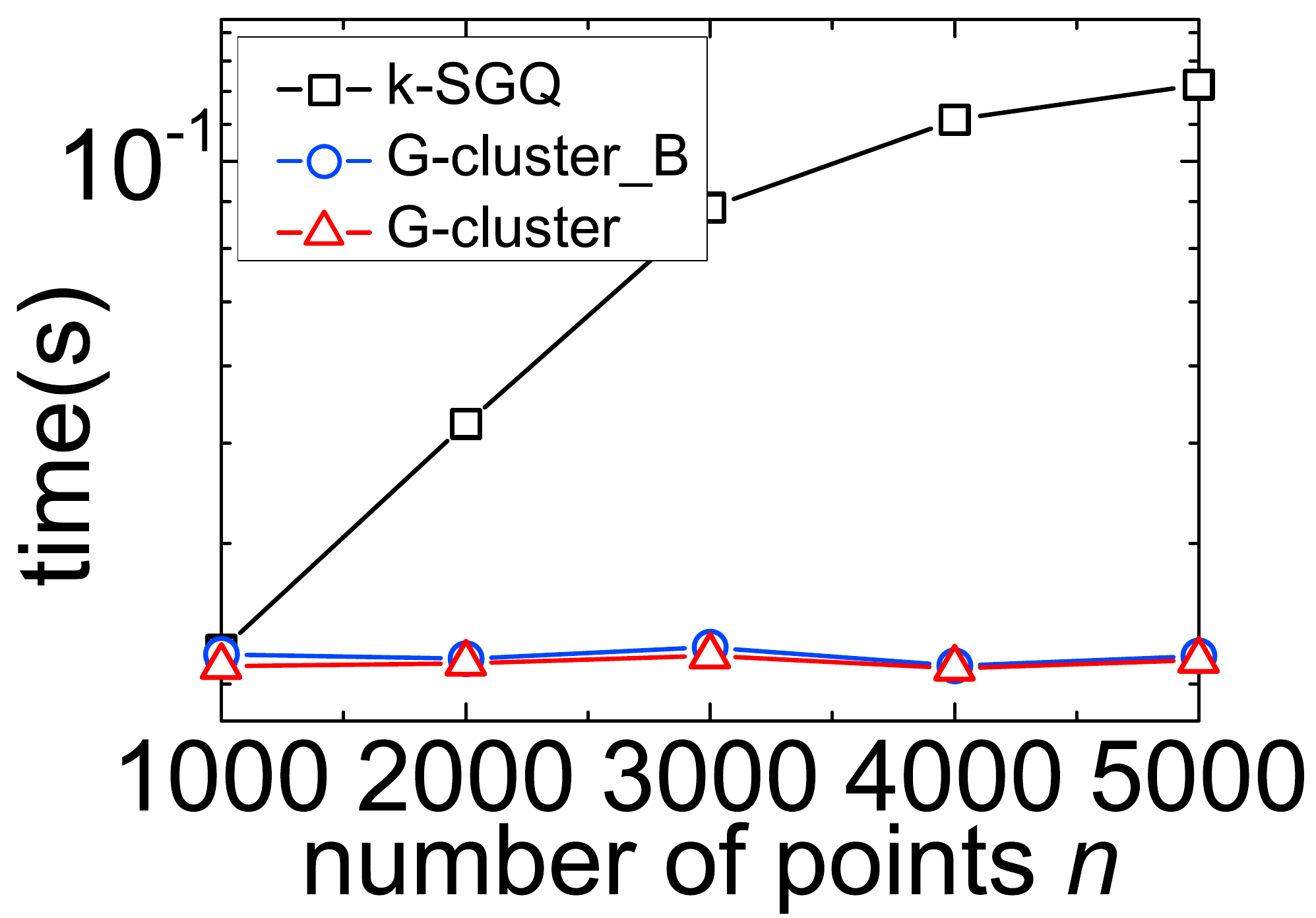}}
  \caption{Representative skyline in NBA datasets of varying parameters.}
  \label{fig:repre_NBA} %% label for entire figure
\end{figure}

Figure \ref{fig:repre_NBA} illustrates the time cost of $k$-SGQ, G-cluster\_B, and G-clustering. The performance with varying $k$, $l$, and $n$ are shown in Figures \ref{fig:repre_NBA}(a)(b)(c) respectively. We can get the similar conclusion with experiments in the synthetic datasets, our G-clustering method performs better when $k$, $l$ are small and $n$ is large. We can see that our method costs more time than the baseline in some cases. Of particular note is that the main superiority of G-clustering is not the efficiency, but the effectiveness --- it returns groups with different tradeoffs thus more representative (presented in Tables \ref{tab:ksgq} and \ref{tab:gclu}).

\begin{table}[ht!]
    \caption{$k$-SGQ}
    \vspace{-5mm}
    \centering
    \begin{center} 
        \label{tab:ksgq}
        \scalebox{0.68}{
        \begin{tabular}{|c|c|c|c|c|c|}
            \hline
            $G_1$&LeBron James&Hakeem Olajuwon&Dirk Nowitzki&Chris Paul&very balanced\\
            \hline
            $G_2$&LeBron James&Hakeem Olajuwon&Dirk Nowitzki&Isiah Thomas&very balanced\\
            \hline
            $G_3$&LeBron James&Hakeem Olajuwon&Dirk Nowitzki&Earl Monroe&very balanced\\
            \hline
            $G_4$&LeBron James&Hakeem Olajuwon&Carmelo Anthony&Chris Paul&very balanced\\
            \hline
            $G_5$&LeBron James&Hakeem Olajuwon&Carmelo Anthony&Isiah Thomas&very balanced\\
            \hline
            $G_6$&LeBron James&Hakeem Olajuwon&Carmelo Anthony&Earl Monroe&very balanced\\
        \hline
        \end{tabular}}
    \end{center} 
    
    \vspace{2mm}
    \caption{RG-skyline}
    \vspace{-7mm}
    \centering
    \begin{center} 
        \label{tab:gclu}
        \scalebox{0.68}{
        \begin{tabular}{|c|c|c|c|c|c|}
            \hline
            $G_1$&Michael Jordan&LeBron James&Kevin Durant&George Gervin&high PTS\\
            \hline
            $G_2$&Pete Myers&Lance Blanks&Luke Hancock&Wayne Turner&high STL\\
            \hline
            $G_3$&Nate Thurmond&Dave Cowens&Wes Unseld&Jerry Lucas&high REB, BLK\\
            \hline
            $G_4$&Michael Jordan&Anthony Davis&Lance Blanks&Allen Iverson&very balanced\\
            \hline
            $G_5$&John Stockton&Magic Johnson&Steve Francis&John Stockton&high REB, AST, STL\\
            \hline
            $G_6$&Michael Jordan&Luke Hancock&Lance Blanks&Pete Myers&high PTS, STL\\
        \hline
        \end{tabular}}
    \end{center} 
    \vspace{-3mm}
\end{table}

Tables \ref{tab:ksgq} and \ref{tab:gclu} present the $k$-SGQ and our RG-skyline respectively. The last column in Table \ref{tab:ksgq} records the dominance score of current group. From the tables we can see that all groups returned by $k$-SGQ are in single pattern, i.e., they are all very balanced (i.e., get satisfactory score in all attributes) and with similar component. Considering that (group-based) skyline is devised to offer different tradeoffs to the users, $k$-SGQ is not competent to represent G-skyline. In contrast, RG-skyline performs much better: Groups in it are pretty different to each other and the coach will have more choices. For example, the coach can choose $G_4$ for a regular competition and $G_1$ for slam dunk competition. To prevent buzzer beater from opponents at the end of the competition, the coach can choose $G_3$ or $G_5$ to strengthen defense.

\section{Conclusion}
\label{sec:conclusion}

In this paper, we proposed several novel structures to address the G-skyline problem. First, we developed a novel algorithmic technique to build MSL using concurrent search and subspace skyline properties.  We investigated all points by searching concurrently in each dimension and for a point, we compared it only with the subspace skyline of current layer. Then, we developed two new methods to find G-skyline by dividing the G-skyline groups into two categories, primary groups and secondary groups. We used a combination queue to enumerate all primary groups and then find the secondary groups in point-wise and unit group-wise algorithms. To mitigate the drawback of too many returned G-Skyline groups, we extended the clustering algorithm from the point level to the group level to propose our G-clustering algorithm, and then use it to establish the RG-skyline by finding out all cluster centers. Experimental results show that the proposed algorithms perform several orders of magnitude better than the baseline method in most situations.

\iffalse
\ifCLASSOPTIONcompsoc
  % The Computer Society usually uses the plural form
  \section*{Acknowledgments}
\else
  % regular IEEE prefers the singular form
  \section*{Acknowledgment}
\fi
This research is supported in part by the AFOSR DDDAS Program under AFOSR grant FA9550-12-1-0240.
\fi

\bibliographystyle{IEEEtran}
% argument is your BibTeX string definitions and bibliography database(s)
% Generated by IEEEtran.bst, version: 1.14 (2015/08/26)

\vspace{-1cm}
\begin{IEEEbiography}[{\includegraphics[width=1in,height=1.25in,clip,keepaspectratio]{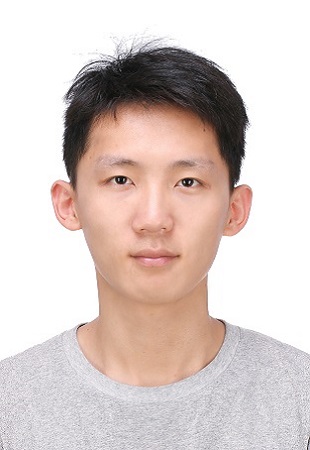}}]{Wenhui Yu}
is a Ph.D. student at Tsinghua University. His research interests include machine learning and data mining. He has published several papers in premier conferences including WWW and CIKM. 
\end{IEEEbiography}
\vspace{-1cm}
\begin{IEEEbiography}[{\includegraphics[width=1in,height=1.25in,clip,keepaspectratio]{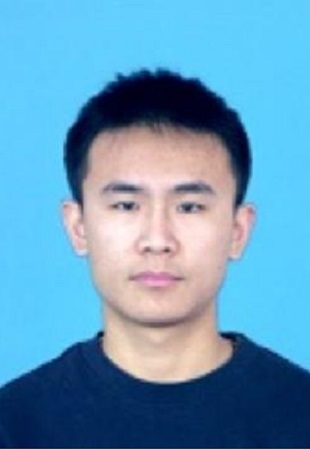}}]{Jinfei Liu}
is a joint postdoctoral research fellow at Emory University and Georgia Institute of Technology. His research interests include skyline queries, data privacy and security, and machine learning. He has published over 20 papers in premier journals and conferences including VLDB, ICDE, CIKM, and IPL.
\end{IEEEbiography}
\vspace{-1cm}
\begin{IEEEbiography}[{\includegraphics[width=1in,height=1.25in,clip,keepaspectratio]{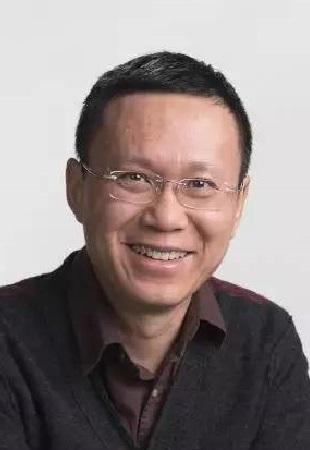}}]{Jian Pei}
is currently Professor at the School of Computing Science, Simon Fraser University, Canada. He is one of the most cited authors in data mining, database systems, and information retrieval. Since 2000, he has published one textbook, two monographs and over 200 research papers in refereed journals and conferences, which have been cited by more than 84,000 in literature. He was the editor-in-chief of the IEEE Transactions of Knowledge and Data Engineering (TKDE) in 2013-2016, is currently the Chair of ACM SIGKDD. He is a Fellow of ACM and of IEEE. In his recent leave-of-absence from the university, he is a Vice President of JD.com and was the Chief Data Scientist, Chief AI Scientist, and a Technical Vice President of Huawei Technologies.
\end{IEEEbiography}
\vspace{-1cm}
\begin{IEEEbiography}[{\includegraphics[width=1in,height=1.25in,clip,keepaspectratio]{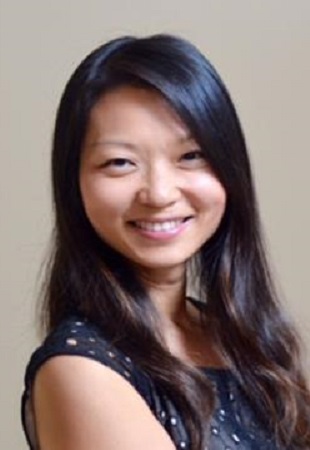}}]{Li Xiong}
is a Professor of Computer Science and Biomedical Informatics at Emory University. She conducts research that addresses both fundamental and applied questions at the interface of data privacy and security, spatiotemporal data management, and health informatics. She has published over 100 papers in premier journals and conferences including TKDE, JAMIA, VLDB, ICDE, CCS, and WWW. She currently serves as associate editor for IEEE Transactions on Knowledge and Data Engineering (TKDE) and on numerous program committees for data management and data security conferences.
\end{IEEEbiography}
\vspace{-1cm}
\begin{IEEEbiography}[{\includegraphics[width=1in,height=1.25in,clip,keepaspectratio]{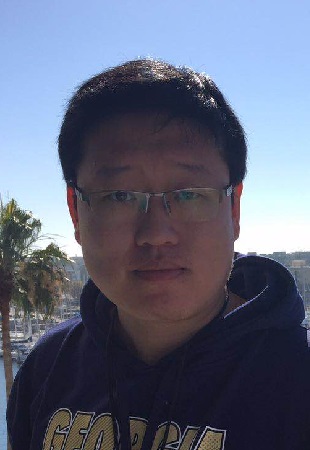}}]{Xu Chen}
is a PhD student in Tsinghua University. His research interests include machine learning, data mining, user behavior analysis. He has published over 10 papers in premier and conferences including SIGIR, WWW, CIKM, WSDM.
\end{IEEEbiography}
\vspace{-1cm}
\begin{IEEEbiography}[{\includegraphics[width=1in,height=1.25in,clip,keepaspectratio]{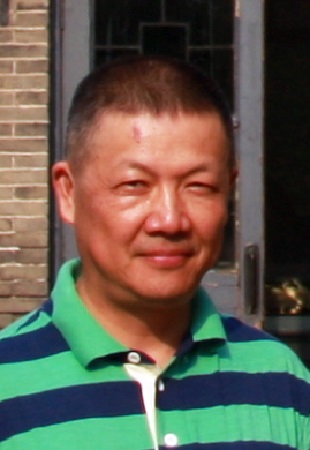}}]{Zheng Qin}
is a Professor in the School of Software and Information Institute of Tsinghua University. He conducts research that addresses both fundamental and applied questions at the formal verification of software, data mining, and computer vision. He has published over 50 papers in premier journals and conferences including WWW, SIGIR, CIKM, WSDM, PCM.
\end{IEEEbiography}

% You can push biographies down or up by placing
% a \vfill before or after them. The appropriate
% use of \vfill depends on what kind of text is
% on the last page and whether or not the columns
% are being equalized.

%\vfill

% Can be used to pull up biographies so that the bottom of the last one
% is flush with the other column.
%\enlargethispage{-5in}
\end{document}